\documentclass[
english,
aps,
a4paper,
prd,
groupedaddress,
nofootinbib,
preprint,
floatfix
]{revtex4-1}

\usepackage{graphicx} 
\usepackage[utf8]{inputenc}
\usepackage{amsmath}
\usepackage{amssymb}
 \usepackage{array}
\usepackage{graphicx}
\usepackage{gensymb}
\usepackage{caption}
\usepackage{subcaption}
\usepackage{float}
\usepackage{geometry}
\usepackage[myheadings]{fullpage}
\usepackage{multirow}
\usepackage{url}
\usepackage{hyperref}
\usepackage{cleveref}
\usepackage[compat=1.1.0]{tikz-feynman}
\usepackage{feynmp}
\usepackage{xcolor}
\usepackage[normalem]{ulem}
\usepackage{tikz}
\usepackage{tikz-3dplot}

\DeclareMathOperator*{\argmin}{arg\,min}
\def\lsim{\raise0.3ex\hbox{$\;<$\kern-0.75em\raise-1.1ex\hbox{$\sim\;$}}}

\newcommand{\AddrLIP}{%
 LIP --- Laborat\'orio de Instrumenta\c{c}\~ao e F\'isica Experimental de Part\'iculas, 
 Departamento de F\'isica,  Escola de Ciências, Universidade do Minho, 
 4701-057 Braga, Portugal}
 
\newcommand{\AddrWurz}{%
    Institut f\"ur Theoretische Physik und Astrophysik,
    Uni W\"urzburg \\
        Campus Hubland Nord, Emil-Hilb-Weg 22,
D-97074 W\"urzburg, Germany
}

\newcommand{\AddrIPPP}{
    Institute for Particle Physics Phenomenology,
    Durham University,
    Durham DH1 3LE,
    United Kingdom
}

\begin{document}


\title{
Exploring Scotogenic Parameter Spaces and Mapping Uncharted Dark Matter Phenomenology with Multi-Objective Search Algorithms
}

\author{Fernando Abreu de Souza}\email{abreurocha@lip.pt}
\affiliation{\AddrLIP}
\author{Nuno Filipe Castro}\email{nuno.castro@fisica.uminho.pt}
\affiliation{\AddrLIP}
\author{Miguel Crispim Rom\~ao}\email{miguel.romao@durham.ac.uk}
\affiliation{\AddrIPPP}
\affiliation{\AddrLIP}
\author{Werner Porod}\email{porod@physik.uni-wuerzburg.de}
\affiliation{\AddrWurz}


\begin{abstract}
We present a novel artificial intelligence approach to explore beyond Standard Model parameter spaces by leveraging a multi-objective optimisation algorithm. We apply this methodology to a non-minimal scotogenic model which is constrained by Higgs mass, anomalous magnetic moment of the muon, dark matter relic density, dark matter direct detection, neutrino masses and mixing, and lepton flavour violating processes. Our results successfully expand on the phenomenological realisations presented in previous work. We compare between multi- and single-objective algorithms and we observe more phenomenologically diverse solutions and an improved search capacity coming from the former.
We use novelty detection to further explore sparsely populated regions of phenomenological interest. These results suggest a powerful search strategy that combines the global exploration of multi-objective optimisation with the exploitation of single-objective optimisation.
\end{abstract}
\maketitle

\section{Introduction}

The Standard Model (SM) of particle physics has turned out to be highly successful in describing phenomena from low energies up to the TeV scale. The Higgs boson, which was discovered in 2012 \cite{ATLAS:2012yve,CMS:2012qbp} has been the last missing particle predicted by the SM.
After this discovery, the particle physics community is eager to discover new phenomena that would imply physics beyond the SM. This expectation rests mainly on two pillars with strong experimental support: the observation of non-baryonic dark matter as well as neutrino masses and neutrino oscillations. The elementary particle responsible for dark matter would be neutral and stable and, to be consistent with structure formation, it must behave as cold dark matter.

A model, addressing both problems, was proposed by E.~Ma \cite{Ma:2006km}, the so-called scotogenic model. This model features three neutral fermions and an additional scalar SU(2)$_L$ doublet beside the usual SM particles. It is assumed that all additional particles are odd with respect to a postulated $Z_2$ symmetry in contrast to the SM particles which are even. In this way, one obtains a dark matter (DM) candidate and neutrino masses are generated at the one-loop level. 
After the first works on minimal scotogenic realisation \cite{Ma:2006km,Toma:2013zsa,Vicente:2014wga,Fraser:2014yha,Baumholzer:2019twf}, more complex models have emerged in recent years, studied mainly at the level of dark matter phenomenology and lepton flavour violating (LFV) observables \cite{Toma:2013zsa,Rocha-Moran:2016enp, Avila:2019hhv,Ahriche:2020pwq,DeRomeri:2021yjo,Boruah:2021ayj} but also for studies of future colliders, e.g.~a muon collider \cite{Liu:2022byu}.
In \cite{Restrepo:2013aga} a systematic classification of scotogenic models has been presented. One of these models received particular attention \cite{Esch:2014jpa,Sarazin:2021nwo}, the so-called `T1-2-A' model, where the SM is extended by a scalar doublet, a scalar singlet, a fermionic Dirac doublet, and a fermionic singlet. There was also the hope that this model could also explain possible deviations in the anomalous magnetic moment of the muon $a_\mu$ \cite{Muong-2:2006rrc,Aoyama:2020ynm,Muong-2:2021ojo}. However, it has then been shown in \cite{Alvarez:2023dzz} that one needs an additional singlet fermion to consistently explain DM, neutrino data and $a_\mu$ while at the same time being consistent with bounds from the non-observation of flavour violation in the charged lepton sector. In ref.~\cite{Alvarez:2023dzz} a Markov Chain Monte Carlo (MCMC) study of this extended model was performed to classify various regions in the parameter space of phenomenological interest which then triggered additional DM studies \cite{Eisenberger:2023tgn}. In ref.~\cite{Alvarez:2023dzz} some parameters were restricted by hand to get a faster convergence of the MCMC runs.

In recent years, Artificial Intelligence (AI) and Machine Learning (ML) have gained considerable interest and traction to solve many data- and computational-intensive tasks in High Energy Physics (HEP)~\cite{hepmllivingreview}, becoming part of the computational toolset of HEP researchers. While the main application of AI/ML to HEP has been leverage the large amounts of data collected by experiments, recent developments have shown that AI/ML can almost trivialise scanning highly constrained multidimensional Beyond the Standard Model (BSM) parameter spaces.

Early attempts at using AI/ML for BSM studies have relied on supervised~\cite{Caron:2016hib,Ren:2017ymm,Staub:2019xhl,Kronheim:2020vct,Caron:2019xkx,Goodsell:2022beo,AbdusSalam:2024obf} and generative~\cite{Hollingsworth:2021sii,Baretz:2023mra} methods that rely on amassing a large amount of valid points before being of any practical use. This is particularly relevant if the calculation of observables is time intensive e.g. of order of seconds. A more promising  alternative was presented by us in~\cite{deSouza:2022uhk}, where AI/ML black-box search algorithms were shown to quickly converge to the subregion of BSM parameter spaces that are phenomenologically viable without any prior requirements of data collection and annotation. This methodology has since been expanded to include a novelty reward to boost exploration~\cite{Romao:2024gjx} and also to find points that maximise the likelihood against experimental data~\cite{Basiouris:2024qfe}. Finally, in~\cite{Diaz:2024yfu,Diaz:2024sxg} it has been adapted with an explicit ranking methodology to address the presence of multiple constraints, which is also one of the focuses of this work.

In this paper, we use AI/ML search algorithms to explore the highly constrained parameter space of a scotogenic model that provides an explanation for the muon anomalous magnetic moment, neutrino data, Higgs mass, dark matter relic density, all while agreeing with bounds from searches for charged lepton flavour violation and from dark matter direct detection, as well as self-consistency requirements such as perturbativity of the underlying couplings.
Our study presented in this work expands on the original~\cite{Alvarez:2023dzz}, where a Monte Carlo Markov Chain scan was performed, by expanding the size of the parameter and assessing the impact of different choices of parametrisations for the neutrino data, in order to increase the scan coverage and generality. Such increase in the problem complexity is only possible to tackle due to the high exploration capabilities of the AI/ML search algorithms employed, which in turn will allow us to find new phenomenological realisations and consequences of the model that were overlooked in the first study.

This paper is organised as follows: we briefly summarize the main
features of the model in \cref{sec:model}. 
We then present our scan strategy in \cref{sec:methodology}.
The corresponding scan results are presented in \cref{sec:scans} whereas the novel phenomenological aspects are discussed in \cref{sec:pheno}. Finally, we draw our conclusions in \cref{sec:conclusion}. Moreover we collect supplementary material in the appendices.

\section{Model}
\label{sec:model}

In this section, we briefly describe the model introduced in~\cite{Alvarez:2023dzz} to define the parameter space and to present the relevant features that will be discussed later. The model is an extension of the `T1-2-A' scotogenic model~\cite{Esch:2014jpa,Sarazin:2021nwo} to which we add an extra fermion singlet. 
The states beyond the SM are presented in~\cref{tab:bsm_states}, where all new states are odd under a $Z_2$-symmetry while the SM states are $Z_2$-even, allowing for a stable DM candidate while neutrino masses are generated at one-loop level.
The additional singlet allows for a rank three neutrino mass matrix implying all neutrino masses to be non-vanishing. 
Additionally, the presence of this new singlet fermion allows for a greater freedom in the couplings between new and SM states, which is leveraged to explain the anomalous magnetic moment of the muon while preventing too large branching ratios for charged lepton flavour violating decays. 
{\renewcommand{\arraystretch}{0.9}
\begin{table}[H]
    \centering
    \begin{tabular}{ccccccc}
        \hline\hline
        \multicolumn{1}{c}{}&\multicolumn{4}{c}{Fermions}&\multicolumn{2}{c}{Scalars}\\
         & $\Psi_1$ & $\Psi_2$ & $F_1$ & $F_2$ & $\eta$ & $S$  \\\hline
         $SU(2)_L$& $\mathbf{2}$ & $\mathbf{2}$ & $\mathbf{1}$ & $\mathbf{1}$ & $\mathbf{2}$ & $\mathbf{1}$ \\
         $U(1)_Y$ & $-1$ & $1$  & $0$ & $0$ & $1$ & $0$ \\\hline\hline
    \end{tabular}
    \caption{Beyond the Standard Model states of the model. All fields are colourless and are odd under a $Z_2$-symmetry, while the Standard Model fields are even.}
    \label{tab:bsm_states}
\end{table}
}

\subsection{The scalar sector}
\label{Sec:ScalarSector}

From~\cref{tab:bsm_states} we see that the scalar sector of the model is extended beyond the SM Higgs field, $H$, to include an additional real singlet $S$, and a $SU(2)_L$ doublet $\eta$. Their charges are given in \cref{tab:bsm_states}. As the SM Higgs acquires a vacuum expectation value (VEV), triggering the electroweak symmetry breaking (EWSB), the doublets components can be represented in the basis
\begin{align}
	H ~=~ \begin{pmatrix} G^+ \\ \frac{1}{\sqrt{2}} \big[ v + h^0 + i G^0 \big] \end{pmatrix}, \qquad
	\eta ~=~ \begin{pmatrix} \eta^+ \\ \frac{1}{\sqrt{2}} \big[ \eta^0 + i A^0 \big] \end{pmatrix} \, ,
\end{align}
where $h^0$ is the SM Higgs boson, $G^0$ and $G^+$ are the would-be Goldstone bosons, and $v = \sqrt{2} \langle H \rangle \approx 246$ GeV the SM Higgs field VEV. Furthermore, $\eta^0$ and $A^0$ are $CP$-even and $CP$-odd neutral scalars, and $\eta^+$ is a charged scalar. We emphasise that $S$ and $\eta$ do not obtain a VEV due to (and preserving) the assumed $Z_2$-symmetry.

The scalar potential is given by
\begin{align}
    \begin{split}
	V_{\rm scalar} ~=&~ M_H^2 \big| H \big|^2 + \lambda_H \big| H \big|^4 + \frac{1}{2} M_S^2 S^2 + \frac{1}{2} \lambda_{4S} S^4 + M_{\eta}^2 \big| \eta \big|^2 + \lambda_{4\eta} \big| \eta \big|^4 \\ &~+ \frac{1}{2} \lambda_S S^2 \big| H \big|^2  +  \frac{1}{2} \lambda_{S\eta} S^2 \big| \eta \big|^2+\lambda_{\eta} \big| \eta \big|^2 \big| H \big|^2 
		+ \lambda'_{\eta} \big| H \eta^{\dag} \big|^2 \\ &~+ \frac{1}{2} \lambda''_{\eta} \Big[ \big( H \eta^{\dag} \big)^2 + {\rm h.c.}\Big] + \alpha \Big[ S H \eta^{\dag} + {\rm h.c.} \Big] \, ,
    \end{split}
	\label{Eq:ScalarPotential}
\end{align}
where first two terms relate to the SM Higgs doublet, $H$, which below the symmetric phase set the Higgs boson mass $m^2_{h^0} = -2 M_H^2 = 2 \lambda_H v^2$, which is experimentally measured at $m_{h^0} \simeq 125$ GeV. Furthermore, we assume for simplicity that $\lambda''_{\eta}$ and $\alpha$ are real.

From the potential, one can immediately infer that the mass matrix of the neutral scalars in the basis $\{ S, \eta^0, A^0 \}$ after EWSB is
\begin{equation}
    {\cal M}_{\phi}^2 ~=~ \begin{pmatrix} M^2_S + \frac{1}{2} v^2 \lambda_S & v \alpha & 0 \\ v \alpha & M^2_{\eta} + \frac{1}{2}v^2 \lambda_L & 0 \\ 0 & 0 & M^2_{\eta} + \frac{1}{2} v^2 \lambda_A \end{pmatrix} \, .
    \label{eq:scalar_mass_matrix}
\end{equation}
Here we have defined $\lambda_{L,A} = \lambda_{\eta} + \lambda'_{\eta} \pm \lambda''_{\eta}$. The mass eigenstates are obtained by a basis rotation
\begin{align}    
\big( \phi^0_1, \phi^0_2, A^0 \big)^T ~=~ U_{\phi} \, \big( S, \eta^0, A^0 \big)^T \,,
\label{eq:def_Uphi}
\end{align}
where $m_{\phi^0_1} < m_{\phi^0_2}$. 

\subsection{The fermion sector}
\label{Sec:fermionSector}

As for the fermions, the Lagrangian for the additional states listed in \cref{tab:bsm_states} reads
\begin{align}
\begin{split}
	{\cal L}_{\rm fermion} ~=~& i  \Big( \overline{\Psi}_j \sigma^{\mu}D_{\mu} \Psi_j + \frac{1}{2} \overline{F}_j \sigma^{\mu} \partial_{\mu} F_j \Big) 
	- \frac{1}{2} M_{F_{ij}} F_i F_j \\ 
	&  - M_{\Psi} \Psi_1 \Psi_2 - y_{1i} \Psi_1 H F_i - y_{2i} \Psi_2 \tilde H F_i \\
	& - g_{\Psi}^k \Psi_2 L_k S - g_{F_j}^k \eta L_k F_j - g_R^k e^c_k \tilde \eta \Psi_1 + \mathrm{h.c.}
	\label{eqn:fermion_lagrangian}
\end{split}
\end{align}
with $i,j=1,2$ running over the two new singlet states and $k=1,2,3$ running over the family indices. $L_k$ and $e^c_k$ denote the left-handed and right-handed SM leptons, respectively. 
We also have introduced the notation $\tilde \phi = i \sigma_2 \phi^*$ for $\phi = H,\eta$. Without loss of generality, we work in a basis where $M_{F_{12}} = 0$ and we impose $|M_1| \le |M_2|$, where we have simplified the notation by setting $M_i = M_{F_{ii}}$ for $i=1,2$. Finally, we adopt the phase convention in which the components of the fermion $SU(2)_L$ doublets read $\Psi_1 =(\Psi^0_1, \Psi^-_1)$ and $\Psi_2 =(\Psi^+_2, -\Psi^0_2)$.

After EWSB, we obtain a massive charged Dirac particle $\Psi^+$ with a mass $M_\Psi$, alongside four neutral Majorana fermions. The mass matrix for the neutral fermions, arranged in the basis $\{F_1, F_2, \Psi^0_1, \Psi^0_2\}$, is expressed as
\begin{align}
{\cal M}_{\chi^0} ~=~ 
\begin{pmatrix}
M_1 & 0 & \frac{v}{\sqrt{2}} \, y_{11} & \frac{v}{\sqrt{2}} \, y_{21} \\
0 & M_2 & \frac{v}{\sqrt{2}} \, y_{12} & \frac{v}{\sqrt{2}} \, y_{22} \\
\frac{v}{\sqrt{2}} y_{11} & \frac{v}{\sqrt{2}} \, y_{12} & 0 & M_\Psi \\
\frac{v}{\sqrt{2}} y_{21} & \frac{v}{\sqrt{2}} y_{22} & M_\Psi &0 
\end{pmatrix} \, ,
\label{eq:fermion_mass_matrix}
\end{align}
which can be diagonalised by a unitary matrix, $U_\chi$, which defines the eigenmass basis $\{\chi_1^0,\chi_2^0,\chi_3^0,\chi_4^0\}$ via
\begin{align}
    \text{diag}\big(m_{\chi^0_1}, m_{\chi^0_2}, m_{\chi^0_3}, m_{\chi^0_4}\big) ~=~ U_\chi {\cal M}_{\chi^0} U^{-1}_\chi \, ,  
\label{eq:def_Uchi}
\end{align}
and where we will use the convention $m_{\chi^0_i} \le m_{\chi^0_j}$ for $i<j$.

\subsection{Neutrino masses}
\label{Sec:NuMasses}

The main novel feature of the model presented in~\cite{Alvarez:2023dzz} and summarised herein is the existence of a second singlet fermion. This extra degree of freedom allows for a rank three neutrino mass matrix implying that all three active neutrinos are massive. After EWSB, in the mass eigenbases a Majorana mass term is generated at the one-loop level via the diagram
\begin{equation} 
    \feynmandiagram[small,layered layout,baseline=(d.base),horizontal = f2 to f3] {
f2[particle=\(\nu_i\)] -- [fermion]b[dot]  -- [ edge label'= \(\chi^0_k\)] d[dot] -- [anti fermion] f3[particle=\(\nu_j\)],
b-- [scalar, half left,looseness=1.55, edge label= \(\phi^0_n\)] d ;
}; \quad \equiv \quad \overline{\nu_j^c} \, \big({\cal M}_{\nu} \big)_{ji} \, \nu_i \, ,
\label{eq:feynman_mnu}
\end{equation}
and the neutrino mass matrix, $\mathcal{M}_\nu$, can be expressed as
\begin{equation} \label{Eq:mnu}
    {\cal M}_\nu ~=~ \mathcal{G}^T \, M_L \, \mathcal{G} \,.
\end{equation}
where the matrix $\mathcal{G}$ contains the couplings defined in Eq.\ \eqref{eqn:fermion_lagrangian} ordered as
\begin{align}
    {\cal G} ~=~ \begin{pmatrix} 
    g_{\Psi}^1 & g_{\Psi}^2 & g_{\Psi}^3 \\
    g_{F_1}^1 & g_{F_1}^2 & g_{F_1}^3 \\
    g_{F_2}^1 & g_{F_2}^2 & g_{F_2}^3 
    \end{pmatrix} \,,
\label{eq:G_matrix}
\end{align}
and $M_L$ is a $3 \times 3$ symmetric matrix encoding the loop function and the neural scalar and fermion mixings, c.f.~\cref{eq:def_Uphi,eq:def_Uchi}. The 
corresponding formula is given in \cref{sec:neutrino_loop_functions}.
Although the entries for the neutrino mass matrix can be computed using the parameters listed in the Langrangian~\cref{eqn:fermion_lagrangian}, one could also use the so-called Casas-Ibarra parametrisation~\cite{Casas:2001sr, Basso:2012voo} in which the couplings in~\cref{eq:G_matrix} are obtained in terms of the neutrino data~\cite{deSalas:2020pgw, Gonzalez-Garcia:2021dve}, see \cite{Alvarez:2023dzz} for details. 
In~\cref{sec:parameter_space_and_constraints} we will discuss these two realisations of the parameter space of the model, which we will explore in detail in our scans in~\cref{sec:scans}, which methodology we discuss in the next section.

\subsection{$a_\mu^\text{BSM}$ and LFV Observables}
\label{Sec:g-2}

In general, BSM contributions to anomalous magnetic moment can be written as one-loop decays in the form $l_{i} \rightarrow l_{j}\gamma$ when $i=j$, while if $i>j$ we get contributions to charged LFV processes. These contributions can be represented by an effective operator $c_{R}^{ij} l_{i} \sigma^{\mu\nu} P_R l_jF^{\mu\nu}$, where $c_{R}$ is the corresponding Wilson coefficient (again, when $i=j$, we get contributions to anomalous magnetic moment, and when $i>j$ we get contributions to charged LFV processes), and it comes from the effective operator $\mathcal{O}_{eB} = (\bar{L}\sigma^{\mu\nu}e_{R})HB^{\mu\nu}$ defined in~\cite{Grzadkowski:2010es}, before EWSB. 

\begin{figure}[H]
    \centering
    \begin{tikzpicture}[baseline=(a1.base)]  
    
        \begin{scope}
            \begin{feynman}
                \vertex (a1) {\(L_i\)};
                \vertex[right=1.5cm of a1, dot] (a2) {};
                \vertex[below=0.35cm of a2] (c2) {\(g^{i}_{\Psi}\)};
                \vertex[right=2cm of a2, label=\(M_{\Psi}\)] (a3);
                \vertex[right=2cm of a3, dot] (a4) {};
                \vertex[below=0.35cm of a4] (c4) {\(g^{j}_{R}\)};
                \vertex[right=1.5cm of a4] (a5){\(e^c_{j}\)};
                \vertex[right=0.35cm of a3] (a6) {};
                \vertex[above=0.14cm of a6] (a7) {};

                \vertex[above=2cm of a3, dot] (b1) {};
                \vertex[above right=0.35cm of b1] (c1) {\(\alpha\)};
                \vertex[above=1.5cm of b1] (b2) {\(H^{\dagger}\)};
                \vertex[below=1.64cm of a7] (b3) {\(V\)};

                \diagram[horizontal=a1 to a5, layered layout] {
                    (a1) -- [fermion] (a2) -- [anti fermion, edge label'=\(\Psi_2\)] (a3) -- [fermion, edge label'=\(\Psi_1\)] (a4) -- [anti fermion] (a5),
                    (a2) -- [insertion=1.0] (a3),
                    (b2) -- [scalar] (b1) -- [scalar, quarter right, edge label'=\(S\)](a2),
                    (b1) -- [scalar, quarter left, edge label=\(\eta\)] (a4),
                    (a7) -- [boson] (b3),
                };
            \end{feynman}
        \end{scope}
        \quad
        \begin{scope}[xshift=8cm]  
            \begin{feynman}
                \vertex (a1) {\(L_i\)};
                \vertex[right=1.5cm of a1, dot] (a2) {};
                \vertex[below=0.35cm of a2] (c2) {\(g^{i}_{F_k}\)};
                \vertex[right=2cm of a2, dot] (a3) {};
                \vertex[above=0.35cm of a3] (c3) {\(y^{*}_{1k}\)};
                \vertex[right=2cm of a3, dot] (a4) {};
                \vertex[below=0.35cm of a4] (c4) {\(g^{j}_{R}\)};
                \vertex[right=1.5cm of a4] (a5){\(e^c_{j}\)};
                \vertex[above=2cm of a3] (a7) {};
                \vertex[right=1.2cm of a7] (c7) {\(\eta\)};

                \vertex[below=1.5cm of a3] (b1) {\(H^{\dagger}\)};
                \vertex[above=1.5cm of a7] (b2) {\(V\)};

                \diagram[horizontal=a1 to a5, layered layout] {
                    (a1) -- [fermion] (a2) -- [anti fermion, edge label'=\(F_k\)] (a3) -- [fermion, edge label'=\(\Psi_1\)] (a4) -- [anti fermion] (a5),
                    (b1) -- [scalar] (a3),
                    (a2) -- [scalar, half left, looseness=1.75] (a4),
                    (a7) -- [boson] (b2),
                };
            \end{feynman}
        \end{scope}

    \end{tikzpicture}
\caption{Feynman diagrams depicting the main contributions to the anomalous magnetic moment of the muon (when $i=j=2$) and LFV decays (when $i>j$). Diagram on the left is $\sim \alpha \hspace{0.25em} g_{\Psi} \hspace{0.25em} g_R$ while diagram on the right is $\sim y_{1} \hspace{0.25em} g_F \hspace{0.25em} g_R$.
The $V$ boson line represents the couplings to the $SU(2)_L\times U(1)_Y$ gauge bosons.
}
\label{fig:feynman_g-2}
\end{figure}
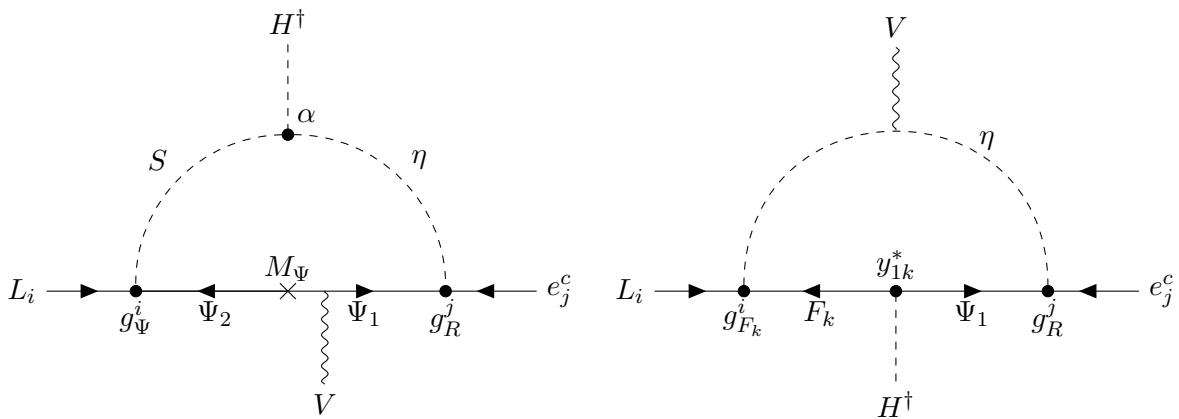

In the model studied in this work, the addition of heavy fermion and scalars presented in~\cref{tab:bsm_states} open new possibilities for decays of the form $l_{i} \rightarrow l_{j}\gamma$ at the one-loop level. The main new contributions to the anomalous magnetic moment of the muon ($a^{\text{BSM}}_{\mu}$) come in the form of the diagrams presented in~\cref{fig:feynman_g-2}. The diagram on the left depends on $g_{\Psi}$ and $\alpha$ and the diagram on the right depends on $g_F$ and $y_1$, and all of these couplings affect the neutrino masses as well, as shown in~\cref{Eq:mnu},~\cref{eq:G_matrix}.

One of the main challenges in finding viable solutions in this scotogenic model parameter space, and an appropriate motivation to apply our methodology later introduced in~\cref{sec:methodology}, comes from the interplay between the constraints coming from the neutrino sector, the anomalous magnetic moment of the muon and the charged LFV decays, in particular the highly constrained $\mu \rightarrow e \gamma$ with an upper limit to its branching ratio of $4.2 \times 10^{-13}$~\cite{MEG:2016leq}. Although we need a significant contribution to the magnetic moment of the muon to accommodate the experimental value of $a^{\text{BSM}}_{\mu} = (251 \pm 59) \times 10^{-11}$~\cite{Aoyama:2020ynm, Muong-2:2021ojo}, at the same time these contributions are likely to be correlated to the generation of neutrino masses at one-loop level through the couplings $g_{\Psi}$, $g_F$, $\alpha$ and $y_{1,2}$ with additional correlation with the strongly constrained charged LFV processes. 

While the constraint coming from $a^{\text{BSM}}_{\mu}$ pushes $g^{2}_{F_k}$ and/or $g^{2}_{\Psi}$ together with $g^{2}_{R}$ to larger absolute values, through the diagrams in~\cref{fig:feynman_g-2} for $i=j=2$, these parameters are also pushed to the opposite direction by the LFV decays constraints, in the same diagrams in~\cref{fig:feynman_g-2} for $i\neq j$. At the same time, the sizeable value for neutrino mixings pushes $g^{i}_{F_k}$ and/or $g^{i}_{\Psi}$ to a similar scale across all flavours, i.e., $g^{1}_{F_k} \sim g^{2}_{F_k} \sim g^{3}_{F_k}$ and/or $g^{1}_{\Psi} \sim g^{2}_{\Psi} \sim g^{3}_{\Psi}$. This tension is an important aspect of this model which motivates the development of an adjustment to our methodology where we introduce a hierarchy in the constraints, as discussed later in~\cref{sec:hierarchy}.

\section{Scan Strategy and Methodology}
\label{sec:methodology}

In this section we discuss the AI scan methodology and the two formulations of the parameter space, with and without Casas-Ibarra parametrisation.

\subsection{The black-box Optimisation Methodology}

We make use of the AI black-box optimisation methodology introduced in~\cite{deSouza:2022uhk}, and expanded in~\cite{Romao:2024gjx}, with the goal of enhancing the exploratory capacity when scanning the parameter space $\mathcal{P}$. The method relies on exploiting the valuable information of \textit{how far} a given point $\theta \in \mathcal{P}$ is from being valid. This information can be conveniently encoded in a constraint function, $C$, defined as

\begin{align}
    C(\mathcal{O}) = \max{(0, - \mathcal{O}+\mathcal{O}_{\textrm{LB}}, \mathcal{O}-\mathcal{O}_{\textrm{UB}})},
    \label{eq:constraint_func}
\end{align}
where $\mathcal{O}$ is the predicted value for the observable considered and $\mathcal{O}_{\textrm{LB}}$ and $\mathcal{O}_{\textrm{UB}}$ are the experimental (or theoretical, when applied) lower and upper bounds, respectively, for the observable considered. One must keep in mind that $\mathcal{O}$ depends solely on the parameter space point $\theta$, that is, $\mathcal{O} = \mathcal{O}(\theta)$, therefore, one can write $C(\mathcal{O}) = C(\mathcal{O}(\theta)) = C(\theta)$, in other words, $C$ also depends only on $\theta$. Regarding the possible values for $C$ two possible scenarios can occur:
\begin{enumerate}
    \item $C(\mathcal{O}) = 0 \iff \mathcal{O}_{\textrm{LB}} \leq \mathcal{O} \leq \mathcal{O}_{\textrm{UB}}$. 
    
    This situation corresponds to the desired outcome when the predicted value for the observable $\mathcal{O}$ is in agreement with the constraint considered for $\mathcal{O}$.
    \item $C(\mathcal{O}) > 0 \iff \left\{\begin{array}{c}
         \mathcal{O} < \mathcal{O}_{\textrm{LB}} ; \\ \mathcal{O} > \mathcal{O}_{\textrm{UB}}. \end{array}\right. $
         
    This corresponds to the outcome when the predicted value for the observable $\mathcal{O}$ is not in agreement with the constraint considered for $\mathcal{O}$.
\end{enumerate}

 And so, our goal is to find the set $\mathcal{V}$ of valid points in the parameter space which can be defined as
 \begin{align}\label{eq:valid-region}
    \mathcal{V} = \left\{ \theta^*: \theta \in \mathcal{P} , \quad C(\theta) = 0 \right\},
\end{align}
or equivalently, given that $C(\theta) = 0$ correspond to the minima of $C(\theta)$,
 \begin{align}
    \mathcal{V} = \left\{ \theta^*: \theta \in \mathcal{P} , \quad \theta^* = \argmin C(\theta) \right\},
    \label{eq:optimisation}
\end{align}
where $\argmin$ is the value of the argument that minimises the function. This way,~\cref{eq:optimisation} successfully reframes the validation task as an optimisation problem in which finding the valid points $\theta^*$ in the parameter space $\mathcal{P}$ is equivalent to minimising $C(\theta)$. It is exactly this change in approach that allows us to leverage the power of AI optimisation algorithms to tackle the challenging task of scanning highly constrained multidimensional BSM parameter spaces. In previous work~\cite{deSouza:2022uhk}, we showcased how four different classes of AI search algorithms perform the exploration task, exhibiting different trade-off between exploration and exploitation. In this work we want to assess whether a multi-objective approach, where the algorithm searches for promising points by populating a Pareto front drawn from all the constraints, is better suited for the task than a single-objective search, where all the constraints are aggregated into a single value to be minimised. To this effect, we will compare the Covariance Matrix Adaptation Evolution Strategy (CMA-ES)~\cite{Hansen2006}, the fastest converging algorithm in~\cite{deSouza:2022uhk} that excels at single objective objective optimisation problems, with the Non-dominated Sorting Genetic Algorithm - III (NSGA-III)~\cite{6595567,6600851}, the state-of-the-art multi-objective genetic algorithm, which we present next.

\subsection{Single-objective Optimisation with the Covariance Matrix Adaptation Evolution Strategy (CMA-ES)}

When facing multiple constraints, one can simply take the constraint function $C(\theta)$ of a parameter space point $\theta$ as the sum of the constraint functions for each observables associated which each constraint, that is
\begin{align}\label{eq:single-ojective-loss}
    C(\theta) = C(\mathcal{O}_1) \wedge C(\mathcal{O}_2) \wedge \ldots \wedge C(\mathcal{O}_n) = \sum_k^n C(\mathcal{O}_k),
\end{align}
where $n$ is total number of constraints considered. This scenario corresponds to a single objective optimisation problem where there is an implied assumption that all the objectives ($C(\mathcal{O}_k))$ are independent of each other and contribute equally to the total cost function, and it was the scenario previously considered in~\cite{deSouza:2022uhk,Romao:2024gjx}.

In previous applications of the black box optimisation methodology for BSM parameter spaces scans~\cite{deSouza:2022uhk,Romao:2024gjx}, CMA-ES has shown to excel at eagerly exploiting the constraints to quickly converge to valid regions of the parameter space, i.e. to quickly minimise~\cref{eq:single-ojective-loss}. CMA-ES is an evolutionary algorithm, which means it is a population-based algorithm in which an initial population of proposed solutions is iteratively adapted following a selection criterion according to a fitness value associated to each individual. Individuals with the best fitness are selected to generate a new population (the offspring) for the next iteration. The process iterates until either the minimum or some other stopping criterion has been reached. For the context of this work, a population is a set of individual points $\{\theta_k\}$ in the parameter space each with an associated fitness value given by~\cref{eq:single-ojective-loss}.

In CMA-ES at each iteration (generation) $g$, the population is sampled according to a multivariate normal distribution given by 
\begin{align}
    \theta^{(g+1)}_i \sim m^{(g)} + \sigma^{(g)} \mathcal{N}(0, \mathcal{C}^{(g)}) \quad \textrm{for } i = 1, \ldots, \lambda,
    \label{eq:cmaes}
\end{align}
where $\lambda$ is the population size, $m^{(g)}$ is the mean of the distribution at generation $g$, $\sigma^{(g)}$ is a scaling factor, called step size, at generation $g$, and $\mathcal{N}(0, \mathcal{C}^{(g)})$ describes a multivariate normal distribution with mean $0$ and covariance matrix $\mathcal{C}^{(g)}$ at generation $g$. At each generation $g$, the individuals $\theta^{(g+1)}_i$ are ranked according to their fitness values, given by the constraint function $C(\theta^{(g+1)}_i)$. Then, the set of $\mu$ best fitted individuals are used to derive the mean $m^{(g+1)}$, covariance matrix $\mathcal{C}^{(g+1)}$ and step size $\sigma^{(g+1)}$ for the next generation $g+1$. This way, the iterative process of updating the mean will push the algorithm towards the direction of the steepest descent of the fitness function, just like a first-order optimisation method would, while the updates on the covariance matrix, besides dictating the changes in the shape of the distribution of the population, can also be interpreted as an approximation of the local Hessian of the fitness function, just like would be used in a second-order optimisation method.

The fact that CMAE-ES traverses the parameter space using only highly localised information can potentially lead to two problems: stagnation at local minima and reduced capacity to explore the whole parameter space. The first problem is mitigated by the fact that CMA-ES can dynamically increase its step size $\sigma^{(g)}$, if it is getting stuck in a local minima. This is in fact one of the reasons why such a localised algorithm is so powerful at finding global minima. The second shortcoming is specific for our needs as we are not only looking for the global minima of~\cref{eq:single-ojective-loss}, but the region defined by~\cref{eq:valid-region}. This challenge was successfully tackled in~\cite{Romao:2024gjx} where CMA-ES was enhanced with a novelty detection algorithm to force the exploration to map the global minima region.

\subsection{Multi-objective Optimisation with the Non-dominated Sorting Genetic Algorithm - III (NSGA-III)}

The single-objective approach presented above assumes that all the objectives are independent of each other and contribute equally to the total cost function. Although this assumption can make the optimisation process simpler and more straightforward, in the context of BSM observables, the objectives are often correlated or, in the worst case, they can conflict with one other, making the optimisation task more challenging and more prone to get stuck at local minima. 

The multi-objective optimisation approach addresses the problem of conflicting objectives by considering each objective independently, that is
\begin{align}
    C(\theta) = C(\mathcal{O}_1) \wedge C(\mathcal{O}_2) \wedge \ldots \wedge C(\mathcal{O}_n) = \left(C(\mathcal{O}_1), C(\mathcal{O}_2), \ldots, C(\mathcal{O}_n) \right).
\label{eq:multi_constraint}
\end{align}
This way, solutions which improve one objective while simultaneously worsening another are treated as equally viable. In such cases, in terms of the multi-objective optimisation terminology, we would say that one solution does not dominate over the other. It is then said that one solution $\theta_A$ dominates over another solution $\theta_B$ if $\theta_A$ improves at least one objective over $\theta_B$ without worsening any other objective, that is
\begin{align}
\begin{split}
    \forall i \in \{1, \ldots, n \}, \quad C(\mathcal{O}_i(\theta_A)) \leq C(\mathcal{O}_i(\theta_B)) \quad \textrm{and} \quad \exists j \in  {\mathcal{O}_1, \ldots, \mathcal{O}_n} \quad \\ \textrm{such that} \quad C(\mathcal{O}_j(\theta_A)) < C(\mathcal{O}_j(\theta_B)),
\end{split}
\end{align}
where $n$ is the number of objectives. When transitioning from a solution $\theta_B$ to a solution $\theta_A$ where $\theta_A$ dominates over $\theta_B$ it is called a Pareto improvement. And so, a solution is considered Pareto optimal if no Pareto improvement can be made. The set of Pareto optimal solutions is called the Pareto optimal front or non-dominated front.

NSGA-III is a genetic multi-objective optimisation algorithm. Genetic algorithms are a family of evolutionary algorithms where each individual in a population corresponds to a unique genetic representation of the candidate solution to the optimisation problem. At each generation, the best fitted individuals are selected (selection process) to undergo a crossover (mating) process where a new pair of solutions is created by combining the genetic information (genes) of two parent solutions. After crossover, the generated individuals undergo mutation processes in which each gene has a probability to randomly change its values. This process is repeated iteratively until a stop criterion is fulfilled, typically when either convergence or a maximum number of generations is reached. In our context, the genetic representation (genes) of a solution $\theta$ corresponds to the values of its parameters. 

The selection procedure for NSGA-III leverages the idea of non-dominated fronts as follows: Given a parent population $P_i$ with size $N$ at the $i$th generation of the algorithm, an offspring population $Q_i$ of size $N$ is produced via crossover and mutation. The population for the next generation is selected from the combined population $R_i = P_i \cup Q_i$ of size $2N$. Each individual from $R_i$ is sorted into non-dominated fronts according to an iterative process as follows: First, the non-dominated front $F_1$ from $R_i$ is identified and designated rank 1. Then, the individuals in $F_1$ are set aside and the non-dominated front $F_2$ for the remaining individuals $R_i \setminus F_1$ is determined and assigned rank 2. This process is repeated until all individuals in $R_i$ are assigned a rank and the parents for the next generation $P_{i+1}$ are selected according to this ranking.  In the context of our work, convergence is achieved when the first non-dominated front collapses to the origin of the objective space.

\begin{figure}
    \centering
\tdplotsetmaincoords{70}{120}

\begin{tikzpicture}[
    scale=5,
    tdplot_main_coords,
    simplex/.style={very thin, gray},
    reference/.style={blue},
    points/.style={teal}
    ]
	\begin{scope}[thick]
		\draw[->] (0,0,0) -- (1.25,0,0) node[anchor=north east]{$C_1$};
		\draw[->] (0,0,0) -- (0,1.25,0) node[anchor=west]{$C_2$};
		\draw[->] (0,0,0) -- (0,0,1.25) node[anchor=south]{$C_3$};
	\end{scope}

	\draw[simplex] (1,0,0) -- (0,1,0) -- (0,0,1) -- cycle;

	\begin{scope}[gray!50, thin]
		\foreach \x in {0,0.33333333333,...,1.0} {
				\draw (1-\x,0,\x) -- (0,1-\x,\x);
                \draw (1-\x,\x,0) -- (1-\x,0,\x);
                \draw (1-\x,\x,0) -- (0,\x,1-\x);
			}
	\end{scope}

	\foreach \x in {0,1,...,3} {
			\foreach \y in {0,1,...,3} {
					\foreach \z in {0,1,...,3} {
							\pgfmathtruncatemacro{\sumxyz}{\x + \y + \z}
							\ifnum\sumxyz=3 
								\coordinate (pt) at (\x/3, \y/3, \z/3);
								\fill[red] (pt) circle (0.5pt);
							\fi
						}
				}
		}

    \draw[reference] (0,0,0) -- (0,1,0.5) node[right]{$r_1$};
    \fill[points] (0.1, 0.8, 0.5) circle (0.5pt);
    \draw[points] (0.1, 0.8, 0.5) node[above]{$p_1$};
    \draw[points] (0.1, 0.8, 0.5) -- (0.0, 0.8, 0.4);

    \draw[reference] (0,0,0) -- (0.5,0,1) node[above]{$r_2$};
    \fill[points] (1.1, 0.2, 1.0) circle (0.5pt);
    \draw[points] (1.1, 0.2, 1.0) node[above]{$p_2$};
    \draw[points] (1.1, 0.2, 1.0) -- (0.425, 0.0, 0.85);

\end{tikzpicture}
    \caption{Illustration depicting the scheme for constructing the reference directions ($r_i$, as blue lines) from the reference points (red points) in NSGA-III, and how points in the constraint space ($p_i$, as green) are assigned to a reference direction.}
    \label{fig:reference_directions}
\end{figure}
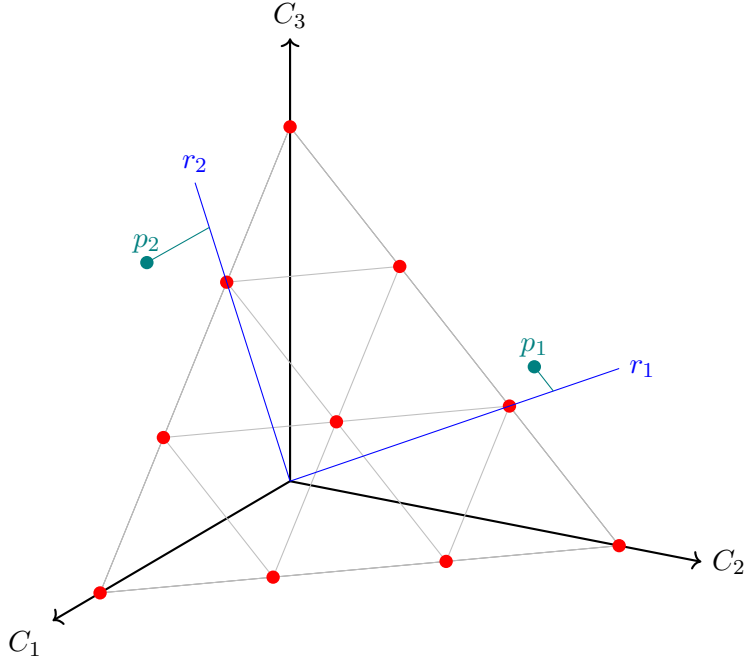

Since the population size is fixed at each generation, in most cases the last rank will exceed the population size $N$ and thus there must be a criterion to select a subset of individuals from the last non-dominated front to fill in the remaining available slots of selected individuals for the next generation. NSGA-III uses previously assigned reference points in the objective space as means of preserving a diversity of solutions in the last non-dominated front. Each individual in the last non-dominated front is associated with the closest reference direction which is defined as the line that connects the origin of the objective space and the reference point, as shown in~\cref{fig:reference_directions}. And so, the individuals from the last non-dominated front are selected according to the criterion of maximizing the diversity in reference directions. This construction forces the algorithm to try to preserve a global picture of the parameter space across each generation by maintaining diversity in the reference directions which in turn is likely to also maintain diversity in the parameter space (since, the functions which are being optimized, the constraint functions $C$ from Eq.\ \eqref{eq:constraint_func}, work as maps from the parameter space to the objective space. And so, diversity in the objective space often means diversity in the parameter space.).

\subsection{The Parameter Space and Constraints\label{sec:parameter_space_and_constraints}}

As explained in~\cref{sec:model}, there are two ways of parametrising the parameter space of the mode: with and without Casas-Ibarra (CI) parametrisation. With CI parametrisation, the neutrino data are inputs, and the Lagrangian parameters related to their origin, i.e. $g_\Psi^k$ and $g_{F_j}^k$, are constrained by perturbativity. Without CI parametrisation, the parameter space is defined by the Lagrangian parameters, which have to fit neutrino data. CI parametrisation simplifies the search for phenomenologically viable regions of the parameter space by reducing the effective dimensionality of the parameter space to the submanifolds where the model is in agreement with neutrino data. In this work, we will explore both parametrisations of the model as to assess whether our methodology is capable of mapping the phenomenologically viable region without the need for simplifications such as using the CI parametrisation.

The parameters used as inputs for the scans for both parametrisations with their allowed ranges as well as the prior mapping are shown in~\cref{tab:parameters}. We notice that ``symlog'' refers to the symmetric log transformation, which allows to map from a single $[0,1]$ interval to the intended range with a logarithmic scale while keeping track of the sign, which would not be possible under a single log transformation.

{\renewcommand{\arraystretch}{0.75}
\begin{table}[H]
	\makebox[\textwidth][c]{
		\begin{tabular}{cccc}
			\hline\hline
			Parameter                                               & Interval                                    & Parametrisation                          & Mapping 
			\\
			\hline
			$\lambda_H$                                             & $[0.1,\ 0.3] $                              & Both                                                & linear     
			\\
			$\lambda_{4S}$, $\lambda_{4\eta}$                       & $[10^{-8},\ 1]$                             & Both                                                & log     
			\\
			$\lambda_{S\eta}$, $\lambda_{S}$,
			$\lambda_{\eta}$, $\lambda_{\eta}'$, $\lambda_{\eta}''$ & $\pm[10^{-8},\ 1]$                          & Both                                                & symlog  
			\\
			$\alpha$ (GeV)                                               & $\pm[1,\ 10^4]$                        & Both                                                     & symlog  
			\\
			$M_S^2$, $M_\eta^2$  (GeV$^2$)                                   & $5\times [10^5,\ 0^6]$             & Both                                                         & linear     
			\\
			$M_{1}$, $M_{2}$ (GeV)                                       & $[100,\ 2000]$                         & Both                                                     & linear     
			\\
			$M_{\Psi}$     (GeV)                                         & $[700,\ 2000]$                         & Both                                                     & linear     
			\\
			$\Re(y_{ij})$, $(i,j=1,2)$                              & $\pm[10^{-8},\ 3]$                          & Both                                                & symlog  
			\\
			$\Im(y_{ij})$, $(i,j=1,2)$                               & $\pm[10^{-8},\ 3]$                          & Both                                                & symlog  
			\\
			$\Re(g^k_R)$, $(k=1,2,3)$                                 & $\pm[10^{-8},\ 3]$                          & Both                                                & symlog  
			\\
			$\Im(g^k_R)$, $(k=1,2,3)$                               & $\pm[10^{-8},\ 3]$                          & Both                                                & symlog  
			\\
                \hline
                $\Re(g^k_\Psi)$, $(k=1,2,3)$                        & $\pm[10^{-8},\ 3]$                          & Non CI                                                & symlog  
			\\
			$\Im(g^k_\Psi)$, $(k=1,2,3)$                        & $\pm[10^{-8},\ 3]$                          & Non CI                                                & symlog  
			\\
			$\Re(g^k_{F_j})$,, $(k=1,2,3)$, $(j=1,2)$                & $\pm[10^{-8},\ 3]$                          & Non CI                                                & symlog  
			\\
			$\Im(g^k_{F_j})$, $(k=1,2,3)$, $(j=1,2)$               & $\pm[10^{-8},\ 3]$                          & Non CI                                                & symlog
                \\
                \hline
                $m_{\nu_1}$                                             & $[10^{-16},\ 10^{-10}]$                      & CI                                               & log     
			\\
			$m_{\nu_2}$                                             & $[\sqrt{m_{\nu_1}^2 + 6.82 \times 10^{-23}},\ \sqrt{m_{\nu_1}^2 + 8.04 \times 10^{-23}}]$ & CI   & linear     
			\\
			$m_{\nu_3}$                                             & $[\sqrt{m_{\nu_1}^2 + 2.435 \times 10^{-21}},\ \sqrt{m_{\nu_1}^2 + 2.598 \times 10^{-21}}]$ & CI & linear     
			\\
			$\theta^{\text{PMNS}}_{12}$                             & $[31.27,\ 35.86] \frac{\pi}{180}$   & CI                                                         & linear     
			\\
			$\theta^{\text{PMNS}}_{13}$                             & $[8.20,\ 8.93] \frac{\pi}{180}$     & CI                                                         & linear     
			\\
			$\theta^{\text{PMNS}}_{23}$                             & $[40.1,\  51.7] \frac{\pi}{180}$    & CI                                                         & linear     
			\\
			$\delta_{\text{CP}}$                                    & $[120,\  369] \frac{\pi}{180}$     & CI                                                          & linear    
			\\
			$\Re(\theta^{\text{R}}_{k})$, $(k=1,2,3)$              & $[0,\ 2 \pi]$               & CI                                                                 & linear     
			\\
			$\Im(\theta^{\text{R}}_{k})$, $(k=1,2,3)$               & $\pm[10^{-8},\ 3]$          & CI                                                                 & symlog  
			\\
			\hline\hline
		\end{tabular}
	}
	\caption{Parameters used as inputs for the scans for both parametrisation and their mapping from the box parameter space to the physical parameter space, see~\cref{sec:implementation_details} for details.}
	\label{tab:parameters}
\end{table}
}

\subsubsection{Non Casas-Ibarra}

We now define the parameter space $\mathcal{P}$ for the scotogenic model without the CI parametrisation. This parametrisation of the model is read directly from the Lagrangian, and we have 
\begin{align}
\begin{split}
    \mathcal{P} = & \left\{ \theta \equiv \{\lambda_H, \lambda_{4S}, \lambda_{4\eta}, \lambda_{S\eta}, \lambda_S, \lambda_{\eta}, \lambda_{\eta}', \lambda_{\eta}'', \right. \\
    & \left. \alpha, M_S^2, M_{\eta}^2, M_1, M_2, M_{\Psi}, \right.\\
    & \left. y_{ij}, g_R^k, g_{\Psi}^k, g_{F_j}^k \}: \theta \in \mathcal{S} \right\},
    \label{eq:parameter_space}
\end{split}
\end{align}
where $\mathcal{S}$ is comprised by the parameter intervals shown in~\cref{tab:parameters} which together define the non-CI case, and $i,j = \{1, 2\}$ runs through the new singlet fermions, $F_i$, and $k = \{1, 2, 3\}$ runs through the SM family indices.
In this case, neutrino data is not used as an input but rather constrain the parameters listed above, as shown later in~\cref{tab:observables}.

\subsubsection{Casas-Ibarra Parametrisation}

In the CI parametrisation \cite{Casas:2001sr, Basso:2012voo} we express the couplings in \cref{eq:G_matrix} in terms of neutrino oscillation data \cite{deSalas:2020pgw, Gonzalez-Garcia:2021dve} according to
\begin{equation} \label{eq:G_param}
    {\cal G} ~=~ U_L \, D^{-1/2}_L \, R \, D_{\nu}^{1/2} \, U^{*}_{\rm PMNS} \, ,
\end{equation}
where $D_L$ is the diagonal matrix defined by
\begin{align}
   D_L ~=~ U_L^T \, M_L \, U_L \,, 
\end{align}
and $D_{\nu}$ is the diagonal matrix containing the neutrino mass eigenvalues. Finally, $U_{\rm PMNS}$ is the usual unitary matrix relating neutrino flavours to their mass eigenstates, assuming that the charged leptons are already in their mass eigenbasis.

Moreover, as even a precise knowledge of all the parameters and observables in $M_L$ and ${\cal M}_\nu$ does not univocally define ${\cal G}$, the extra degrees of freedom are encoded in the complex orthogonal $3 \times 3$ matrix $R$, parametrised as
\begin{equation} \label{eq:Rmat}
    R ~=~ \begin{pmatrix}
    c_2 c_3 ~&~ - c_1 s_3 - s_1 s_2 c_3 ~&~ s_1 s_3 - c_1 s_2 c_3 \\
    c_2 s_3 ~&~ c_1 c_3 - s_1 s_2 s_3 ~&~ -s_1 c_3 - c_1 s_2 s_3 \\
    s_2 ~&~ s_1 c_2 ~&~ c_1 c_2
\end{pmatrix} \, ,
\end{equation}
depending on three complex angles $\theta_i$ with $s_i = \sin\theta_i$ and $c_i = \sqrt{1 - s^2_i}$. Note the importance of these degrees of freedom since they modify the flavour structure of the Yukawa matrix. The latter is of great relevance when considering charged lepton flavour violation and the anomalous magnetic moment, as we will show in the next sections.

The parameter space under the CI parametrisation, $\mathcal{P_{\textrm{CI}}}$, is then defined as
\begin{align}
\begin{split}
    \mathcal{P_{\textrm{CI}}} = & \left\{ \theta_{\textrm{CI}} \equiv \{\lambda_H, \lambda_{4S}, \lambda_{4\eta}, \lambda_{S\eta}, \lambda_S, \lambda_{\eta}, \lambda_{\eta}', \lambda_{\eta}'', \right. \\
    & \left. \alpha, M_S^2, M_{\eta}^2, M_1, M_2, M_{\Psi}, \right. \\
    & \left. y_{ij}, g_R^k, m_{\nu_k}, \theta_{qr}^{\textrm{PMNS}}, \delta_{\textrm{CP}}, \theta_k^R \}: \theta_{\textrm{CI}} \in \mathcal{S}_{\textrm{CI}} \right\},
    \label{eq:parameter_space_ci}
\end{split}
\end{align}
where $\mathcal{S}_{\textrm{CI}}$ is comprised by the relevant parameter intervals shown in~\cref{tab:parameters} which together define the CI case ,  and $i,j = \{1, 2\}$ and $k,q,r = \{1, 2, 3\}$ with $q \neq r$.

With the CI parametrisation, the couplings $g^k_{F_j}$ and $g^k_\Psi$ are derivable from neutrino data instead of input parameters. In addition to the common constraints, for the CI we need to constrain these couplings to retain perturbativity, as shown in~\cref{tab:observables}.

\subsubsection{Constraints}
\label{sec:constraints}

In~\cref{tab:observables} we list the common constraints to both parameterisations. These constraints include measure observables, such as the Higgs boson mass, $m_h$, the BSM contribution to the muon anomalous magnetic moment, $a^{\text{BSM}}_{\mu}$, and the dark matter relic density, $\Omega_{DM}h^2$, as well as upper limits on lepton flavour changing charged currents and spin-independent DM cross-section bounds set by the LZ experiment. 

For the Higgs boson mass and the dark matter relic density, the theoretical uncertainties are far greater than the measured quantities and as such we increase the range of allowed values based on \cite{Boudjema:2011ig,Harz:2016dql,Slavich:2020zjv}. Whereas, for $a^{\text{BSM}}_{\mu}$ we take the combined uncertainty from experimental results and the SM prediction. The allowed ranges for neutrino sector observables represent the $3\sigma$ experimental uncertainty for normal ordering among the neutrino masses~\cite{neutrino_data}. For the LZ experimental bounds on the dark matter (DM) spin-independent cross-section, we used their $95\%$ confidence limits in the DM mass vs. spin-independent cross-section plane~\cite{lz_data, lz_hepdata}. From these limits, we fit a spline that allows us to determine whether a given DM mass value lies above (excluded) or below (allowed) the exclusion boundary. The distance to this boundary is incorporated into the total $C$ function in the same way as other constraints.

One advantage of our methodology, is that the upper limits can be implemented using the $C(\theta)$ function instead of the one-sided Normal usually used in MCMC scans, which implicitly imprints these as likelihoods that impact the MCMC posterior probability estimation when conceptually these limits should act as a prior in such scans.

{\renewcommand{\arraystretch}{0.75}
\begin{table}[t]
	\makebox[\textwidth][c]{
		\begin{tabular}{ccc}
			\hline\hline
			Observable                & Allowed Values                            & Parametrisation
			\\
			\hline
			$m_h$                     & $[124.25, 126.25]$ GeV                   & Both                   
			\\
			$a^{\text{BSM}}_{\mu}$    & $[74,\ 428] \times 10^{-11}$                 & Both               
			\\
			$\Omega_{\text{DM}} h^2$ & $[0.11,\ 0.13]$               & Both                             
			\\
            BR( $\mu^-$ $\to$ $e^- \gamma$ ) & $< 4.2 \times 10^{-13}$              & Both
            \\
            BR( $\tau^-$ $\to$ $e^- \gamma$ ) & $< 3.3 \times 10^{-8}$              & Both
            \\
            BR( $\tau^-$ $\to$ $\mu^- \gamma$ ) & $< 4.2 \times 10^{-8}$                & Both
            \\
            BR( $\mu^-$ $\to$ $e^- e^+ e^-$ ) & $< 1.0 \times 10^{-12}$             & Both
            \\
            BR( $\tau^-$ $\to$ $e^- e^+ e^-$ ) & $< 2.7 \times 10^{-8}$             & Both
            \\
            BR( $\tau^-$ $\to$ $\mu^- \mu^+ \mu^-$ ) & $< 2.1 \times 10^{-8}$               & Both
            \\
            BR( $\tau^-$ $\to$ $e^- \mu^+ \mu^-$ ) & $< 2.7 \times 10^{-8}$             & Both
            \\
            BR( $\tau^-$ $\to$ $ \mu^- e^+ e^-$ ) & $< 1.8 \times 10^{-8}$              & Both
            \\
            BR( $\tau^-$ $\to$ $ \mu^- e^+ \mu^-$ ) & $< 1.7 \times 10^{-8}$                & Both
            \\
            BR( $\tau^-$ $\to$ $ \mu^+ e^- e^-$ ) & $< 1.5 \times 10^{-8}$              & Both
            \\
            BR( $\tau^-$ $\to$ $  e^- \pi $ ) & $< 8.0 \times 10^{-8}$              & Both
            \\
            BR( $\tau^-$ $\to$ $ e^- \eta$ ) & $< 9.2 \times 10^{-8}$               & Both
            \\
            BR( $\tau^-$ $\to$ $ e^- \eta'$ ) & $< 1.6 \times 10^{-7}$              & Both
            \\
            BR( $\tau^-$ $\to$ $  \mu^- \pi $ ) & $< 1.1 \times 10^{-7}$                & Both
            \\
            BR( $\tau^-$ $\to$ $ \mu^- \eta$ ) & $< 6.5 \times 10^{-8}$             & Both
            \\
            BR( $\tau^-$ $\to$ $ \mu^- \eta'$ ) & $< 1.3 \times 10^{-7}$                & Both
            \\
            CR$_{\mu \to e}$(Ti) & $< 4.3 \times 10^{-12}$              & Both
            \\
            CR$_{\mu \to e}$(Pb) & $< 4.3 \times 10^{-11}$              & Both
            \\
            CR$_{\mu \to e}$(Au) & $< 7.0 \times 10^{-13}$              & Both
            \\
            BR( $Z^0$ $\to$ $ e^\pm \mu^\mp  $ ) & $< 7.5 \times 10^{-7}$               & Both
            \\
            BR( $Z^0$ $\to$ $ e^\pm \tau^\mp $ ) & $< 5.0\times 10^{-6}$                & Both
            \\
            BR( $Z^0$ $\to$ $ \mu^\pm \tau^\mp$ ) & $< 6.5 \times 10^{-6}$              & Both
            \\
            $\sigma^{\text{DM}}_{\text{SI}}$ & LZ bounds            & Both
            \\
            \hline
            $\Delta m^2_{21}$         & $[6.82,\ 8.04] \times 10^{-5} \text{eV}^2$      & Non CI 
            \\
            $\Delta m^2_{31}$         & $[2.435,\ 2.598] \times 10^{-3} \text{eV}^2$    & Non CI
            \\
            $\sin^2{(\theta_{12})}$   & $[0.269,\ 0.343]$       & Non CI                       
            \\
            $\sin^2{(\theta_{13})}$   & $[0.02032,\ 0.02410]$       & Non CI                    
            \\
            $\sin^2{(\theta_{23})}$   & $[0.415,\ 0.616]$       & Non CI                        
            \\
            $\delta_{\text{CP}}$      & $[120,\ 369] \times \frac{2 \pi}{360}$      & Non CI   
            \\
            \hline
            $|g^k_{F_j}|$, $(k=1,2,3)$, $(j=1,2)$          & $[0,\ 4\pi]$       & CI
            \\
            $|g^k_{\Psi}|$, $(k=1,2,3)$          & $[0,\ 4\pi]$     & CI
            \\
            \hline\hline
        \end{tabular}
    }
    \caption{Ranges for the observable constraints considered for each parametrisation. We use theoretical uncertainties when those are greater than experimental ones, see text for details. }
    \label{tab:observables}
\end{table}}

\subsection{Implementation details}
\label{sec:implementation_details}

For the computation of physical observables, the modified ``T1-2-A'' scotogenic model was first implemented in \texttt{SARAH-4.14.5} to generate the computational routines for \texttt{SPheno} and \texttt{micromegas-5.3.41}. We use \texttt{SPheno} to compute all the observables except for the DM relic density, $\Omega_{DM}h^2$, for which we use \texttt{micromegas}.
For each parametrisation, the parameter space is defined by the set of relevant parameters in~\cref{tab:parameters} and constrained by the bounds on the relevant observable in~\cref{tab:observables}. For the CI parametrisation we have 41 real parameters, while for the non-CI we have 46. 
The apparent mismatch in parameter count has to do with unphysical phases in the neutrino sector which are unspecified in the CI parametrisation but can appear in the non-CI case. We notice that we are allowing for a much larger parameter space than that considered in~\cite{Alvarez:2023dzz} as we want to perform a minimal bias scan that goes beyond the one produced before to map the whole parameter space and uncover novel phenomenology that could have been missed.

The sampling of the optimisation algorithms is performed in the 46 (41) dimensional box space ($\mathcal{P}_{\text{box}}$) for the non-CI (CI) with size of length one, i.e,
\begin{equation}
    \mathcal{P}_{\text{box}} = \left\{ \theta_{\text{box}} \in \mathbb{R}^{N}, \quad 0 \leq \theta_{\text{box}}^{(i)} \leq 1 \right\},
    \label{eq:box_parameter_space}
\end{equation}
where $N$ is the dimensionality of the parameter space.
Each point in the box space is uniquely mapped into a single point in the physical parameter space. We also appropriately allow for different kinds of mapping between the sampling box parameter space and the physical parameter space according to \cref{tab:parameters}. This choice allows the optimisation algorithms to better address the different orders of magnitude allowed for each parameter in the parameter space. This is akin to priors in global fits, but we must note that our methodology does not provide a statistical interpretable result, and so these ``priors'' should only be seen as a way of facilitating the convergence of the search algorithm.

One of the goals of our work is to introduce the use of multi-objective optimisation algorithms (using NSGA-III) to scan the parameter space of the model and compare it with a single-objective approach (using CMA-ES). Here, both algorithms have the goal of minimising a constraint function $C(\theta)$ given by~\cref{eq:constraint_func}. 
In order to prevent constraints which produce nominally large values, e.g. consider the different nominal orders of magnitude between $a^{\text{BSM}}_{\mu}$ and $m_h$, to dominate the total constraint function for CMA-ES,~\cref{eq:single-ojective-loss}, we perform a scaling in two steps. First, we scale each one of the different contributions $C(\mathcal{O}_k)$ to the total constraint function $C(\theta)$ as
\begin{equation}
    C(\mathcal{O}_k) \rightarrow \log_{10} (1 + C(\mathcal{O}_k)).
\end{equation}
While this step helps set the nominal scales of all quantities to similar values, its main benefit is to allow the introduction of a ``numerical infinity'' for points which produce unphysical results for observables  without producing overflows\footnote{By using the log, the maximum value a constraint can now take would be the log of the maximum number allowed at a given floating point precision. For a $64$ bit computer, the maximum floating point number is around $\mathcal{O}(10^{308})$, while its log is $\gtrsim 700$.}. This will be of special importance when we develop a hierarchical approach to the loss function in~\cref{sec:hierarchy} where some constraints are conditionally forced to this ``numerical infinity''.
Afterwards, each contribution $C(\mathcal{O}_k)$ to the total constraint function is rescaled to the range $[0,1]$ across the generation population. This rescaling ensures that, within each generation, no single constraint dominates the total cost due to differing numerical scales, enabling a fair comparison of candidates across all constraints.\footnote{As a result, the scaled values act as implicitly ranking the candidates by overall constraint satisfaction. This ranking-based behaviour was first observed in~\cite{Romao:2024gjx} while studying a highly constrained 3HDM model, where CMA-ES successfully converged to the viable region. We argue that this emergent ranking from normalised constraint values is what allows the single-objective approach to remain effective even when handling many constraints, which would otherwise lead to convergence issues due to local minima. A complementary approach that uses explicit ranking was explored in~\cite{Diaz:2024yfu}.} The NSGA-III algorithm performs this rescaling step automatically to guarantee an equal distribution of Pareto optimal candidates in its Pareto front.

One problem that naturally arises in our methodology is that of runs where the black-box algorithm gets stuck in a local minima away from the desired global minima that defines a valid point. To address this, we have implemented early stopping criteria to stop non-promising runs to save computational resources and time for more promising runs. For the CMA-ES runs, we monitored the number of generations without finding a good point, not improving on the average total number of satisfied constraints, and the observed minimum of the constraint function~\cref{eq:single-ojective-loss}. For these we set a patience of $1000$, $1000$, and $100$ respectively, which have to be all jointly triggered in order to stop the run. On the other hand, a priori, NSGA-III runs would not have a similar value to monitor for early stopping since, due to its multi-objective nature, its constraint function is multi-dimensional~\cref{eq:multi_constraint}. This setback is addressed when we introduce runs with hierarchy among the constraints as discussed later in~\cref{sec:hierarchy}. Nevertheless, we set a maximum number o 2000 generations for each NSGA-III run.

We implemented CMA-ES using the \texttt{cmaes} \text{0.10.0}~\cite{nomura2024cmaes}, while NSGA-III was implemented using \texttt{deap} \text{1.3.3}~\cite{DEAP_JMLR2012}.

\subsubsection{Hyperparameters}

Here we briefly discuss the hyperparameters present in each optimisation algorithm used. CMA-ES main hyperparameters are the population size at the first generation (we used the default from the \texttt{cmaes} implementation), the maximum number of generations (we used 1000) and the step size at the first generation, i.e. $\sigma^{(0)}$ from~\cref{eq:cmaes} (we used 1.0). 

For NSGA-III the hyperparameters are the population size, the maximum number of generations and a set of parameters related to the processes of crossover and mutation. Although there are some freedom in the choice of these hyperparameters, there exist some guiding principles to be consider. The population size should be large enough to properly populate each reference direction in order to allow for enough diversity in the population and sufficiently cover the parameter space. Being so, the population size should be, in principle, comparable to the number of reference points. However, it should be noted that the typical method for generating reference points~\cite{nsga3_ref_points} is such that the number of reference points grows exponentially with the number of dimensions in the objective space. In this work we use the method in~\cite{nsga3_ref_points}, however, the reference points can also be provided by the user, when deemed appropriate.

While CMA-ES is not largely dependent on hyperparameter choices and should perform reasonably well out of the box with default settings, NSGA-III can in principle be sensitive to the choice of hyperparameters. In order to investigate if this hypothesis holds for our implementation, we conducted a preliminary study using \texttt{optuna} \texttt{3.2.0}~\cite{optuna} with the goal of optimizing the algorithm performance with respect to the hyperparameters values. The hyperparameters and their optimal values are given in~\cref{tab:hyperparameters}.

{\renewcommand{\arraystretch}{0.75}
\begin{table}[H]
	\makebox[\textwidth][c]{
		\begin{tabular}{cccc}
			\hline\hline
			Hyperparameter & Value & Description & Algorithm \\
                \hline
                $\sigma^{(0)}$ & $1.0$ & Initial step size & CMA-ES \\
                $N$ & $1000$ & Maximum number of generations & CMA-ES \\\hline
                $N$ & $2000$ & Maximum number of generations & NSGA-III \\
                $\mu$ & $400$ & Population size & NSGA-III \\
                $P_{\textrm{cx}}$ & $0.9$ & Crossover probability & NSGA-III \\
                $\eta_{\textrm{cx}}$ & $30$ & Crossover crowding factor & NSGA-III \\
                $P_{\textrm{mut}}$ & $0.5$ & Mutation probability & NSGA-III \\
                $\eta_{\textrm{mut}}$ & $40$ & Mutation crowding factor & NSGA-III \\
                $P_{\textrm{ind}}$ & $\frac{4}{N_{\textrm{dim}}}$ & Independent mutation probability & NSGA-III  \\
			\hline\hline
		\end{tabular}
  }
    \caption{CMA-ES and NSGA-III hyperparameters and their choice of values. $P_{\textrm{ind}}$ refers to the independent probability of mutation for each parameter in a given mutated individual in a population and $N_{\textrm{dim}}$ is the number of dimensions in the parameter space, meaning each mutated individual will have 4 of its parameters mutated, in average. For the CI parametrisation, from~\cref{eq:parameter_space_ci}, $N_{\textrm{dim}}(\theta_{\textrm{CI}}) = 41$, and for the non CI parametrisation, from~\cref{eq:parameter_space}, $N_{\textrm{dim}}(\theta) = 46$.}
	\label{tab:hyperparameters}
\end{table}}

\subsection{Hierarchical Constraints}
\label{sec:hierarchy}

Motivated by the tension between the anomalous magnetic moment of the muon, neutrino masses, and the charged LFV decays, as discussed in~\cref{Sec:g-2}, we performed preliminary runs for CMA-ES and NSGA-III to assess the performance of the algorithms when faced with this tension. The results showed a suboptimal convergence rate ($\approx 10\%$ convergence rate for CMA-ES and $\approx 75\%$ convergence rate with convergence after $1000$ generations on average for NSGA-III). Nevertheless, we deem NSGA-III to be computationally expensive since: (i) it requires on average ten times more evaluations before convergence compared to CMA-ES; (ii) it fails to converge for around 25\% of the runs despite a large maximum number of allowed generations.

Furthermore, analysis of converging runs for NSGA-III shows that, in most cases, the algorithm is able to quickly converge to accommodate a large portion of the constraints ($\approx 70\%$) while getting stuck at a non-global  minimum in $a^{\text{BSM}}_{\mu}$. At this point, in order to try to escape the local minimum, the algorithm seems to start to slowly converge to the $a^{\text{BSM}}_{\mu}$ constraint at the cost of losing convergence in other constraints. If the algorithm is able to escape the local minimum, it is then able to recover the convergence to the remaining constraints and eventually converge to the global minimum. This slow process is caused by the tension discussed in~\cref{Sec:g-2} and is the reason NSGA-III takes a large number of generations to fully converge.

To address this suboptimal behaviour, we introduce a hierarchy in the objective space by setting 
\begin{equation}
C(\mathcal{O}) \rightarrow{} \infty, \quad \text{for} \quad \mathcal{O} \neq a^{\text{BSM}}_{\mu} \quad \text{if} \quad C(a^{\text{BSM}}_{\mu}) > 0,
\label{eq:hierarchy}
\end{equation}
where $C$ is the constraint function defined in~\cref{eq:constraint_func}. This forces the algorithm converge to the $a^{\text{BSM}}_{\mu}$ constraint first and thus avoids the local minima discussed above. This idea can be similarly applied for any objective space where a black-box search algorithm gets stuck at a local minima in one or more particularly difficult objectives.

In preliminary runs of a hierarchical NSGA-III (h-NSGA-III) we observed a change in the behaviour of the algorithm, being now able to quickly converge the $a^{\text{BSM}}_{\mu}$ constraint and then converge the remaining constraints while retaining the $a^{\text{BSM}}_{\mu}$ convergence across all individuals in the population in each generation. 
This behaviour change has the added benefit that, since now the satisfaction of the $a^{\text{BSM}}_{\mu}$ constraint is maintained throughout each generation, the average number of constraints satisfied grows monotonically with the generation number for converging runs, while for non-converging runs it does not. This distinction turns the 
average number
of constraints satisfied in each generation
into a good metric to assess the overall convergence rate of the run.
In turn, this allows us to monitor this metric for early stop criteria for h-NSGA-III and thus, non-promising runs, which are likely to be stuck in local minima, can be safely stopped early before reaching the maximum number of generations allowed, saving computational time which can in turn be allocated to more promising runs. For this we set a patience of 300 generations without improvement to trigger the early stop. In this way, we solve one of the challenges of identifying NSGA-III runs that get stuck at local minima which was discussed in~\cref{sec:implementation_details}.
After this adjustment with the early stop criteria for h-NSGA-III, we observe a convergence rate of $\approx 50\%$, while it takes on average 500 generations to converge. This means h-NSGA-III converges with less evaluations, on average, than NSGA-III, given that for h-NSGA-III, the $\approx 50\%$ non-converging runs are stopped before reaching the maximum number of generations allowed ($2000$), overall leading to more successfully converged runs for the same computational budget.

Similarly, for the CMA-ES, we also developed the hierarchy-driven ansatz. The main benefit of hierarchical CMA-ES (h-CMA-ES) over the original counterpart was the convergence rate, which improved greatly from $10\%$ to $40\%$. Since CMA-ES (and h-CMA-ES) is a single-objective algorithm, it is naturally endowed with a global convergence metric given by the total loss function, as discussed in~\cref{sec:implementation_details}, to be used to trigger an early stop. Therefore, no new improvements over this were gained from its hierarchical variation in the same way that NSGA-III benefited. However, a four-fold increase in the number of converged runs greatly improves the overall efficiency of CMA-ES.
In addition, we observe better coverage of the parameter space for both hierarchical algorithms, as discussed in~\cref{sec:scans}.

\section{Scan Results\label{sec:scans}}

In this section, we present the results of our analysis. Here we focus our discussion on the hierarchical algorithm scans, since those are the ones with the most interesting results. 
We will also compare them to those of ref.~\cite{Alvarez:2023dzz} which used a Markov Chain Monte Carlo (MCMC)\footnote{One of the present authors is also author of ref.~\cite{Alvarez:2023dzz}.}. 
In ref.~\cite{Alvarez:2023dzz} the Casas-Ibarra parametrisation was augmented to include $g_R$ such that not only neutrino data could be included right from the start, but also the $a_\mu^{\text{BSM}}$ constraint. This has been done in such a way that the $y_{ij}$ couplings were kept very small, implying that the second diagram of~\cref{fig:feynman_g-2} did not contribute to this observable. 
In this way, values for the $g_R$ couplings could be chosen such that the $a_\mu^{\text{BSM}}$ constraint is fulfilled, while respecting at the same time the constraints due to the LFV decays.
Although this procedure allows us to  find valid points with interesting phenomenology more easily in relatively short time, it does obviously not cover the complete parameter space, missing potential new phenomenological realisations.
One aim of the present study is to find additional valid regions in parameter space which is made possible by the overall fast convergence of the hierarchical scans discussed in the previous section. 
This clearly demonstrates that the proposed algorithms have great  exploratory capacity.
We begin the discussion with the results of the parameters and proceed with the discussion of the phenomenologically relevant observables. 

\subsection{Parameters}

\begin{figure}[t]
\centering
\includegraphics[scale=0.375]{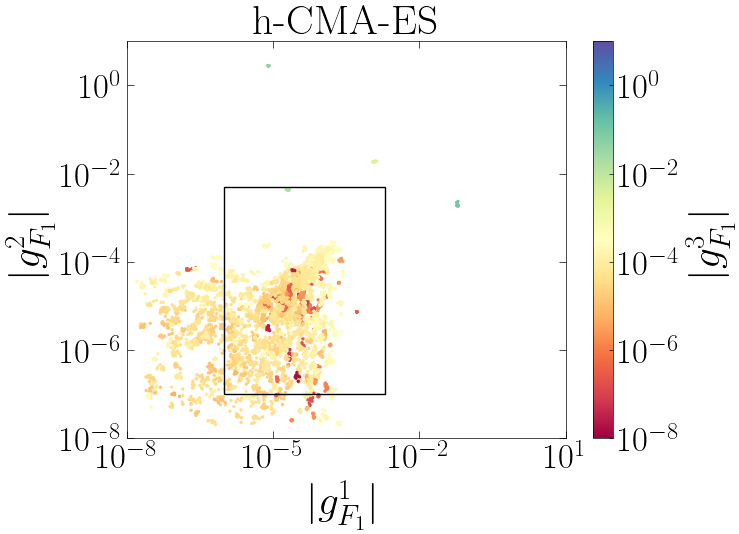}
\includegraphics[scale=0.375]{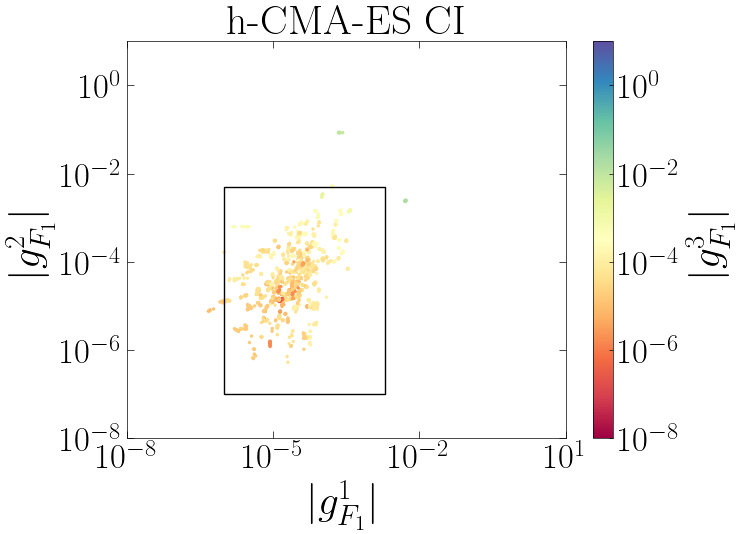}\\
\includegraphics[scale=0.375]{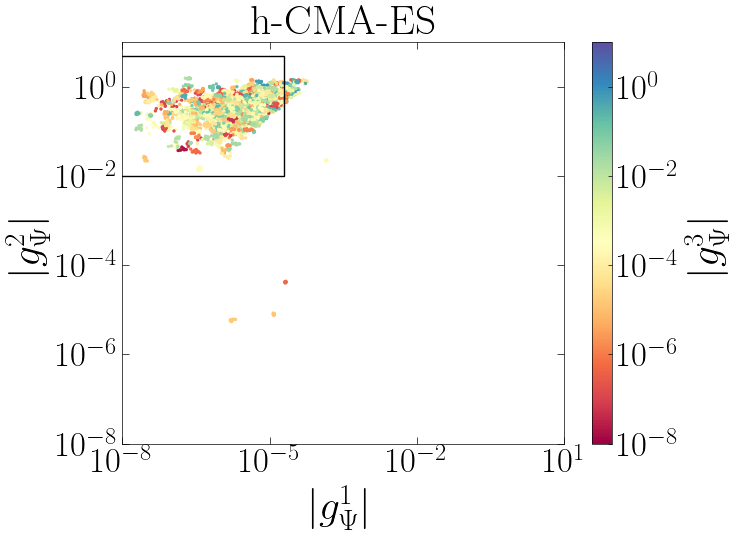} 
\includegraphics[scale=0.375]{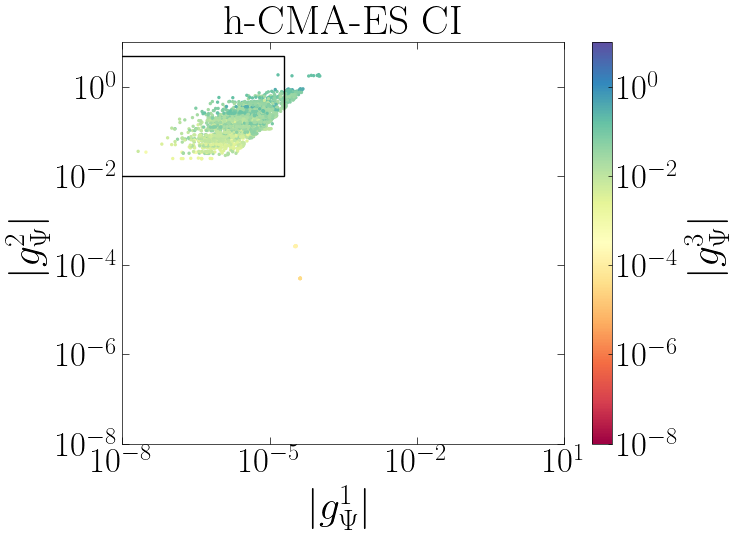}
\caption{Absolute values for the components of the couplings $g_{F_1}$ (upper left and upper right) and $g_{\Psi}$ (lower left and lower right) resulting from h-CMA-ES (left) and h-CMA-ES-CI (right) scans. The box shows the region which encompasses the distribution obtained from MCMC in~\cite{Alvarez:2023dzz}.}
\label{fig:parameters_cmaes}
\end{figure}

\begin{figure}[t]
\centering
\includegraphics[scale=0.375]{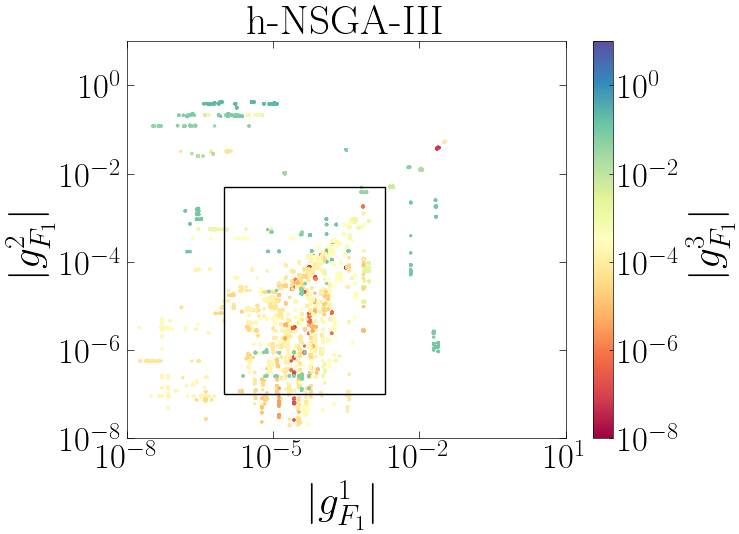}
\includegraphics[scale=0.375]{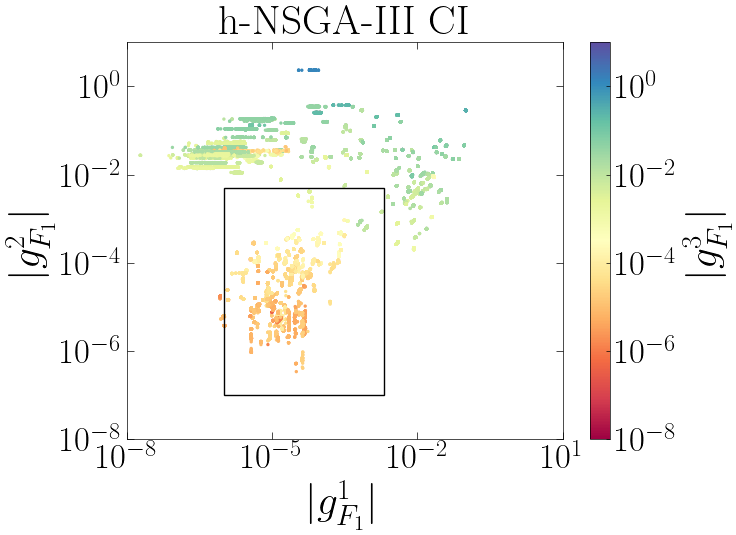}\\
\includegraphics[scale=0.375]{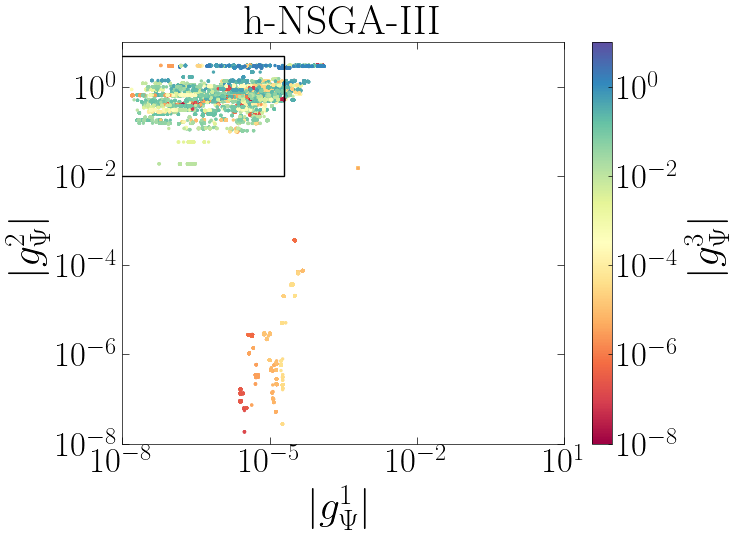}
\includegraphics[scale=0.375]{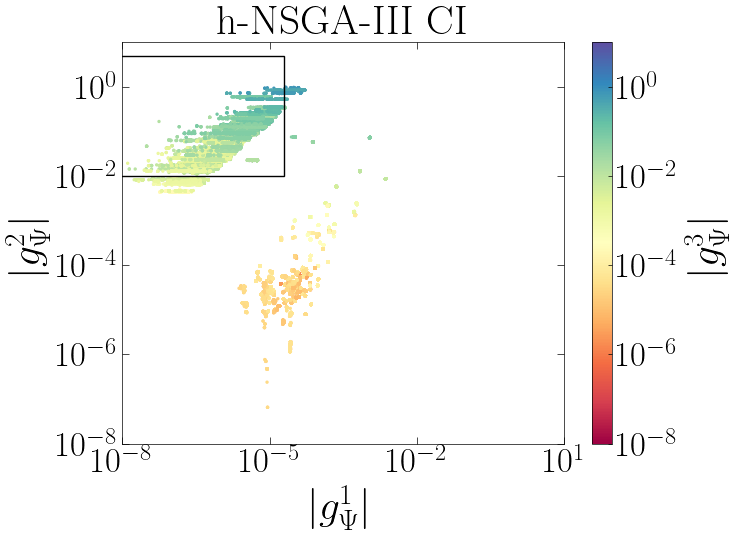}
\caption{Absolute values for the components of the couplings $g_{F_1}$ (upper left and upper right) and $g_{\Psi}$ (lower left and lower right) resulting from h-NSGA-III (left) and h-NSGA-III-CI (right) scans. The box shows the region which encompasses distribution obtained from MCMC in~\cite{Alvarez:2023dzz}.}
\label{fig:parameters_nsga3}
\end{figure}

As discussed in~\cref{Sec:NuMasses,Sec:g-2}, the main parameters of interest are the couplings $g_{F_1}$, $g_{F_2}$, $g_{\Psi}$, $g_{R}$, $y_{ij}$ and $\alpha$ as these have relevant contributions to the neutrinos masses, the anomalous magnetic moment of the muon, and the LFV decays.
The resulting distributions of $g_{F_1}$ and $g_{\Psi}$ are shown in~\cref{fig:parameters_cmaes} and~\cref{fig:parameters_nsga3} for the h-CMA-ES and h-NSGA-III scans, respectively, with and without the CI parametrisation. 
As can be seen, there are notable differences in the scans, whether CI parametrisation has been used or not. These differences arise mainly from numerical issues related to the CI: in~\cref{eq:G_param} quantities spanning different orders of magnitude are multiplied and summed. In our numerical procedure, we check if we can re-obtain the input for the CI using the calculated couplings and we dismiss points for which this fails within the required relative numerical precision, which typically is $10^{-5}$. This has led to numerical problems in finding regions of small couplings using the CI.

Moreover, the results of CI scans seem to indicate a correlation
among $|g^{1}_{\Psi}|$, $|g^{2}_{\Psi}|$, $|g^{3}_{\Psi}|$ and to some extent also between $|g^{1}_{F_1}|$, $|g^{2}_{F_1}|$, $|g^{3}_{F_1}|$ as indicated
by the colour gradients. In contrast, this is not realised in the non-CI scans for the $|g^{2}_{\Psi}| \gtrsim 10^{-3}$ subregion. This is likely due to the numerical precision imposed on the CI operations, which will bias towards the couplings being roughly of the same order of magnitude.

\begin{figure}[t]
\centering
\includegraphics[scale=0.375]{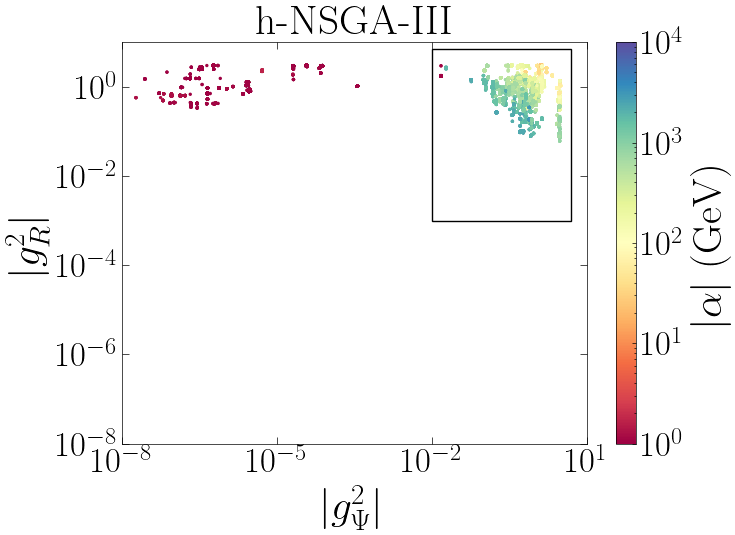}
\includegraphics[scale=0.375]{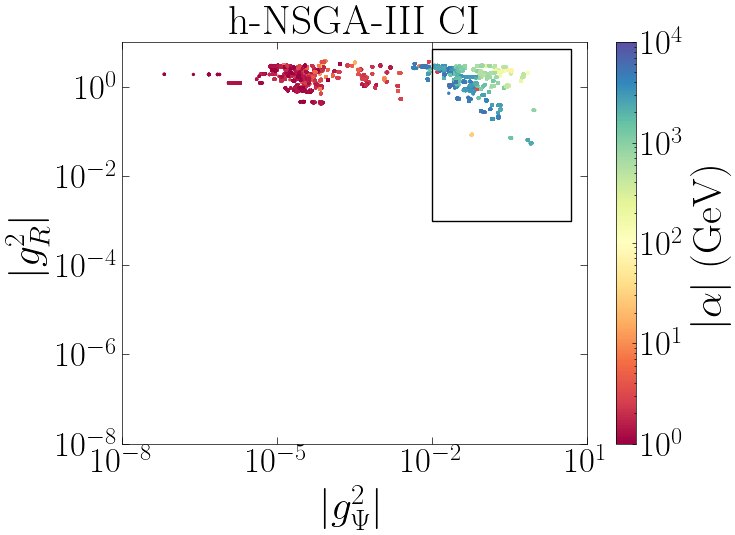}\\
\includegraphics[scale=0.375]{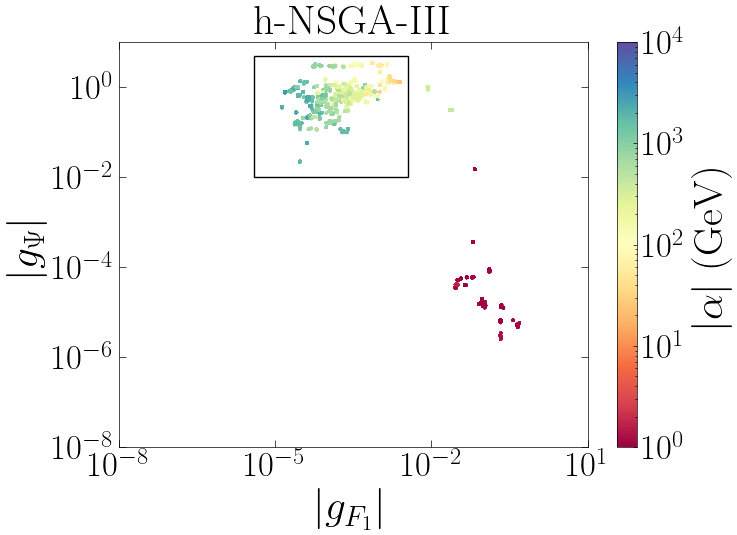}
\includegraphics[scale=0.375]{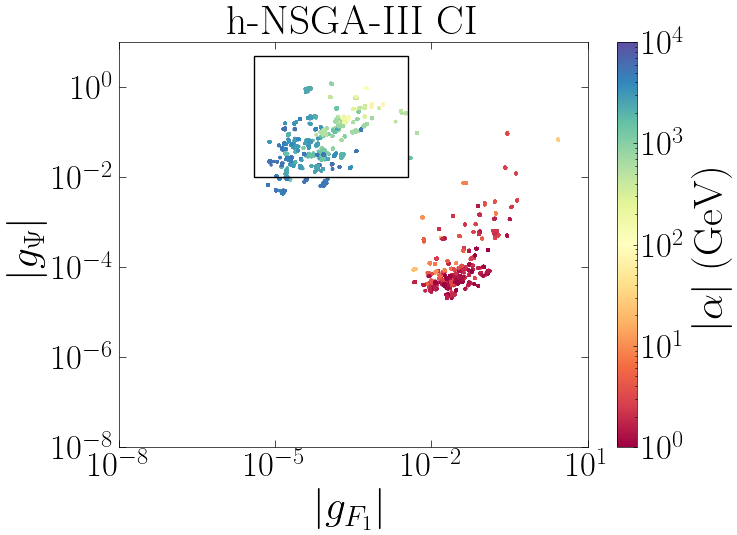}
\caption{Trilinear coupling correlation with absolute values for the second component of the couplings $g_{\Psi}$ and $g_R$ (upper left and upper right) and the absolute value of $g_{\Psi}$ and $g_{F_1}$ couplings (lower left and lower right) resulting from h-NSGA-III (left) and h-NSGA-III-CI (right) scans. The box show region which encompass distribution obtained from MCMC in~\cite{Alvarez:2023dzz}.}
\label{fig:parameter_alpha}
\end{figure}

\begin{figure}[th]
\centering
\includegraphics[scale=0.375]{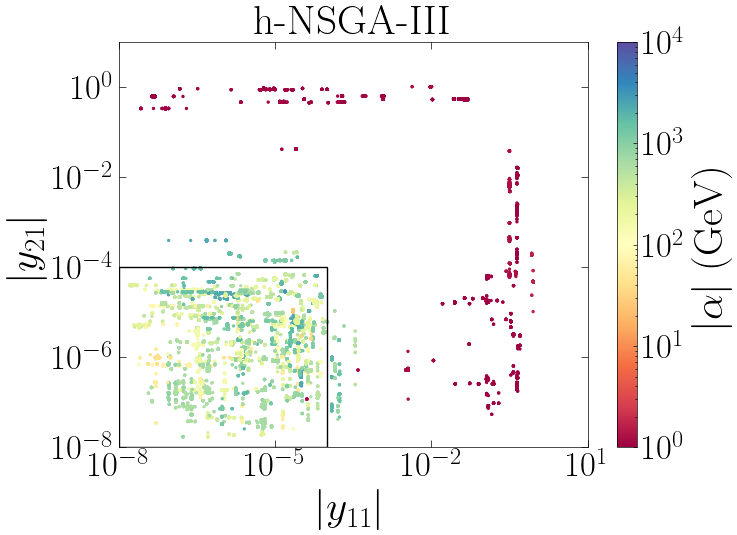}
\includegraphics[scale=0.375]{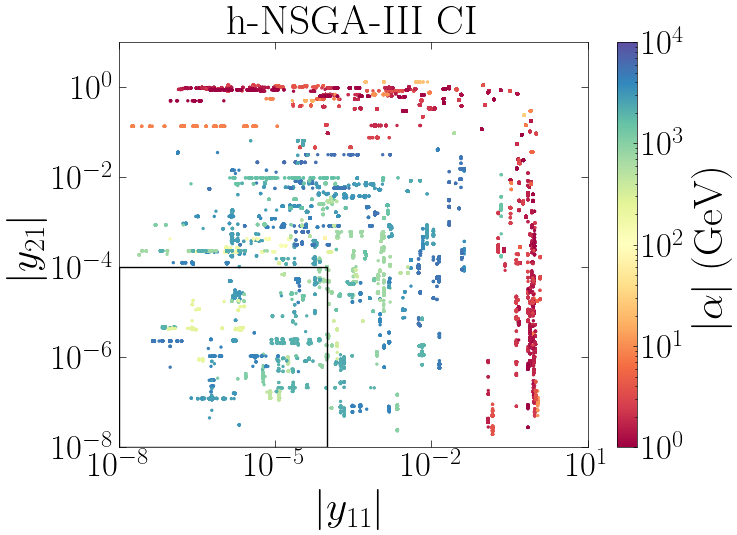}
\caption{Trilinear coupling correlation with the absolute value of $y_{11}$, $y_{12}$ couplings resulting from h-NSGA-III (left) and h-NSGA-III-CI (right) scans. The box show region which encompass distribution obtained from MCMC in~\cite{Alvarez:2023dzz}. 
}
\label{fig:parameter_y}
\end{figure}

Furthermore, \cref{fig:parameter_alpha,fig:parameter_y} show that h-NSGA-III is able to find valid points with large $|y_{ij}|$ couplings ($\sim 1$), small $|\alpha|$ ($\lesssim 10$ GeV) and $|g_F| \gg |g_{\Psi}|$. In this part of the parameter space the right diagram of~\cref{fig:feynman_g-2} gives the dominating contribution for $a_\mu$ and the LFV decays. Likewise, the algorithm also finds parameter regions where the left diagram gives the dominant contribution to these observables: small $|y_{ij}|$ couplings ($\lesssim 10^{-1}$), large $|\alpha|$ ($\gtrsim 100$ GeV) and $|g_F| \ll |g_{\Psi}|$. These two scenarios can have interesting implications for the phenomenology of DM as discussed in the following.


\subsection{Dark Matter}

The model studied in this work features three potential dark matter candidates: the lightest neutral scalar $\phi^{0}_1$, the lightest neutral fermion $\chi^{0}_1$ and the pseudo-scalar $A^{0}$. The actual DM particle is the lightest of these states for a given set of parameters.  EWSB induces mixing between the SU(2)$_L$ singlet and doublet neutral states in, both, the scalar sector (\cref{eq:def_Uphi}) and the fermion sector (\cref{eq:def_Uchi}), see ~\cref{sec:model}. 
The composition of the DM candidate -- whether it is predominantly singlet-like or doublet-like -- depends on the extent of this mixing.
For scalar DM, the singlet contribution of the mixing is given by $|U_{\phi,11}|^{2}$ while for fermionic DM it is given by the sum $|U_{\chi, 11}|^{2} + |U_{\chi, 12}|^{2}$. This determines to which extent gauge boson contributions are important for dark matter annihilation. For example, a DM candidate from a fermionic SU(2)$_L$ doublet should have a mass of about 1.1~TeV if gauge interactions dominate dark matter annihilation. Moreover, the DM particle will be in the decay product of every $Z_2$-odd particle produced. Its nature will also impact the corresponding decay chains that will be discussed in~\cref{sec:pheno}.

\begin{figure}[H]
\centering
\subfloat{\includegraphics[width = 0.47\textwidth]{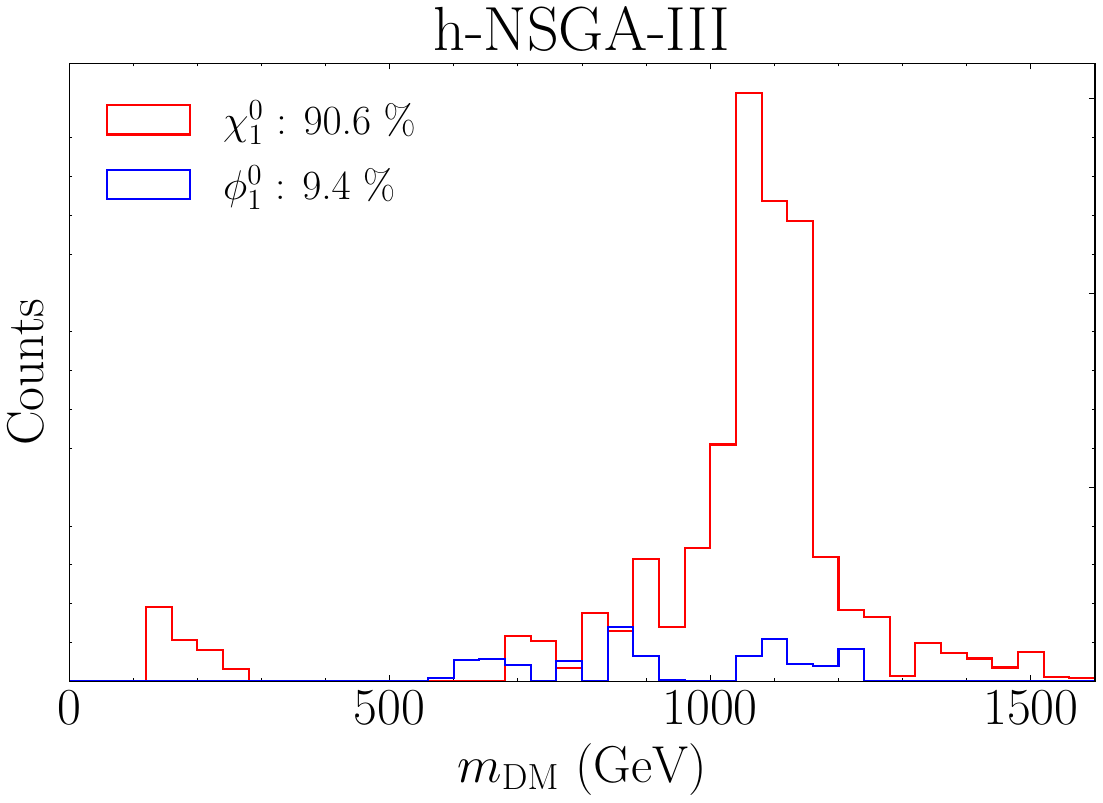}} \quad
\subfloat{\includegraphics[width = 0.47\textwidth]{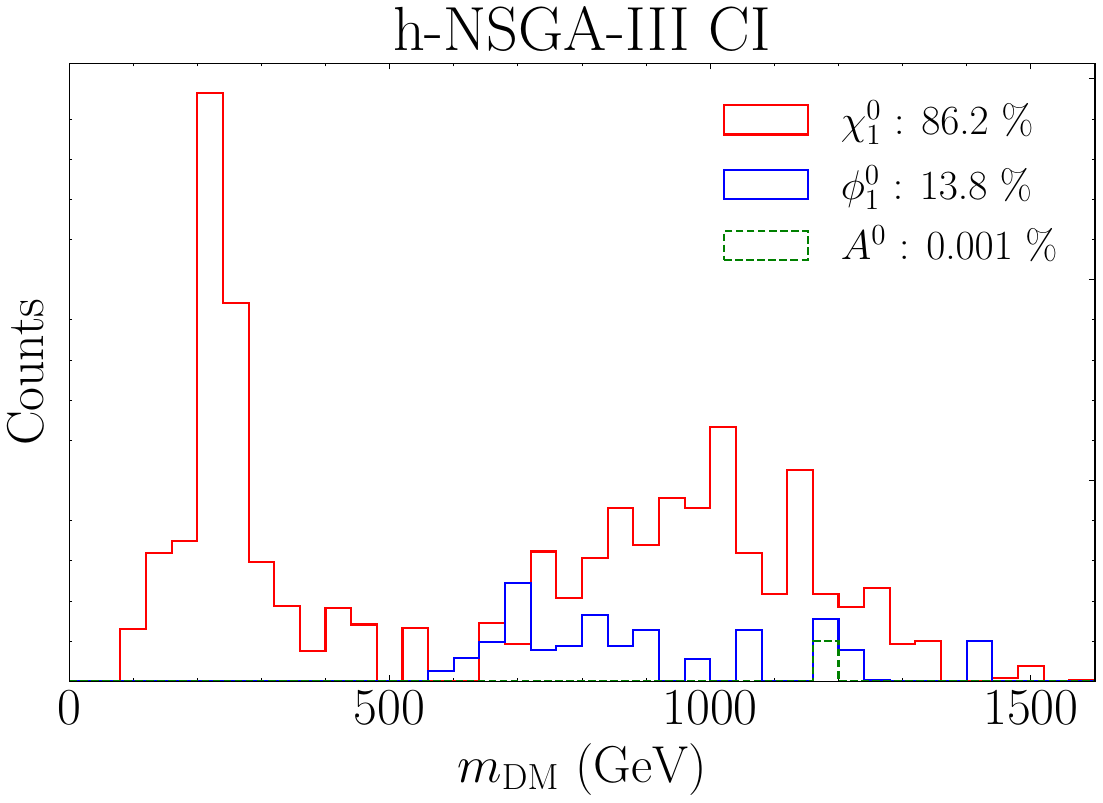}}
\caption{DM nature and mass for valid points found by h-NSGA-III scans. On the left not using CI parametrization and on the right using CI parametrization. The bin corresponding to the pseudoscalar DM was artificially enhanced to help visualization.
}
\label{fig:dm_mass_hist}
\end{figure}

We focus the discussion of the results on the viable points obtained by h-NSGA-III,  as these are the most interesting ones, encompassing a more phenomenologically diverse set of solutions that satisfy all the constraints. 
The results of the remaining scans are provided in~\cref{sec:appendix_scans} and serve as a complement to the findings presented here. 
The values of the DM mass obtained from the h-NSGA-III scans are shown in~\cref{fig:dm_mass_hist} for both scalar and fermionic DM, with and without CI parametrisation. For fermionic DM, h-NSGA-III frequently finds $m_{\text{DM}} = m_{\chi_{1}^{0}} \sim 1100$ GeV (without CI) and $m_{\text{DM}} = m_{\chi_{1}^{0}} \sim 250$ GeV (with CI). In contrast, scalar DM solutions show no strong preference for a specific mass value. This phenomenological realisation of light fermionic DM is a novel aspect of this study compared to the findings in~\cite{Alvarez:2023dzz}. 
Moreover, the algorithm even finds a single point with pseudo-scalar DM, which is another new result for this study. It has a mass of about 1.2 TeV, and the masses of the neutral and charged scalars are the SU(2)$_L$ doublet are also close. 
In addition, this point features a doublet-like pair of neutral and charged fermions with a mass of about 1.28 TeV which in turn helps getting the correct DM relic density.

While the distribution of DM mass per DM type presented in~\cref{fig:dm_mass_hist} hints to different DM realisations, it is important to emphasize that the algorithms used in this work are not designed to produce statistically meaningful probability distributions. 
Therefore, these results should be interpreted as reflecting the relative ``ease'' with which each type of solution is found by the algorithm, rather than how likely or probable (in the posterior language) they are. 
In~\cref{sec:nd}, we will also see that we can perform a novelty detection scan that can produce smooth distributions filling seemingly empty regions of mass distributions or cross sections for direct dark matter detection.

\begin{figure}[t]
\centering
\subfloat{\includegraphics[width = 0.47\textwidth]{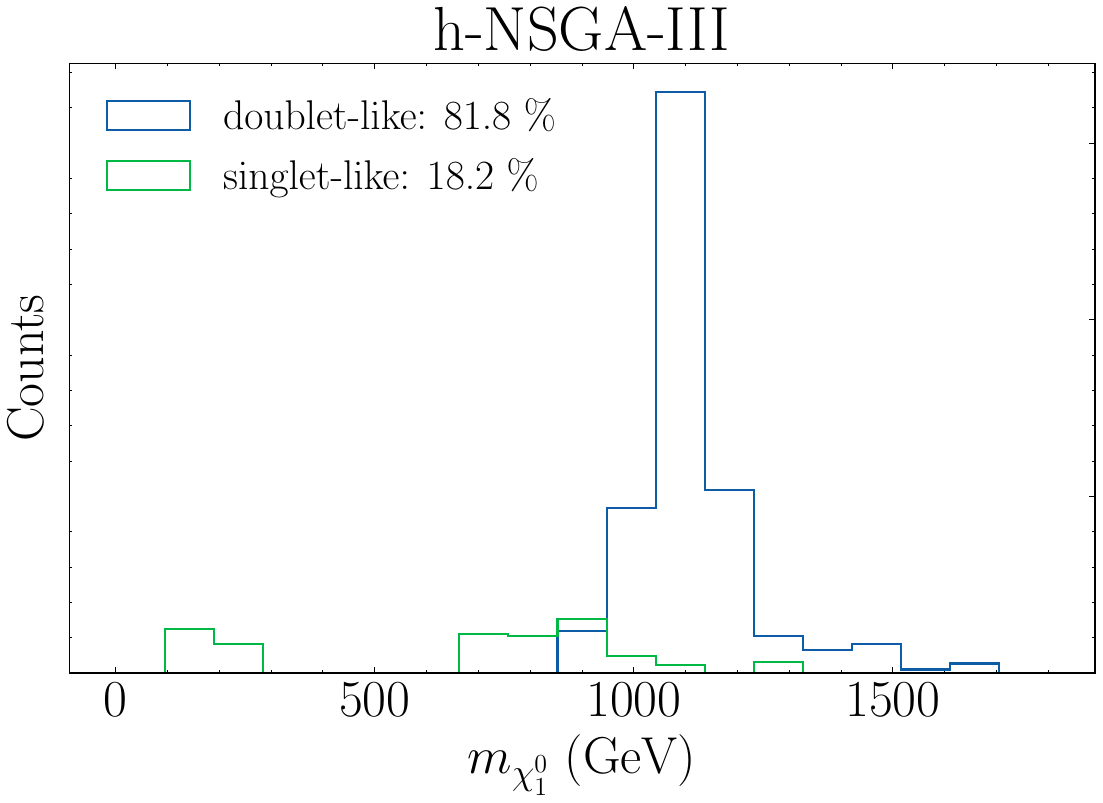}} \quad
\subfloat{\includegraphics[width = 0.47\textwidth]{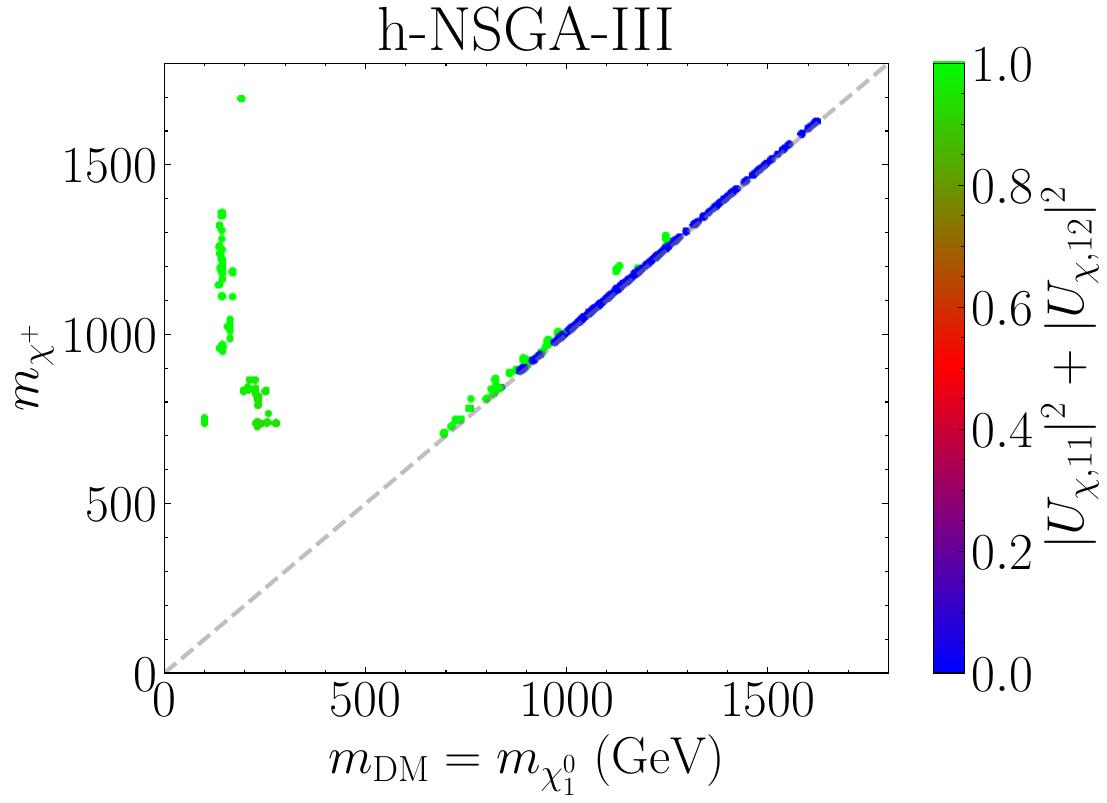}}\\
\subfloat{\includegraphics[width = 0.47\textwidth]{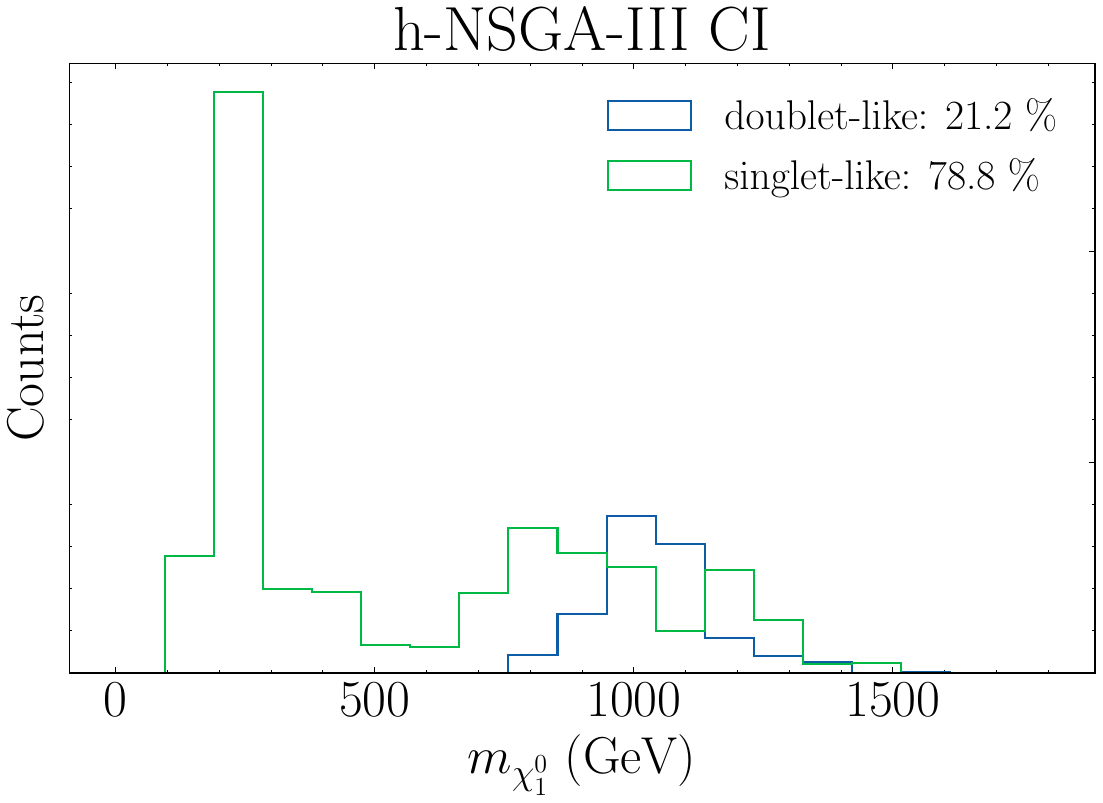}} \quad
\subfloat{\includegraphics[width = 0.47\textwidth]{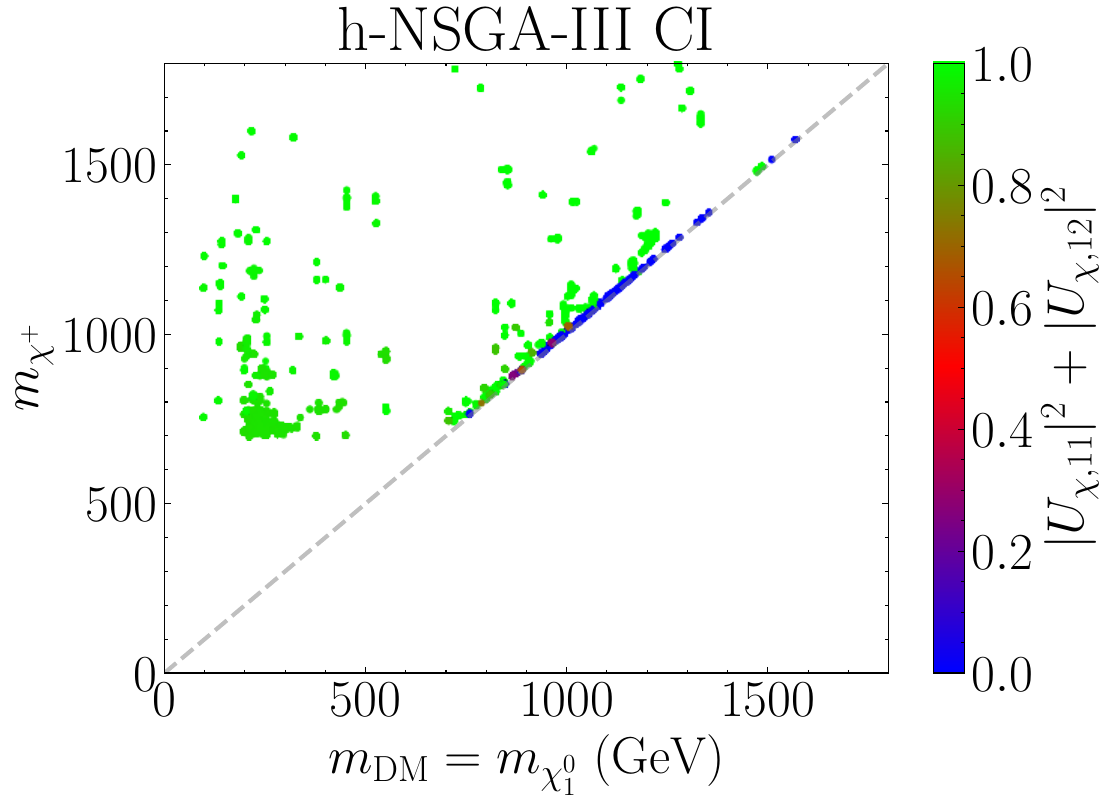}}
\caption{Mixing and mass split of fermionic DM for h-NSGA-III. Top: without CI parametrization, bottom: with CI parametrization. On the right: fermionic DM mass plotted against charged fermion state and colour coded by the singlet amount  of the DM fermion.}
\label{fig:dm_mixing}
\end{figure}

For fermionic DM, h-NSGA-III results for the mixing with singlet and doublet states are presented in~\cref{fig:dm_mixing} plotted against the DM and the heavy charged fermion masses. 
In case of heavy fermionic DM one has in general only a small mass splitting. 
Doublet-like DM with masses above about 1 TeV are only slightly lighter than the charged fermion with a mass splitting of about 1 GeV. 

\begin{figure}[t]
\centering
\includegraphics[scale=0.375]{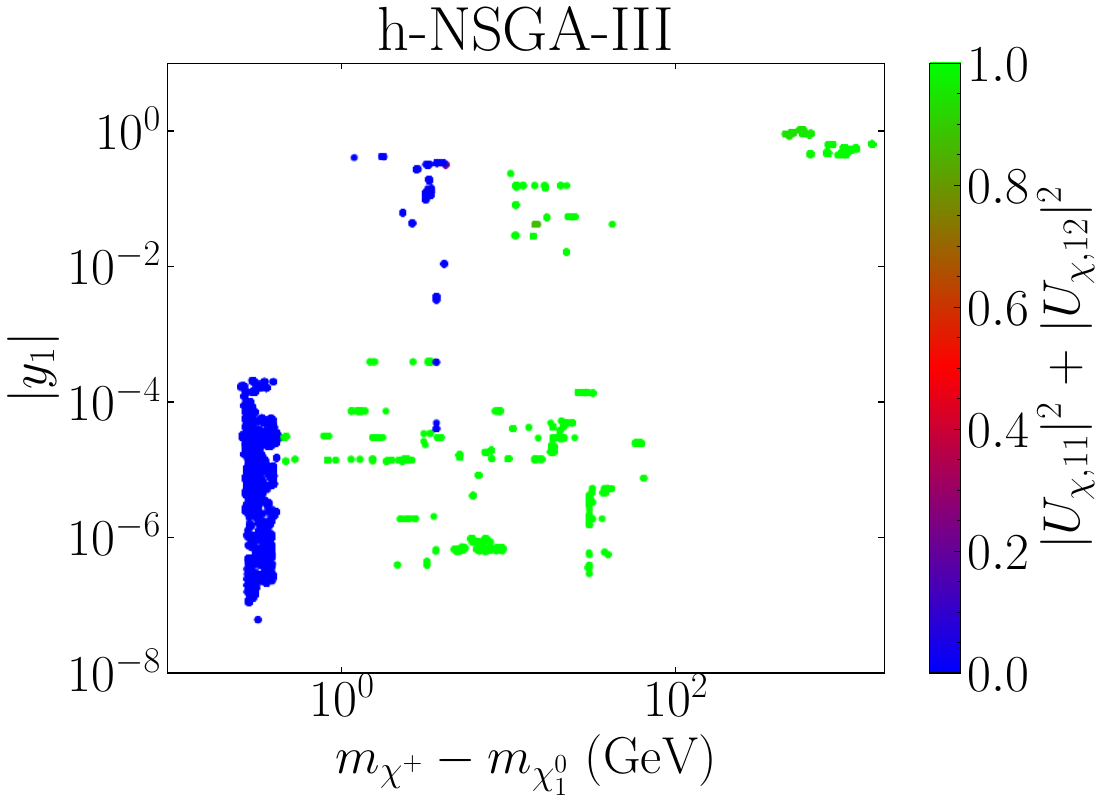}
\includegraphics[clip,scale=0.375]{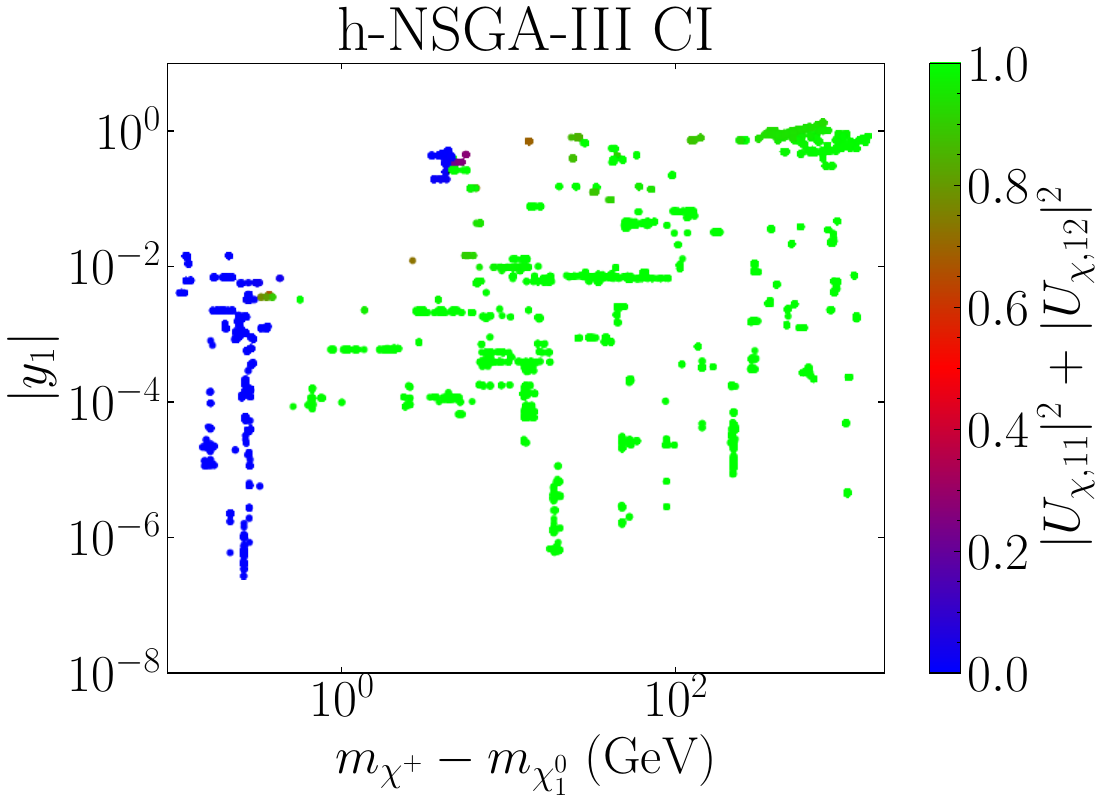}\\
\includegraphics[clip,scale=0.375]{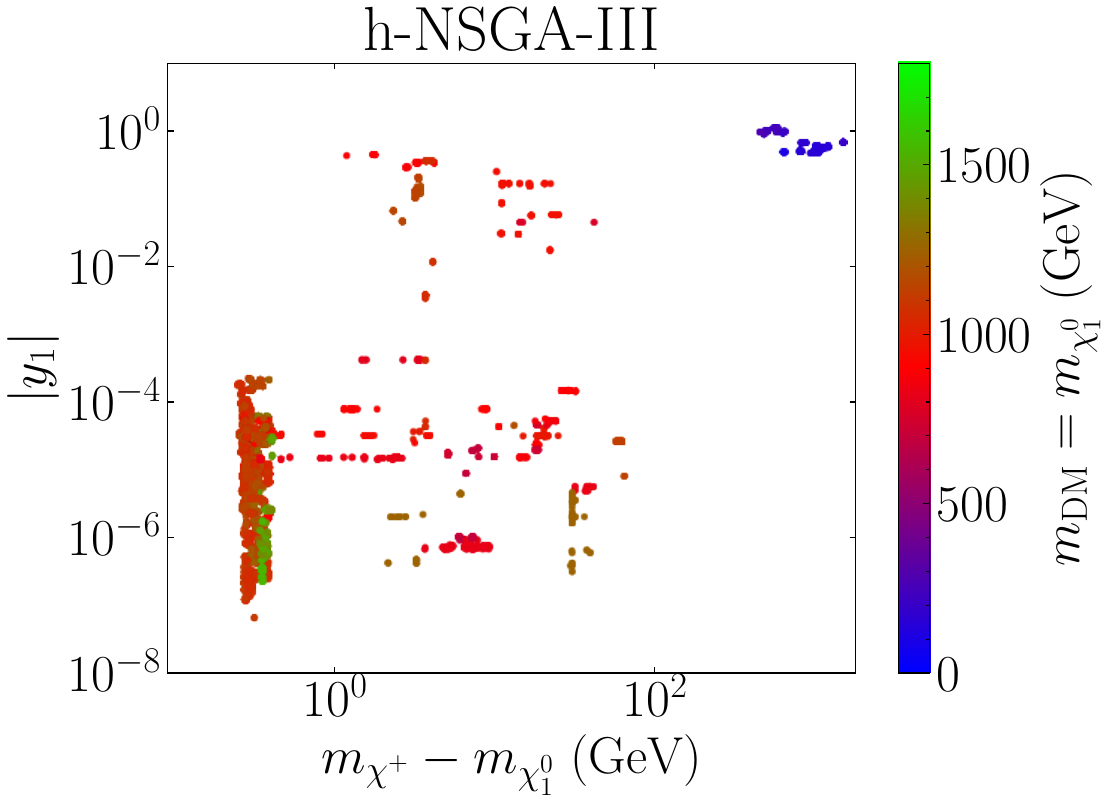}
\includegraphics[clip,scale=0.375]{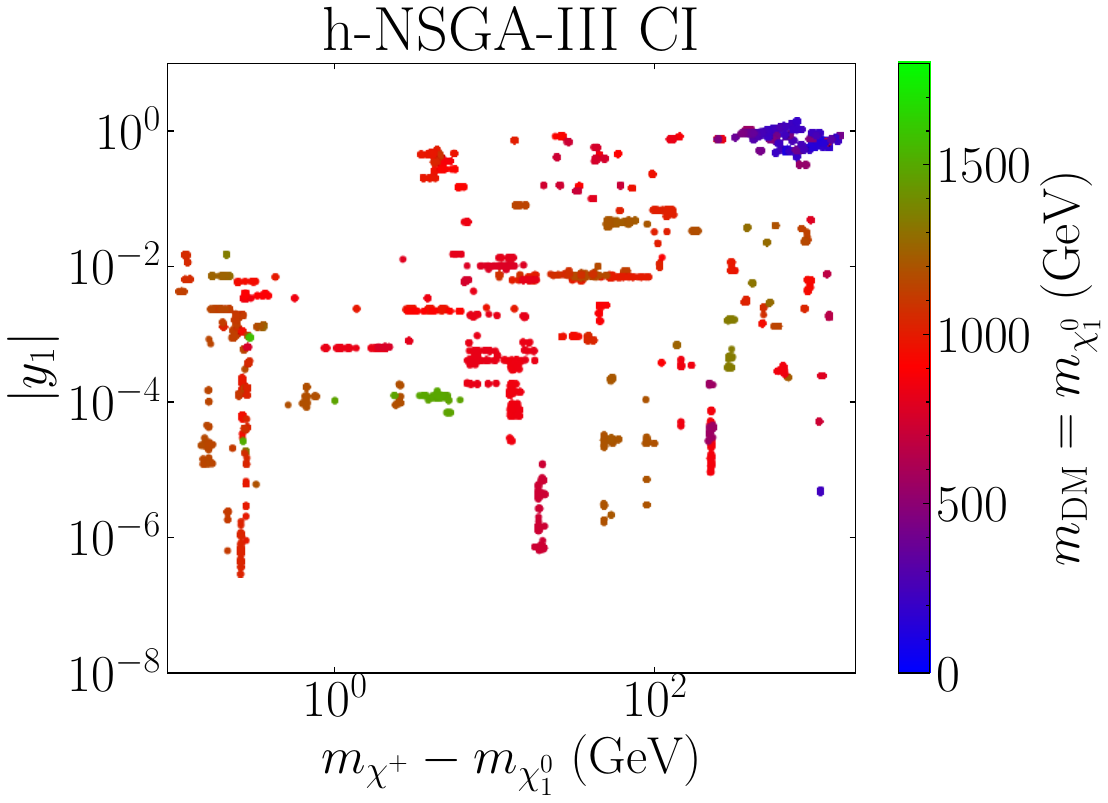}
\caption{Mass difference between charged heavy fermion and fermionic DM plotted against the vector of the absolute value of $y_{11}$, $y_{12}$ and colour coded by the amount of mixing between DM and the singlet neutral heavy fermions (top) and DM mass (bottom).
}
\label{fig:alpha_mass_diff}
\end{figure}

In contrast, for light DM we have only singlet-like DM with a considerable mass splitting which is a consequence of our lower bound on $M_\Psi$. 
Using the mass insertion technique \cite{Gabbiani:1996hi} we find in case of the BSM contributions to $a_\mu$ that the ratio of the left diagram versus the right diagram in~\cref{fig:feynman_g-2} is given by
\begin{align}
\left( \frac{g_\Psi^2 g^2_R \alpha^2 v^2}{M_S^2 M^2_\eta M_\Psi} \right) : 
\left( \sum_{k=1}^2 \frac{g_{F_k}^2 g^2_R M^2_\eta v}{M^2_{F_k} M^2_\Psi}  \right) \,.
\end{align}
We see from \cref{fig:alpha_mass_diff} that scenarios with heavy DM are associated with 
lower values of $|y_1|$ ($\lesssim 10^{-4}$), and therefore higher values of $\alpha$ ($\gtrsim 100$ GeV), c.f.~\cref{fig:parameter_y}. In turn, this implies that in general the left diagram in~\cref{fig:feynman_g-2} dominates over the right one. 
In contrast, the scenarios with light fermionic DM are in the region with large $|y_1|$ ($\sim 1$), corresponding to small $|\alpha|$ ($\lesssim 10$ GeV), c.f.~\cref{fig:parameter_y}, implying that the right diagram dominates in general. In both cases, the extent of the dominance depends on the ratios $|g_\Psi^2/g_{F_k}^2|$.

However, one must keep in mind that points with $m_{\chi^+} \lsim 900$~GeV and  considerable mass difference between the charged fermionic state and the singlet fermionic state DM ($m_{\chi^{+}} - m_{\chi^{0}_1} \gtrsim 250 \textrm{ GeV}$) are likely excluded by charginos and neutralinos searches at the LHC through the decay $\chi^{+} \rightarrow \chi^{0}_1 \textrm{ } W^{+}$~\cite{ATLAS:2024qmx}. 
This particular region of this scotogenic parameter space is therefore essentially excluded.
The full extent depends on the mixing in the neutral fermionic sector as well as on hierarchy of the fermionic to the scalar sector as this impacts the possible decay chains; see also the discussion in \cref{sec:pheno}.
Therefore, the inclusion of this additional constraint would require a detailed reinterpretation of the LHC analysis, which is outside the scope of this study.
Finally, we also note the algorithm found fermionic DM with high mixing between singlet and doublet states (i.e., $|U_{\chi, 11}|^{2} + |U_{\chi, 12}|^{2} \approx 0.5$), another novel result for this model. While only a few points have been found during this scan, we will revisit them in~\cref{sec:nd} where we will explore the parameter space around them.

\begin{figure}[H]
\centering
\includegraphics[trim={0 0 3cm 0},clip,scale=0.4]{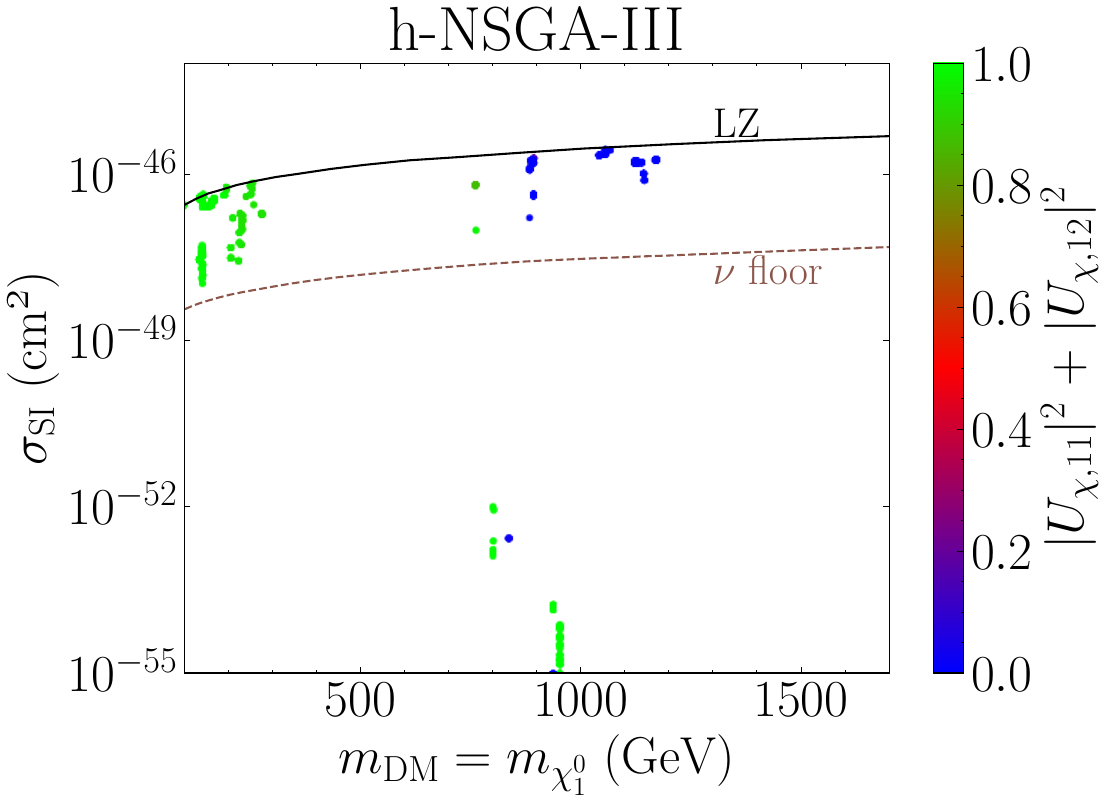}
\includegraphics[trim={3cm 0 0 0},clip,scale=0.4]{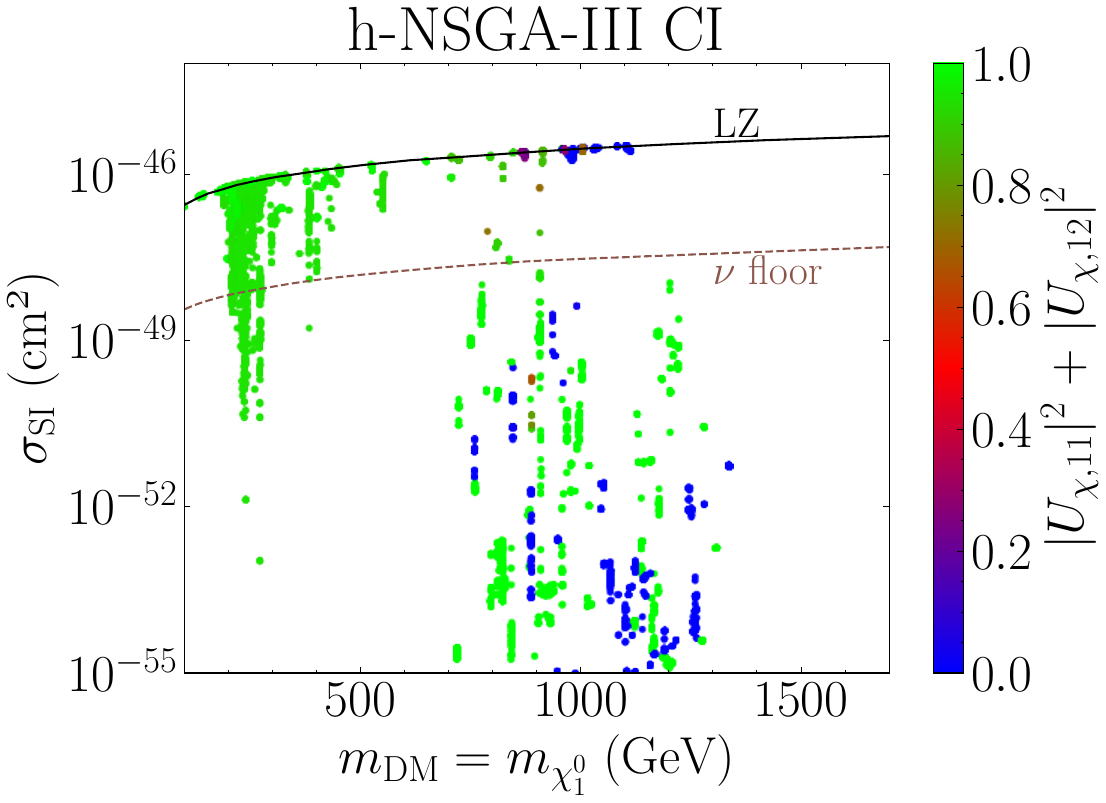}
\caption{spin-independent cross section for fermionic DM for h-NSGA-III plotted against the DM mass and colour coded by the amount of mixing between DM and the singlet neutral heavy fermions. Left: without CI parametrization. Right: with CI parametrization.}
\label{fig:dm_xsec}
\end{figure}

The spin-independent fermionic DM cross section for h-NSGA-III is shown in~\cref{fig:dm_xsec}. The algorithm was able to find fermionic DM with cross section above the neutrino floor and even extremely close to the LZ upper bound. These kind of solutions fall outside the parameter space considered in~\cite{Alvarez:2023dzz}. This was only possible by allowing for a wider parameter space, which considerably raises the challenge of performing a throughout scan. This finding demonstrates the power of the methodology presented in this work at finding novel phenomenological realisations.

Regarding scalar DM, our results show no new insights when compared to the findings in~\cite{Alvarez:2023dzz}. The results for scalar DM can be seen in~\cref{sec:appendix_scalar_dm}.

\subsection{Novelty Detection Scan}
\label{sec:nd}

So far we have shown how the methodologies developed for this scan produce a more complete picture of the phenomenological realisation of the scotogenic model studied in this work. Of particular interest, we have uncovered new regions in the DM parameter space that had been overlooked by previous studies. In this section, we go further to explore the DM phenomenology, by implementing 
the ideas developed in~\cite{Romao:2024gjx}, where CMA-ES was enhanced with additional exploration capabilities. This is achieved by incorporating a \emph{novelty detection reward} into the total loss function. We base this part of the study on CMA-ES instead of NSGA-III because, although CMA-ES employs a more localised search approach, it has a lower computational footprint due to its smaller population size, allowing for a faster exploration. Additionally, we will also seed some of the new runs with points of interest by initializing the mean of the CMA-ES multivariate Gaussian to a point that already satisfies all constraints. This approach was shown to find regions beyond the ones found after the initial convergence~\cite{Romao:2024gjx}, allowing a more complete picture of the model and uncovering new phenomenological realisations and predictions. In this work, it will allow us to further explore around regions of interest in the parameter or phenomenological spaces that were highlighted in~\cref{sec:scans}, namely the novel fermionic dark matter realisation where the cross-section falls between the LZ bounds and the neutrino floor identified in~\cref{fig:dm_xsec}, \footnote{However, one should take note that this region is essentially Higgsino/Singlino DM which is known from SUSY that it should be filled and therefore it should have been covered in~\cite{Alvarez:2023dzz}.} and in pseudo-scalar DM.

The novelty reward is implemented as follows. First, we modify the total single-objective loss function~\cref{eq:single-ojective-loss} by adding $1$ when a point is invalid:
\begin{equation}
	\tilde C(\theta) =\begin{cases}
		1 +C(\theta) & \text{if } C(\theta) > 0 \\
		0            & \text{if } C(\theta) = 0
	\end{cases} \ .
\end{equation}
This ensures that the loss function remains zero if and only if a point is valid while breaking continuity between valid and invalid points. Next, we introduce the novelty reward as a ``density penalty'', $p(\theta)$,
\begin{equation}
	C_T(\theta) = \tilde C(\theta) + p(\theta) \ ,
	\label{eq:loss-new}
\end{equation}
which measures the similarity of a new point, $\theta$, to the valid points already found. To prevent interference with the constraint component of the loss function, $C(\theta)$, the density penalty is normalised to unity, ensuring {$0 \leq p(\theta) \leq 1$}. This normalisation prevents the introduction of new local minima that would arise in inviable regions if $p(\theta)>\tilde C(\theta)$. 
\footnote{A point, $\theta$, is valid if and only if $\tilde C(\theta)=0$ irrespective of the density penalty. $C_T(\theta)$ is what is given to CMA-ES to explore and is very unlikely to be $0$, as that would require the point to both be valid and a complete novelty, which is seldom the case.}
Additionally, we compute the penalty only for valid points, as its purpose is to encourage further exploration of viable regions once CMA-ES has converged.\footnote{Alternatively, one could leave $C(\theta)$ unchanged and define the penalty as $1$ for invalid points and $0 \leq p(\theta) \leq 1$ for valid points, leading to the same final loss function~\cref{eq:loss-new}.} This mechanism effectively ``pushes'' CMA-ES into unexplored areas of the parameter space post-convergence, helping to map contiguous valid regions.\footnote{While this methodology maps the valid region around a CMA-ES convergence region, it may not capture the entire valid parameter space, which could consist of multiple disjoint regions. However, multiple CMA-ES runs from different starting points could identify all viable disconnected subregions.}

In high-dimensional spaces, density estimation, which we want to compute to use as the penalty $p(\theta)$, is particularly challenging due to the ``curse of dimensionality,'' which affects most traditional machine learning methods. In~\cite{Romao:2024gjx}, a simple histogram-based method was employed, which we also use in this current work. Despite its susceptibility to the ``curse of dimensionality'', this approach significantly improved CMA-ES’s exploration capabilities. This method, known as the Histogram-Based Outlier Score (HBOS), was originally introduced in~\cite{CrispimRomao:2020ucc} in the context of signal-agnostic new physics searches at colliders.

For this study we chose to use novelty detection to further explore around the more interesting points presented in~\cref{sec:scans}. These are found by h-NSGA-III CI with fermionic DM, with a considerable amount of mixing with singlet and doublet states and a spin-independent DM cross section close to the LZ experimental upper bound. 
Additionally, we enforce the density penalty to be computed in the observable subspace encompassing the DM mass, the amount of DM mixing with singlet-doublet states, and the spin-independent DM cross section.\footnote{The fact that the density penalty can be computed in the observable space is one of the strengths of this methodology, which allows on to search for novel phenomenological realisations of a model that might have been overlooked before.}

The starting step size $\sigma^{(0)} = 3 \times 10^{-3}$ (~\cref{eq:cmaes}) was used for CMA-ES in this scan, this choice is to force the algorithm to explore locally around the points which are used as seeds. 
The results are presented in~\cref{fig:nd_results} and show that h-CMA-ES with novelty detection successfully expanded around the more phenomenological interesting region found by the NSGA-III scans, mapping the region of DM cross-section between the LZ bounds and the neutrino floor when the DM mass is around $1$ TeV, which is where the initial seeds were located. We also observe that the new points with higher cross section tend to have sizeable if not even dominant SU(2)$_L$ doublet component. Additionally, we notice that h-CMA-ES seems to have explored mainly the cross section direction as highlighted by the vertical path patterns. This suggests that it is more difficult for CMA-ES to change the mass of the DM candidate than it is to change its couplings (this has been hinted to certain extend in~\cref{fig:dm_xsec} as the NSGA-III results produced some clustering around some masses, known as ``schemas'' in the genetic algorithm literature).
These results suggest a potentially powerful strategy which leverages the global exploration of NSGA-III and combine it with the fast local exploitation of CMA-ES, augmented by novelty detection, where potential regions of phenomenological interest {are first identified by NSGA-III, which can be subsequently used as seeds from where CMA-ES with novel detection can start to further investigate.
\begin{figure}[H]
\makebox[\linewidth][c]{
\includegraphics[trim={0 0 3cm 0},clip,scale=0.375]{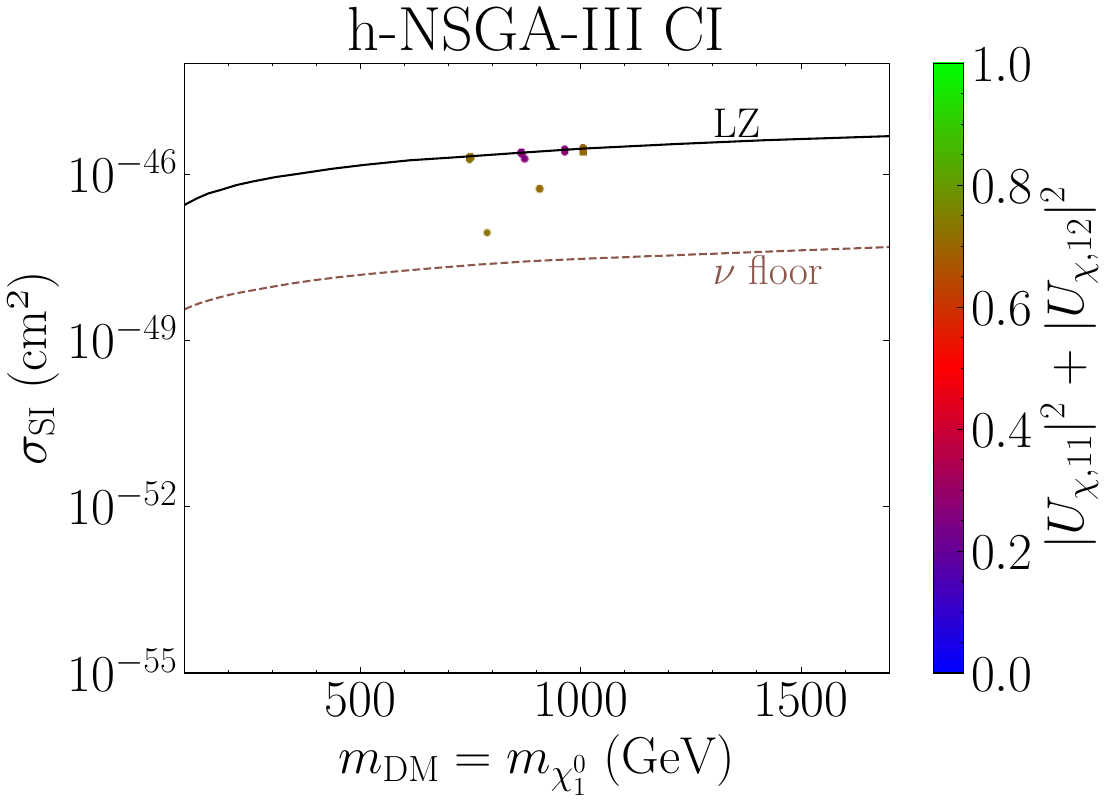}
\includegraphics[trim={3cm 0 0 0},clip,scale=0.375]{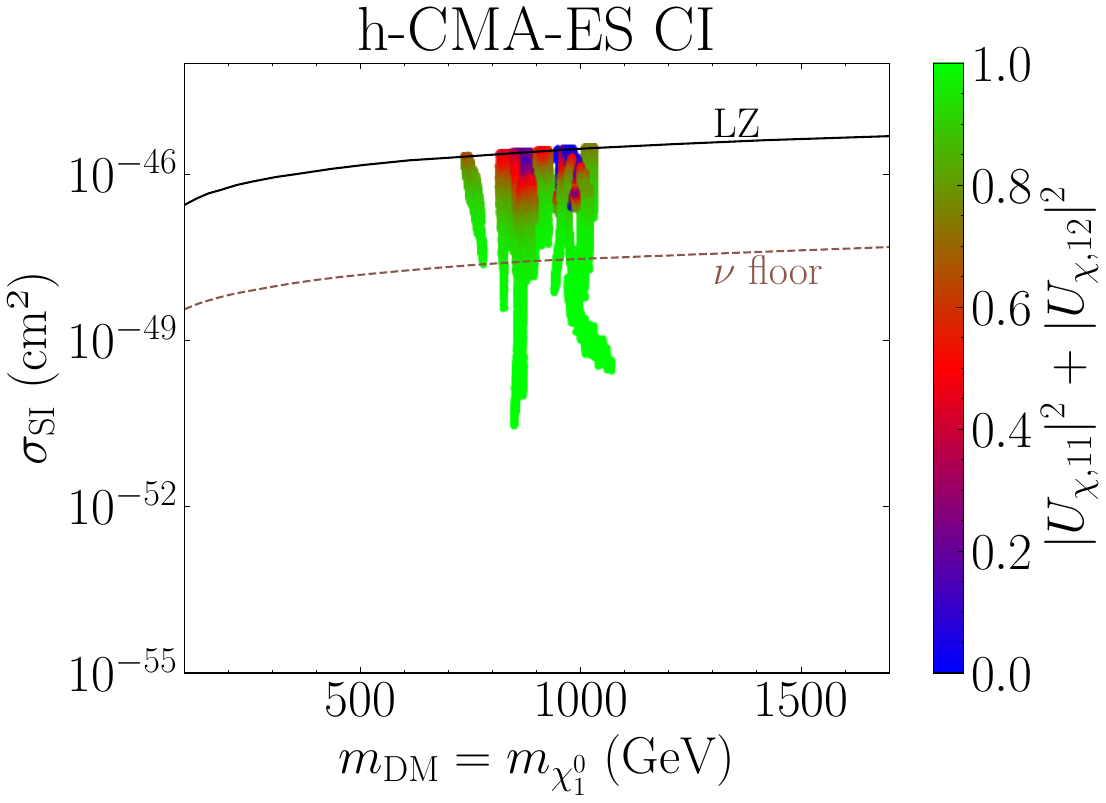}
\includegraphics[trim={3cm 0 0 0},clip,scale=0.375]{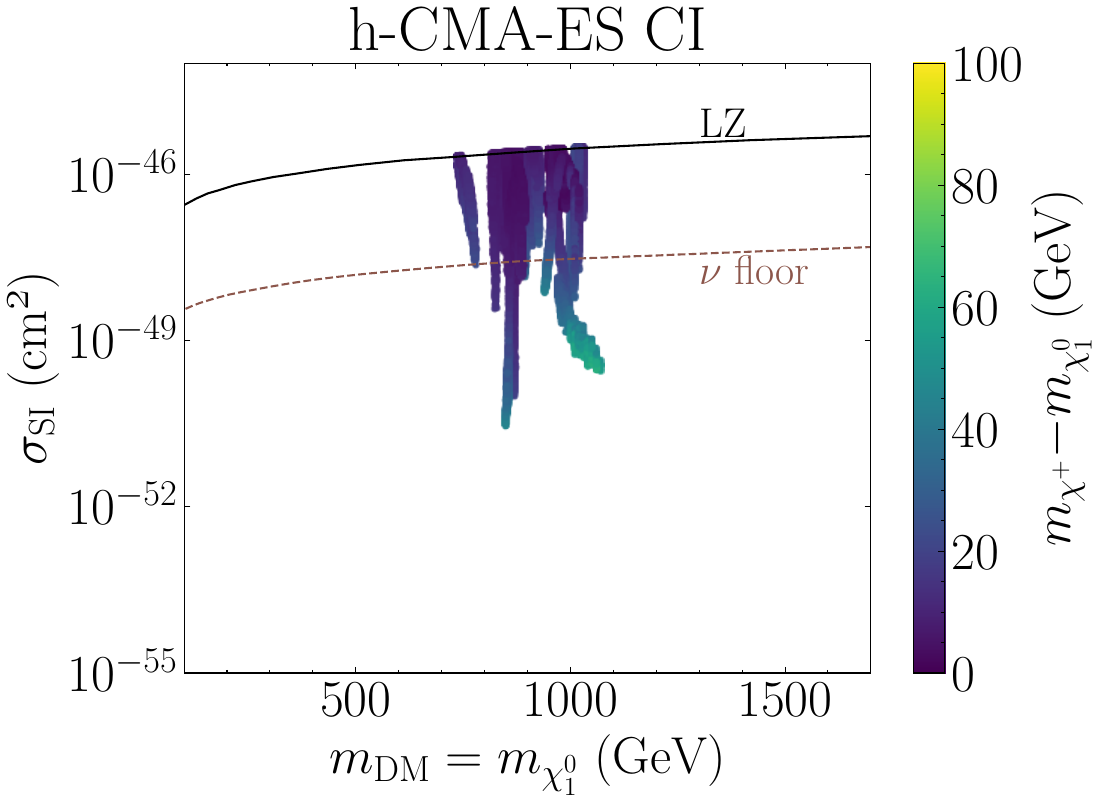}
}
\caption{Results for the fermionic DM novelty detection scan. spin-independent cross section for fermionic DM plotted against the DM mass. Left: Points of the h-NGSA-III-CI run used as seeds for the novelty detection scan carried out by h-CMA-ES-CI colour coded by the amount of mixing between DM and the singlet neutral heavy fermion. Middle: New points obtained from the novelty detection scan using h-CMA-ES-CI colour coded by mixing. Right: The same as the middle, but colour coded by the mass difference between DM and the charged heavy fermion.}
\label{fig:nd_results}
\end{figure}

Another interesting observation in~\cref{fig:dm_mass_hist} was the possibility of pseudo-scalar DM, which is a novel result of our scan. However, only one converged run yielded pseudo-scalar DM, for which a seeded run would only explore around a single convergence region. To explore the pseudo-scalar DM nature, we re-run h-CMAES from scratch, i.e. without seeds, but with the added conditions to the total loss function~\cref{eq:single-ojective-loss}:
\begin{align}
    m_{\phi^0_1} &> m_{A^0}  \nonumber \\
    m_{\chi^0_1} &> m_{A^0} \ ,
    \label{eq:pseudo-scalar-new-conditions}
\end{align}
which, c.f.~\cref{eq:scalar_mass_matrix,eq:fermion_mass_matrix}, will force h-CMAES to explicitly look for pseudo-scalar DM solutions. Additionally, we turn on the novelty detection in the DM parameter space plane, i.e. DM mass and spin-independent direct detection cross-section, once the runs converge to immediately explore the region instead of performing a seeded run a posteriori as we did above. Unsurprisingly, the scan with the new conditions of~\cref{eq:pseudo-scalar-new-conditions} is considerably more difficult to converge to viable regions of the parameter space, with only $35$ out of a $1000$ runs converging. The results are presented in~\cref{fig:nd_results_pseudoscalar}, where we can see that it is possible to virtually fill the region $m_{\text{DM}} \in [500,\ 1000]$ GeV and $\sigma_{\text{SI}} \in [\nu \text{ floor},\  \text{ LZ bounds}]$ as we did with the fermionic dark matter. Furthermore, we also found three separate islands of points where the $m_{A^0}$ and $m_{\phi^0_1}$ are degenerate within our numerical precision. More importantly, we see that for most of these points the mass splitting between the pseudo-scalar DM and the lightest neutral scalar and charged scalar is smaller than $1$ GeV, leading to the possibility of distinct collider phenomenology that is further discussed in~\cref{sec:pheno}.
\begin{figure}[H]
\makebox[\linewidth][c]{
\includegraphics[scale=0.375]{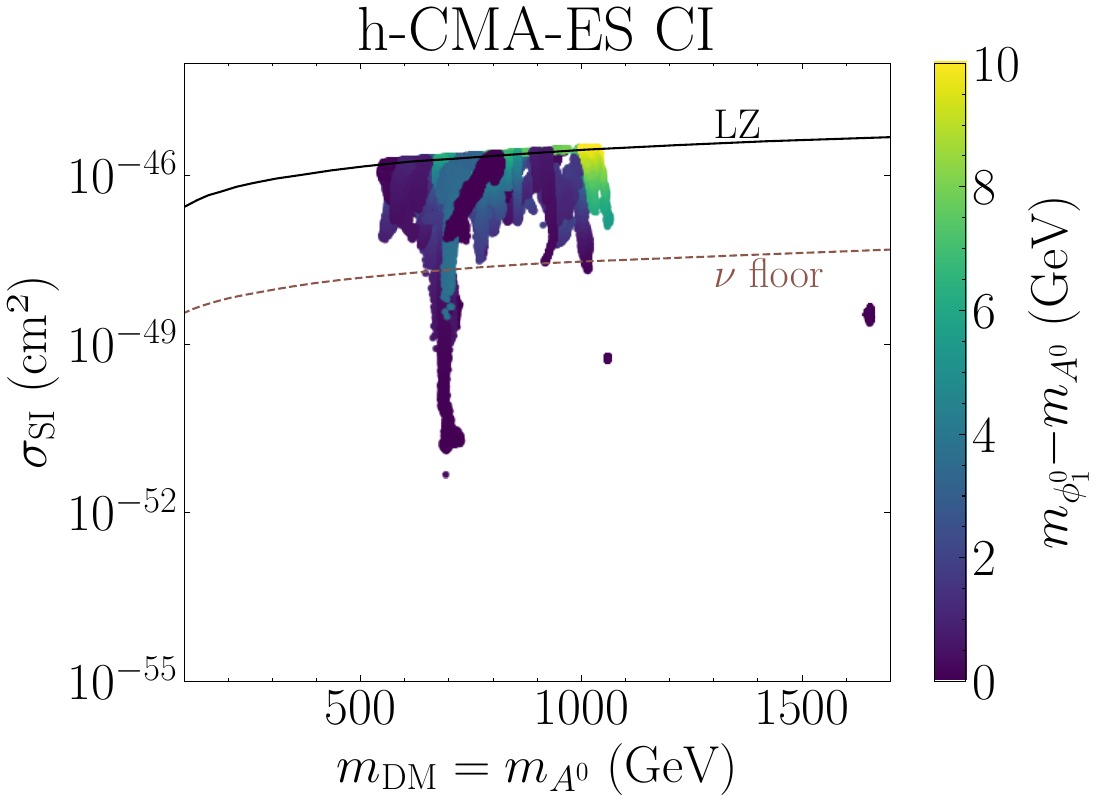} 
\includegraphics[trim={3cm 0 0 0},clip,scale=0.375]{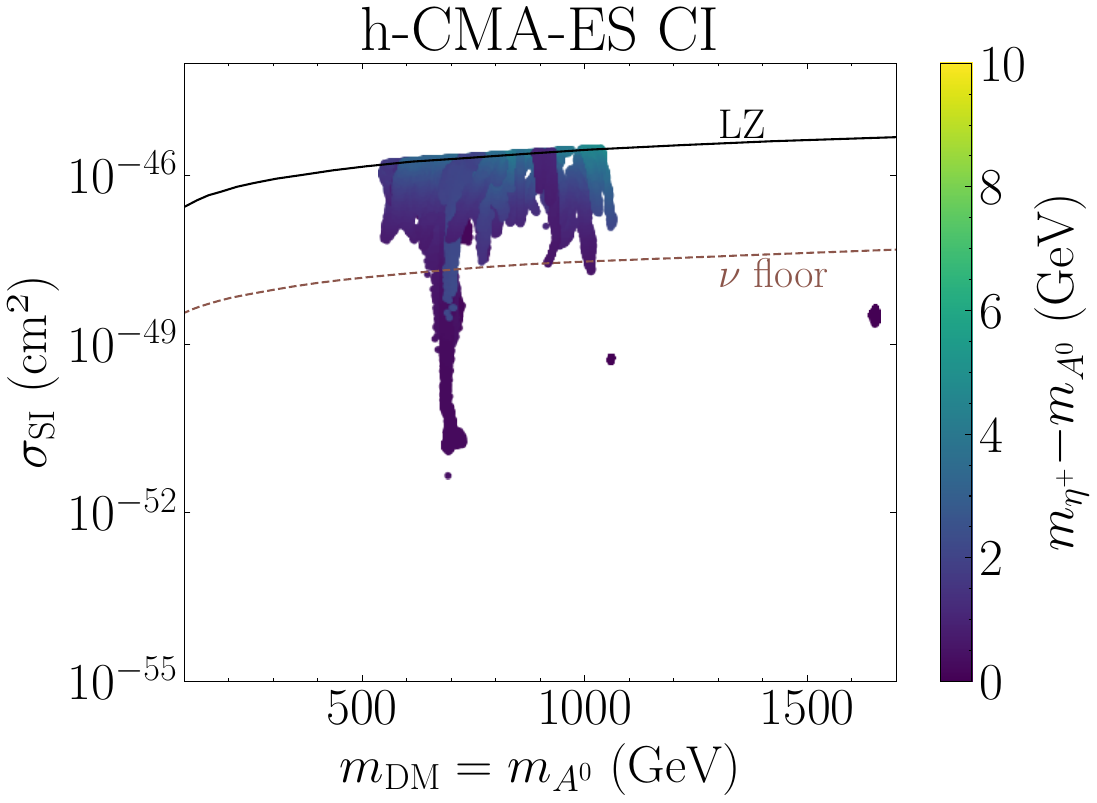}
}
\caption{Results for the pseudoscalar DM novelty detection scan. spin-independent cross section for fermionic DM plotted against the DM mass and colour coded by mass splitting. Left: mass splitting between the pseudoscalar DM and lightest neutral scalar. Right: mass splitting between the pseudoscalar DM and charged scalar.
}

\label{fig:nd_results_pseudoscalar}
\end{figure}

The capacity of the novelty detection runs to fill the $(m_{\text{DM}},\ \sigma_{\text{SI}})$ plane relatively uniformly highlights another feature of this methodology besides targeted exploration. As discussed in~\cite{deSouza:2022uhk,Romao:2024gjx} and in the previous section, the resulting distribution of points obtained by the search algorithms lack statistical interpretability. However, using the novelty detection approach seemingly allows one to produce near uniform distributions of a desired quantity, producing a comprehensive coverage of the phenomenological allowed region. While outside the scope of this work, this in turn could be revisited in terms of per-point likelihoods in order to produce a statistical interpretation. Closing the gap between our results and a statistical interpretation is left to future work.

\section{LHC phenomenology}
\label{sec:pheno}

We have seen that there is a wide mass range for fermionic dark matter between 200 GeV to 1.5 TeV. 
The main production channels at the LHC will be Drell-Yan processes via the SU(2)$_L$ components of the fermions. 
The SU(2)$_L$ doublets have the same quantum numbers as higgsinos in supersymmetric models. We can thus infer from ref.~\cite{LHCxsection}
the corresponding cross sections assuming that the corresponding fermions are pure SU(2)$_L$ states with a common mass. We list in \cref{tab:xsec_fermion} the cross sections at 13.6~TeV for the cases
$M_\Psi=800$, 1000 and 1200 GeV to give an orientation of the order magnitude. The corresponding cross sections for $\sqrt{s}=14$~TeV are only slightly larger. 

\begin{table}[t]
    \begin{center}
    \begin{tabular}{cccccc}
    \hline \hline
       Mass & $\sigma(\chi^0_1 \chi^0_2)$  & $\sigma(\chi^+ \chi^-$)& $\sigma(\chi^0_{1,2} \chi^-)$& $\sigma(\chi^0_{1,2} \chi^+)$ & $\sum \sigma_i$ \\ \hline
        800 & 0.69 & 0.76 & 0.73 & 2.08 & 4.26\\
       1000  & 0.20 & 0.22 & 0.20 & 0.62 & 1.24\\
       1200  & 0.06 & 0.07 & 0.06 & 0.20 & 0.39\\
    \hline \hline
    \end{tabular}
    \end{center}
    \caption{Total LHC cross sections 
    $\sigma(pp\to X_i)\equiv\sigma(X_i)$ in fb for a pure SU(2)$_L$ doublet fermion at 13.6 TeV assuming near mass degeneracy between the different states, numbers taken from \cite{LHCxsection}. $\chi^0_{1,2} \chi^\pm$ is the sum of the cross sections for $\chi^0_{1} \chi^\pm$ + $\chi^0_{2} \chi^\pm$ production.}
    \label{tab:xsec_fermion}
\end{table}

In scenarios in which the DM is dominantly a doublet-like fermion, one has a small mass splitting with respect to the charged fermion as discussed in the previous section. 
The dominant decay modes will be $\chi^+ \to \pi^+ \chi^0_1$ and $\chi^0_2 \to \pi^0 \chi^0_1$.
We can infer from corresponding supersymmetric scenarios that the LHC will not be able to discover the corresponding states, see e.g.~\cite{Barducci:2015ffa}
and references therein.

In scenarios in which the DM candidate is a neutral singlet-like fermion, the situation is more diverse. 
In case of a heavy DM fermion, e.g.~above 700 GeV, the mass difference to the charged fermion is between a few GeV and at most 200 GeV. In case of small mass differences of about 10 GeV one can still have a sizeable doublet component implying decays into off-shell vector-bosons
\begin{align}
\chi^+ \to W^{+*} \chi^0_1 \quad \text{ and } \quad  \chi^0_{j} \to Z^{*} \chi^0_1 
\end{align}
with $\chi^0_j$ being a doublet like neutral fermion.
This will give rise to soft leptons and jets which are again difficult to detect at the LHC. In case of larger mass differences and sizeable values for the Yukawa couplings in \cref{eqn:fermion_lagrangian} off-shell decays via neutral scalars will become important 
\begin{align}
\chi^+ &\to l^+_R \nu \chi^0_1 
\quad \text{ and } \quad 
\chi^0_{j} \to l^+ l^- \chi^0_1 \,.
\end{align}
The requirement of a significant contribution to $a_\mu^{\text{BSM}}$ implies that muons will dominate this final state.
In scenarios in which the mass difference between the doublet-like 
fermions is between $m_W$ and $m_h$ two-body decays into on-shell $W$- and $Z$-bosons become important. However, there is no generic picture and depending on the details of a parameter point either these 2-body decays or the 3-body decays via scalars dominate and in some cases they are even of equal importance. 
In scenarios for even larger mass splittings, the neutral doublet-like fermion $\chi^0_j$ can decay as
\begin{align}
    \chi^0_j \to h \chi^0_1  \ ,
\end{align}
if the corresponding $y_{ij}$ couplings are sizeable, e.g. at least a few times $10^{-2}$.

In the scenarios with singlet-like fermionic DM with masses below 500 GeV
one has two generic scenarios:
(i) the additional scalars are heavier than the doublet-like fermions. In this case, the phenomenology is as described before. 
(ii) The additional scalars are lighter than the doublet-like fermions. In these scenarios one has the following decay possibilities
\begin{align}
  \chi^+ & \to l^+_R \eta^0 \to l^+_R \nu \chi^0_1 \\
     & \to l^+_L S \to l^+_L \nu \chi^0_1 \\
  \chi^0_j & \to  h \chi^0_1 \\
    & \to \nu S \to \nu \nu \chi^0_1 \\
    & \to \nu \eta^0 \to \nu \nu \chi^0_1 \\
    & \to l^\pm \eta^\mp \to l^+ l^- \chi^0_1 \,.
\end{align}
Here, we have indicated the doublet- and singlet-like scalars by $\eta$ and $S$, respectively, and $\eta^0$ can be either the scalar or the pseudoscalar component. We again expect that the leptons in the above states are mainly muons. The resulting phenomenology will depend mainly on the size of the $y_{ij}$ couplings with respect to the other Yukawa couplings as this determines the relative importance of the decay channel $h \chi^0_1$.

We have seen in the previous section, that several scenarios exist with a small mass splitting between the pseudoscalar
$A^0$, the scalar $\phi^0_1$ and $\eta^+$ implying that the later two can only decay via off-shell vector-bosons to $A^0$ plus either pions or leptons. 
All of these decay products are rather soft.
Combining this with the fact, that the pseudoscalar dark matter is always heavier than about 500 GeV as shown in \cref{fig:nd_results_pseudoscalar}, that the LHC will not be sensitive. The reason is that the cross sections are smaller than the ones for higgsinos with the same mass and the LHC is not sensitive to higgsino dark matter in this mass range if the mass splitting is of the same size as in our case. The only exceptions are points for which the mass difference between $\eta^+$ and $A^0$ is so small that only the decay $\eta^+ \to e^+ \nu A^0$ is allowed because in these cases $\eta^+$ will be collider stable leading to a charged track.
The corresponding life times are above $10^{-4}$~s. A recent ATLAS study \cite{ATLAS:2025fdm} shows that some of the scenarios with masses in the 500 GeV range might be discoverable in future searches.

This, however, does not imply that these scenarios cannot be discovered because there is the additional production mechanism via decays of the heavy fermions, e.g.
\begin{align}
\chi^+ &\to \eta^+ \nu \\
       &\to \phi^0_1/A^0 l^+ \\
\chi^0_i &\to \eta^\pm l^\mp  \\
&\to \phi_1/A^0 \nu \,.
\end{align}
The details will depend heavily on the scenario, e.g.\ on the size of the underlying couplings, the mixing details of the neutral particles, and the mass differences between the fermions and bosons. The latter are given in
\cref{fig:dm_massdiff_fermion_pseudoscalar}.
Most likely, the scenarios are already constrained by existing LHC analyses for which one has (i) a sizeable mass difference and (ii) 
$M_ \Psi \lsim 1.2$~TeV. However, the details require a recast of the recent analysis \cite{ATLAS:2025fdm} which is beyond the scope of this paper.
\begin{figure}[ht]
    \centering
    \includegraphics[width=0.5\linewidth]{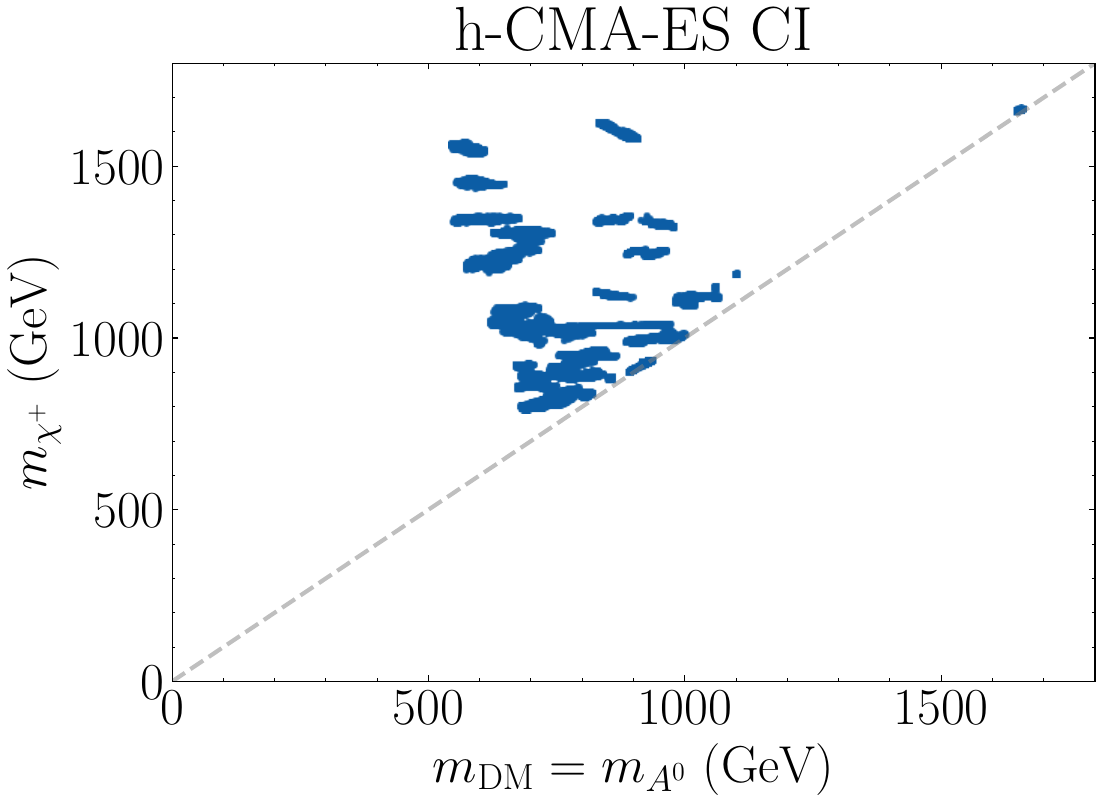}
    \caption{Charged fermion mass as a function of $m_{A^0}$. Results from pseudoscalar DM novelty detection scan.}
    \label{fig:dm_massdiff_fermion_pseudoscalar}
\end{figure}

For scalar DM we briefly summarize here the findings of \cite{Alvarez:2023dzz} for completeness: at the LHC only the scenarios in which $M_ \Psi \lsim 1.2$~TeV are of interest, as the cross sections for the scalars are significantly smaller than the ones for the fermions. The corresponding final states consist of muons and missing transverse energy.

\section{Conclusions}
\label{sec:conclusion}

In this work we used AI black-box search algorithms proposed in~\cite{deSouza:2022uhk,Romao:2024gjx} to explore the parameter space of the scotogenic model first introduced in~\cite{Alvarez:2023dzz}, which is highly constrained by Higgs mass, anomalous magnetic moment of the muon, DM relic density, DM spin-independent cross sections for direct searches, neutrino masses and mixing, and LFV processes. 
In~\cref{sec:model} we highlight a tension amongst some of those constraints, namely the anomalous magnetic moment of the muon, neutrino observables and bounds on LFV decays. To address this, in~\cref{sec:methodology} we proposed a new approach using multi-objective optimisation with the NSGA-III algorithm and compared the results with single-objective optimisation from CMA-ES. 

We identify methodological and phenomenological advances as the main outcomes of this work. For the methodological part, we performed the first multi-objective parameter space scan and introduce the notion of constraint hierarchy to mitigate the suboptimal convergence rate caused by opposing tensions between constraints. The methodology introduced here is not limited to the model at hand and can be applied to any physics case where a multidimensional parameter space is highly constrained by experimentally data or phenomenological consistency like avoiding charge and colour breaking minima.
With these methodological improvements, we produced the scan results presented in~\cref{sec:scans}, where we showed that
the algorithms were very efficient at finding phenomenologically viable solutions with and without the CI parametrisation. Our results also show a more diverse set of solutions from NSGA-III compared to CMA-ES, reinforcing the interpretation of~\cite{deSouza:2022uhk} that genetic algorithms have a wider search capacity than CMA-ES, justifying the combination of both algorithms. Moreover, we successfully used the CMA-ES enhanced by novelty detection methodology introduced in~\cite{Romao:2024gjx} to further explore around the most phenomenologically interesting points found by NSGA-III. The results suggest a powerful search strategy which combines the exploration of NSGA-III with the exploitation of CMA-ES.

The second main contribution of this work is the expansion on the possible dark matter phenomenological scenarios that was possible to discover due to the methodologies developed herein, and which were outside the parameter space considered in the previous study of this scotogenic parameter space~\cite{Alvarez:2023dzz}. More precisely, we discovered a far richer fermionic dark matter phenomenology, where solutions with fermionic DM with cross section above the neutrino floor and extremely close to the LZ upper bound were found, which were not accessible in the study conducted previously in~\cite{Alvarez:2023dzz}. Additionally, we were able to find singlet fermionic dark matter which the mass is split from the new charged heavy fermions. Most of these points yield a light dark matter, $m_{\chi^0_1}\lesssim500$ GeV, which are likely excluded by collider searches if $M_\Psi \lesssim 1$ TeV. Some points included heavier dark matter, at $m_{\chi^0_1}\sim 1$ TeV, that remain out of reach of collider bounds. We then used these points as seeds for a novelty detection scan to paint a complete picture of the fermionic dark matter phenomenology and were able to find fermionic dark matter with any cross-section between the LZ bounds and the neutrino floor for that mass range. Interestingly, the points found can have any amount of mixing between the singlet and doublet states, providing a rich and heterogeneous description of the fermionic dark matter candidate realisation. 
Finally, we perform an explicitly guided scan for pseudo-scalar DM, a possibility that was unknown before this work and that was discovered by our initial scans. In this last focused scan, we found many solutions with pseudo-scalar DM, yielding DM masses and spin-independent cross-sections similar to the previous scans, painting a larger picture of the possible viable DM realisations of this scotogenic model.

Another interesting observation worth mentioning is that the efficiency and coverage of our methodology are largely unaffected by whether we parametrise the model using the CI parametrisation or not. This stands as a favourable feature of our methodology, as it shows that one does not need to rely on such parametrisations, which in our case we have found that can imprint certain patterns in the results due to the numerical precision of the associated mathematical operations.

In conclusion, we were able to expand on the possible phenomenological realisations of the scotogenic model first presented in~\cite{Alvarez:2023dzz} by exploring its highly constrained multidimensional parameter space using a combination of multi-objective and single objective black-box search algorithms with novelty detection. One of the main features of our methodology is its versatility, and we aim to use it to tackle even more challenging multi-objective problems, such as SMEFT studies.

\section*{Acknowledgements}

We thank A.~Karle for contributions in the early stage of this paper.
FAS is supported by FCT under the research grant with reference No. UI/BD/153105/2022.
MCR is supported by the STFC under Grant No.~ST/T001011/1. NC and FAS are supported by FCT under LA/P/0016/2020 and UID/50007/2023.

This work is supported in part by the Portuguese
Fundação para a Ciência e Tecnologia (FCT) under
Contract 2024.01362.CERN; this project is partially funded through
POCTI (FEDER), COMPETE, QREN, and the EU. 

\appendix

\section{Neutrino Loop Functions}
\label{sec:neutrino_loop_functions}

The neutrino mass matrix can be written as $\mathcal{M}_\nu = \mathcal{G}^T M_L \mathcal{G}$, where $M_L$ is a $3 \times 3$ symmetric matrix which encodes the information of the loop function, and the mixing in the neutral scalar and fermion sectors, defined in Eqs.\ \eqref{eq:def_Uphi} and \eqref{eq:def_Uchi}, respectively. For completeness, we explicitly write the expressions for the components of $M_L$,
\begin{eqnarray} 
    (M_L)_{11} &=& \sum_{k,n} b_{kn}  (U_{\chi}^{\dagger})^2_{4k} (U_{\phi}^{\dagger})^2_{1n} \, ,
    \\
    (M_L)_{22} &=& \frac{1}{2} \, \sum_{k,n} b_{kn} (U_{\chi}^{\dagger})^2_{1k}  \left[ (U_{\phi}^{\dagger})^2_{2n} - (U_{\phi}^{\dagger})^2_{3n} \right] \, ,
    \\
    (M_L)_{33} &=& \frac{1}{2} \, \sum_{k,n} b_{kn} (U_{\chi}^{\dagger})^2_{2k}  \left[ (U_{\phi}^{\dagger})^2_{2n} - (U_{\phi}^{\dagger})^2_{3n} \right] \, ,
    \\
    (M_L)_{12} = (M_L)_{21} &=& \frac{1}{\sqrt{2}} \, \sum_{k,n} b_{kn} (U_{\chi}^{\dagger})_{1k} (U_{\chi}^{\dagger})_{4k} (U_{\phi}^{\dagger})_{1n} (U_{\phi}^{\dagger})_{2n} \, ,
    \\
    (M_L)_{13} = (M_L)_{31} &=& \frac{1}{\sqrt{2}} \, \sum_{k,n} b_{kn} (U_{\chi}^{\dagger})_{2k} (U_{\chi}^{\dagger})_{4k} (U_{\phi}^{\dagger})_{1n} (U_{\phi}^{\dagger})_{2n} \, ,
    \\
    (M_L)_{23} = (M_L)_{32} &=& \frac{1}{2} \, \sum_{k,n} b_{kn} (U_{\chi}^{\dagger})_{2k} (U_{\chi}^{\dagger})_{1k} \left[ (U_{\phi}^{\dagger})^2_{2n} - (U_{\phi}^{\dagger})^2_{3n} \right] \, ,
\end{eqnarray}
where $k=1,2,3,4$ and $n=1,2,3$. Moreover, the loop integrals are encompassed in the functions
\begin{align}
    b_{kn} ~=~ \frac{1}{16 \pi^2} \frac{m_{\chi^0_k}}{m_{\phi^0_n}^2 - m_{\chi^0_k}^2 } \left[ m_{\chi^0_k}^2  \log m_{\chi^0_k}^2- m_{\phi^0_n}^2 \log m_{\phi^0_n}^2  \right] \,.
\end{align}

\section{All Scans} \label{sec:appendix_scans}

In this section we provide, for completion, the results for the h-CMA-ES and h-NSGA-III runs that were absent from the main body of text in~\cref{sec:scans}. 

\subsection{Parameters}

Here we provide the results for the Lagrangian parameters. Results for $|g_{F_2}|$ and $|g_R|$ couplings for h-CMA-ES and h-NSGA-III are shown in~\cref{fig:parameters_h_cmaes_appendix} and~\cref{fig:parameters_h_nsga3_appendix}, respectively. Moreover, results for trilinear coupling correlation with $|g_{\Psi}|$, $|g^{2}_R|$ and $|g_{F_1}|$ are shown in~\cref{fig:parameter_alpha_appendix}. Lastly, the trilinear coupling correlation with $|y_{11}|$ and $|y_{12}|$ for h-CMA-ES results is presented in~\cref{fig:parameter_y_appendix}.

\begin{figure}[H]
\centering
\subfloat{\includegraphics[width = 0.47\textwidth]{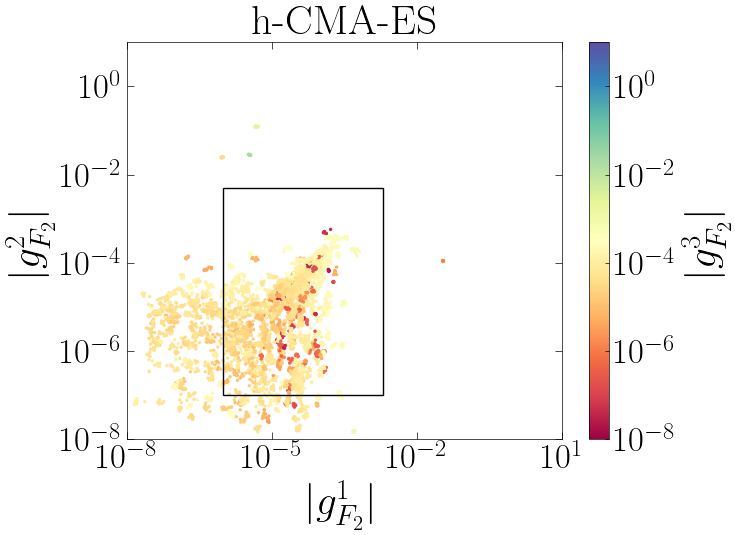}}
\subfloat{\includegraphics[width = 0.47\textwidth]{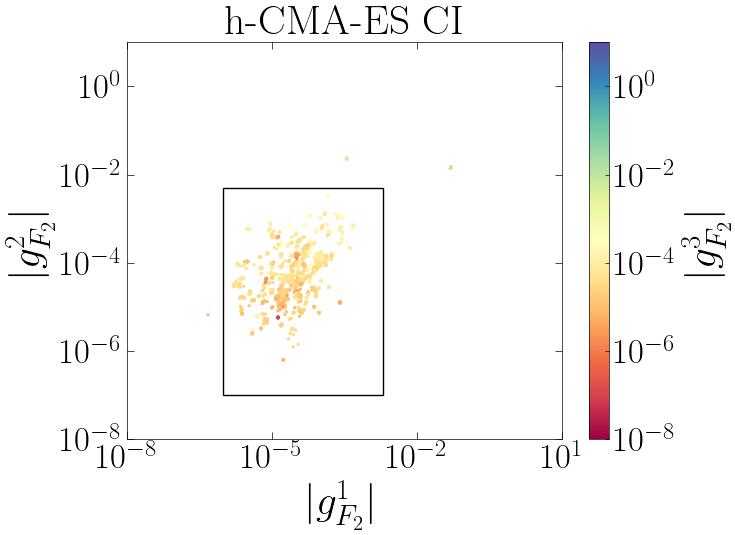}}\\
\subfloat{\includegraphics[width = 0.47\textwidth]{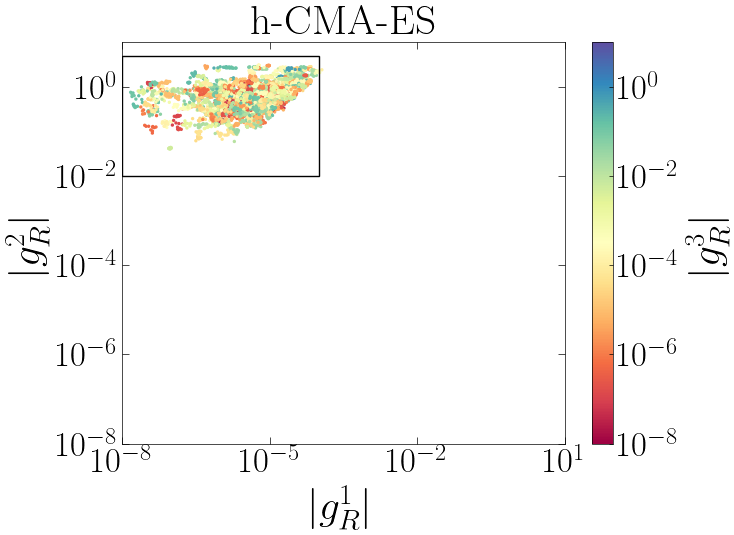}}
\subfloat{\includegraphics[width = 0.47\textwidth]{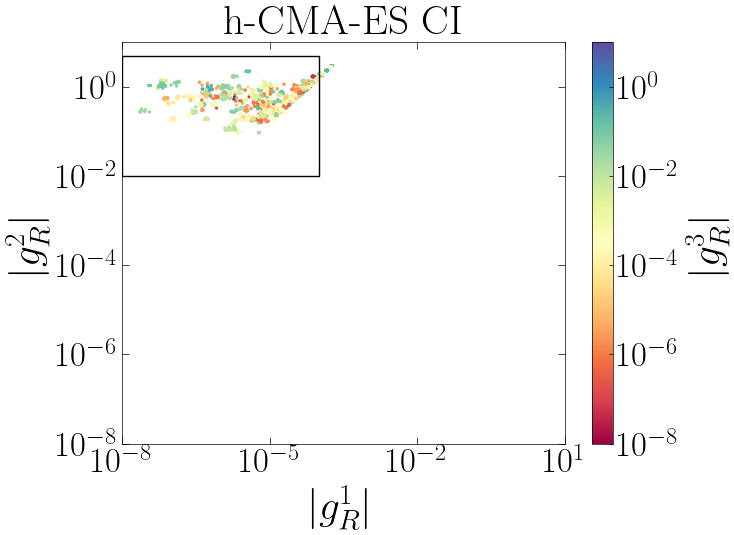}}
\caption{Absolute values for the components of the couplings $g_{F_2}$ (upper left and upper right) and $g_{R}$ (lower left and lower right) resulting from h-CMA-ES (left) and h-CMA-ES CI (right) scans. The box shows the region which encompasses the distribution obtained from MCMC in~\cite{Alvarez:2023dzz}.}
\label{fig:parameters_h_cmaes_appendix}
\end{figure}

\begin{figure}[H]
\centering
\subfloat{\includegraphics[width = 0.47\textwidth]{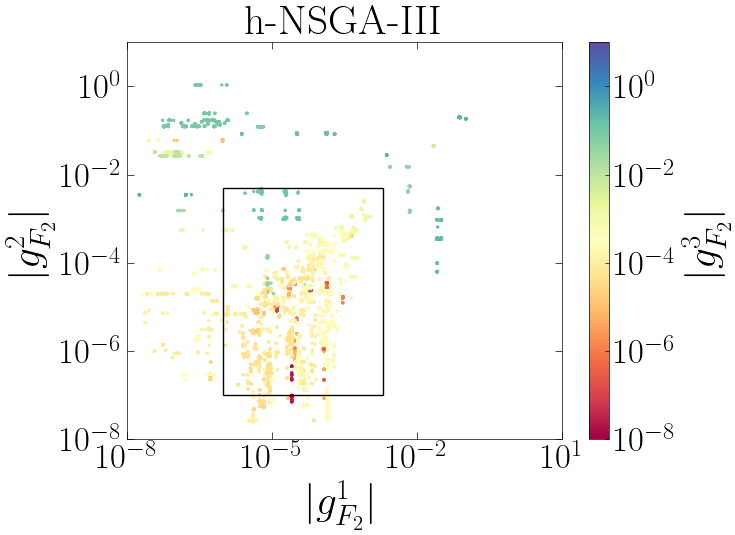}}
\subfloat{\includegraphics[width = 0.47\textwidth]{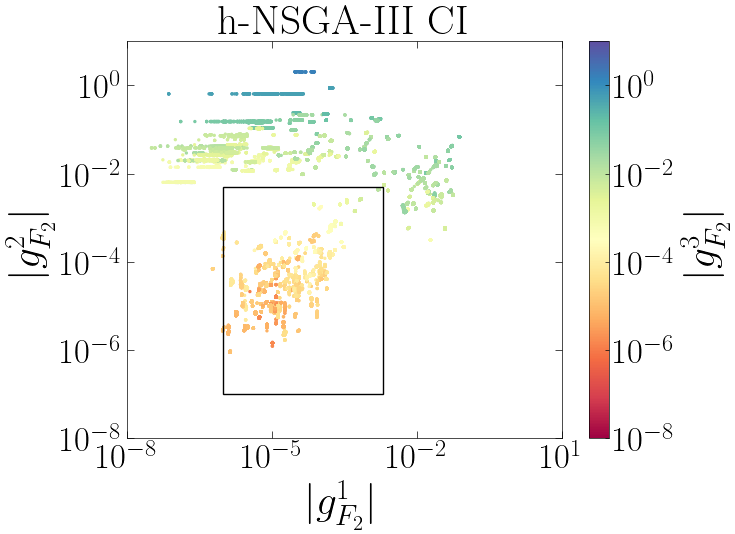}}\\
\subfloat{\includegraphics[width = 0.47\textwidth]{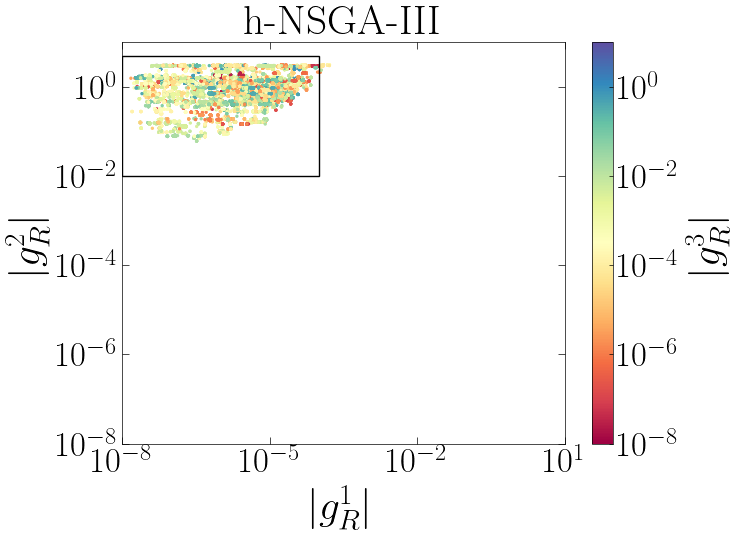}}
\subfloat{\includegraphics[width = 0.47\textwidth]{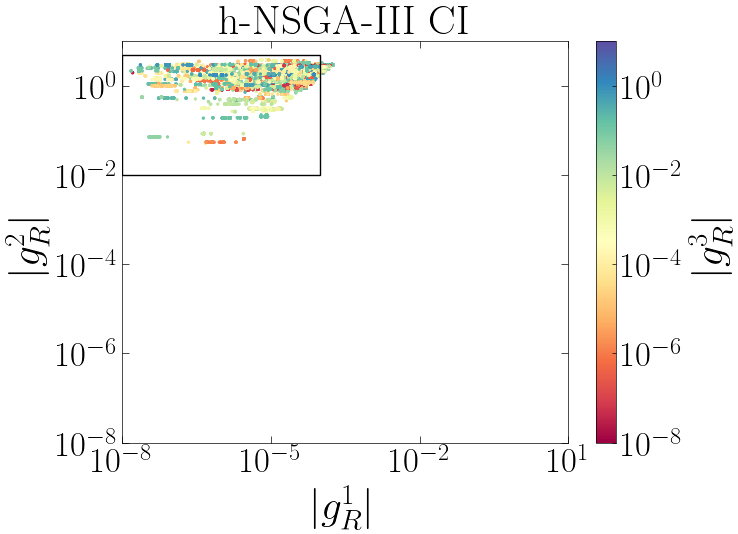}}
\caption{Absolute values for the components of the couplings $g_{F_2}$ (upper left and upper right) and $g_{R}$ (lower left and lower right) resulting from h-NSGA-III (left) and h-NSGA-III CI (right) scans. The box shows the region which encompasses the distribution obtained from MCMC in~\cite{Alvarez:2023dzz}.}
\label{fig:parameters_h_nsga3_appendix}
\end{figure}

\begin{figure}[H]
\centering

\subfloat{\includegraphics[width = 0.47\textwidth]{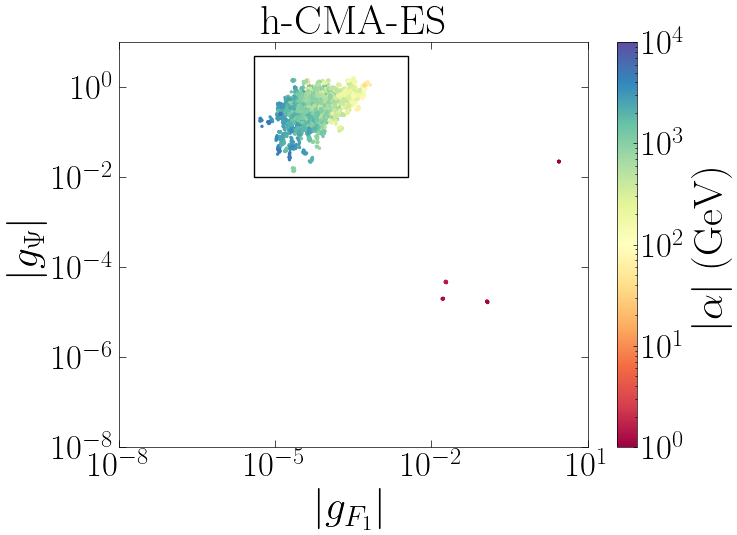}}
\subfloat{\includegraphics[width = 0.47\textwidth]{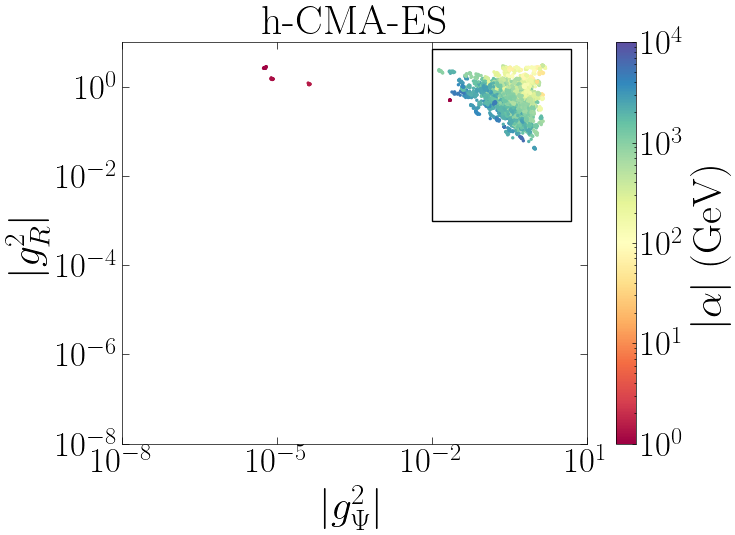}}\\
\subfloat{\includegraphics[width = 0.47\textwidth]{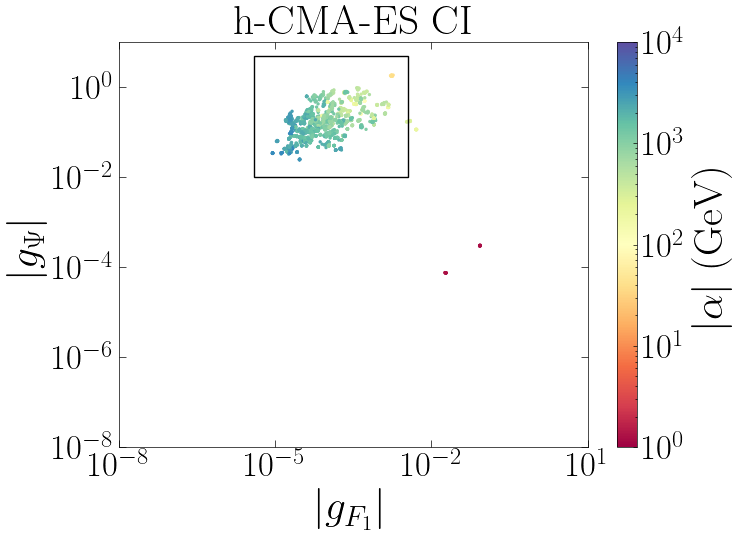}}
\subfloat{\includegraphics[width = 0.47\textwidth]{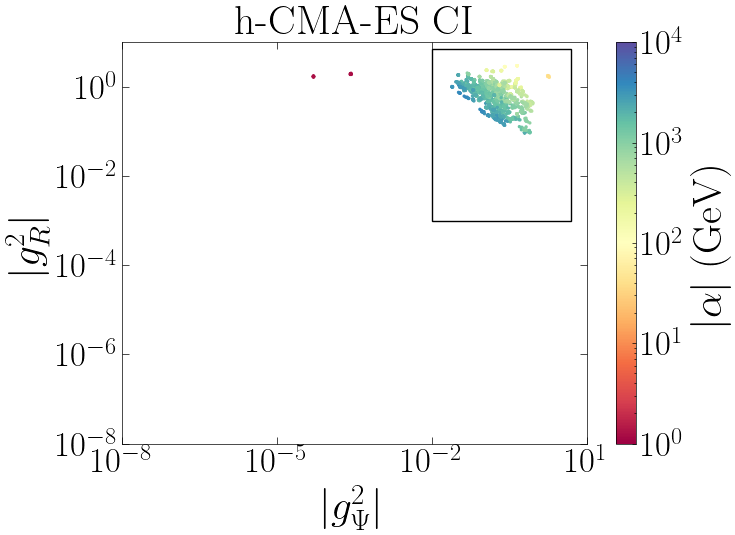}}\\
\caption{Trilinear coupling correlation with absolute values for the second component of the couplings $g_{\Psi}$ and $g_R$ (upper left and upper right) and the absolute value of $g_{\Psi}$ and $g_{F_1}$ couplings (lower left and lower right) resulting from h-CMA-ES (left) and h-CMA-ES CI (right) scans. The box show region which encompass distribution obtained from MCMC in~\cite{Alvarez:2023dzz}.}
\label{fig:parameter_alpha_appendix}
\end{figure}

\begin{figure}[H]
\centering
\subfloat{\includegraphics[width = 0.47\textwidth]{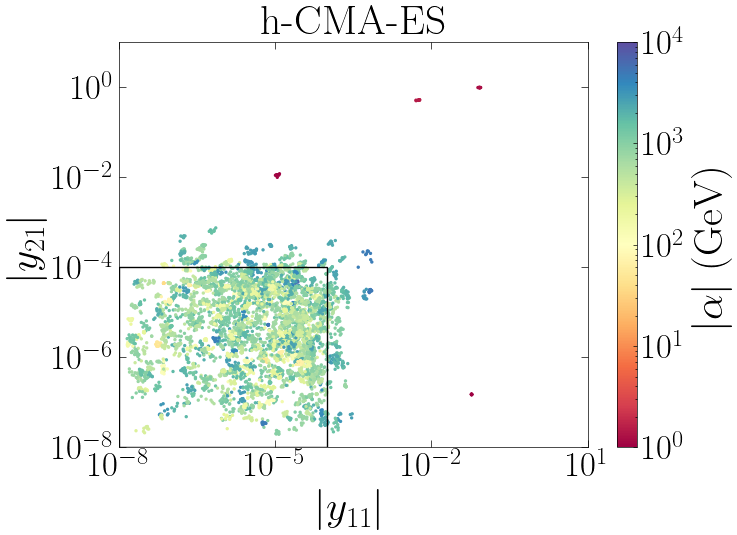}}
\subfloat{\includegraphics[width = 0.47\textwidth]{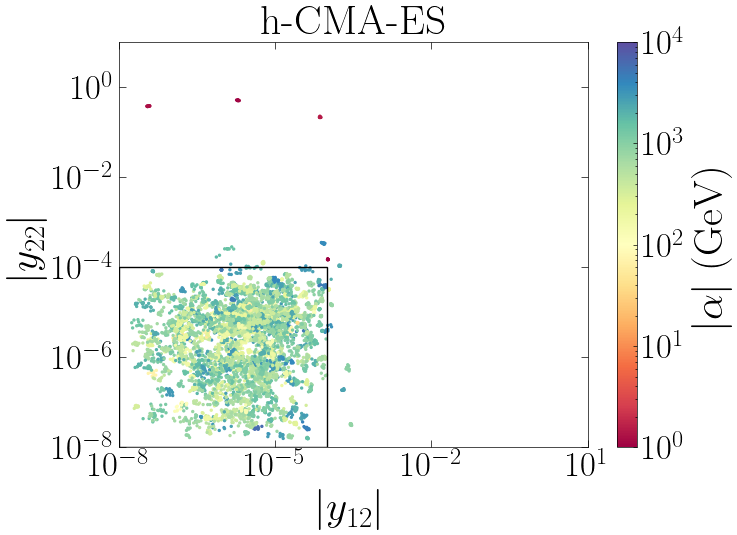}}\\
\subfloat{\includegraphics[width = 0.47\textwidth]{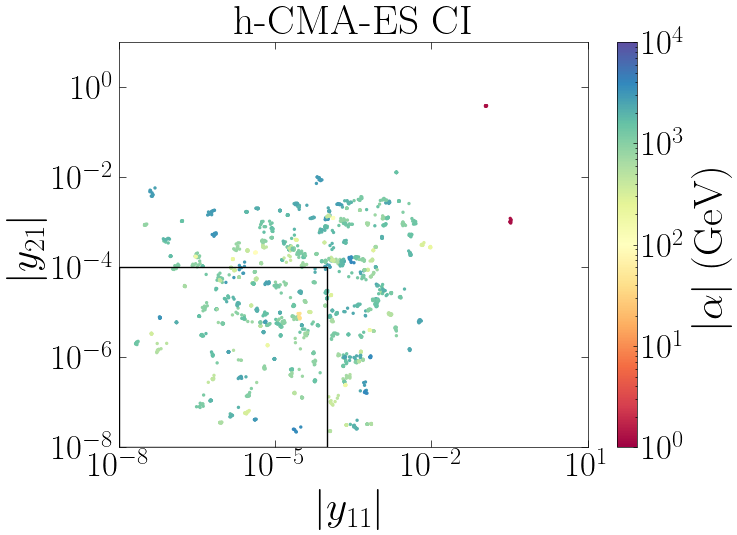}}
\subfloat{\includegraphics[width = 0.47\textwidth]{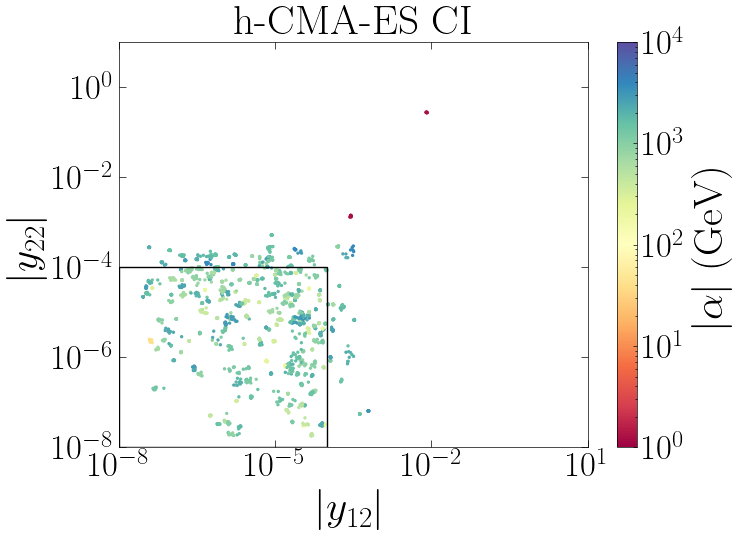}}\\
\caption{Trilinear coupling correlation with the absolute value of $y_{11}$, $y_{12}$ couplings resulting from h-CMA-ES (left) and h-CMA-ES CI (right) scans. The box show region which encompass distribution obtained from MCMC in~\cite{Alvarez:2023dzz}.}
\label{fig:parameter_y_appendix}
\end{figure}

\subsection{Dark Matter}

Here, for completion, we present the results for fermionic dark matter obtained by h-CMA-ES runs. The distribution of DM mass and its nature from h-CMA-ES results is presented in~\cref{fig:dm_mass_appendix}. Additionally, the mixing and mass split for fermionic DM and h-CMA-ES runs is shown in~\cref{fig:fermionic_dm_mixing_appendix}.

\begin{figure}[H]
\centering
\subfloat{\includegraphics[width = 0.47\textwidth]{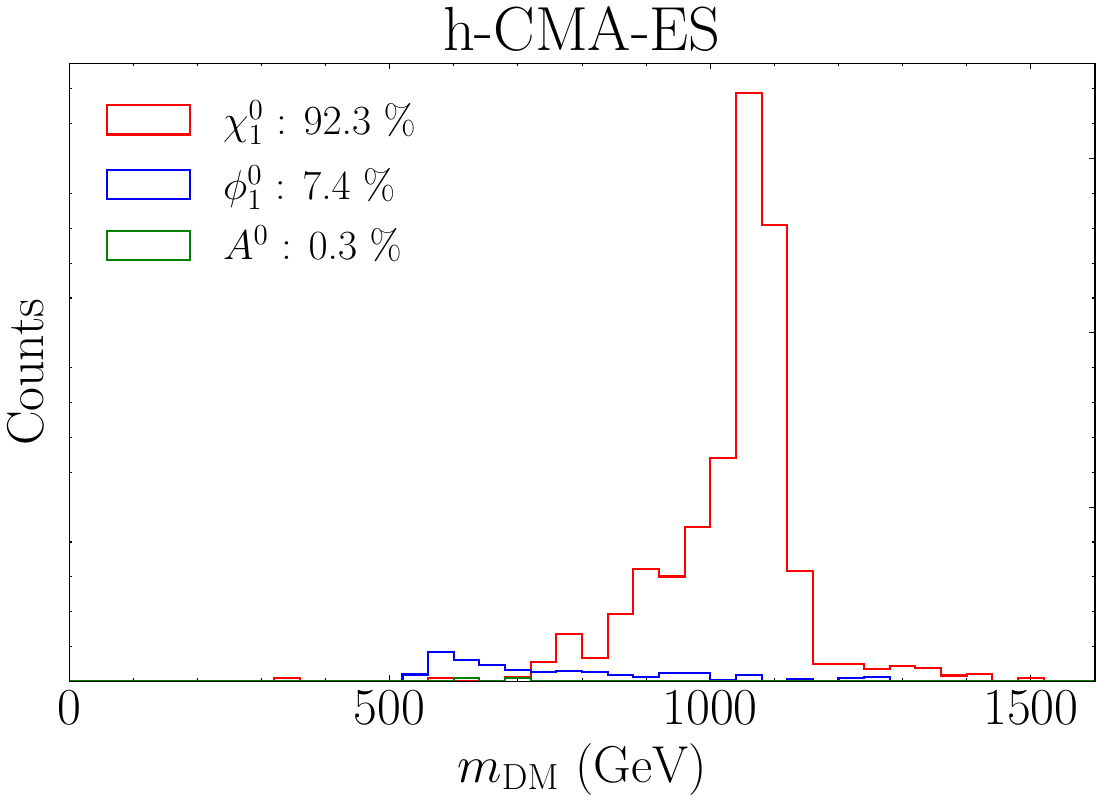}}
\subfloat{\includegraphics[width = 0.47\textwidth]{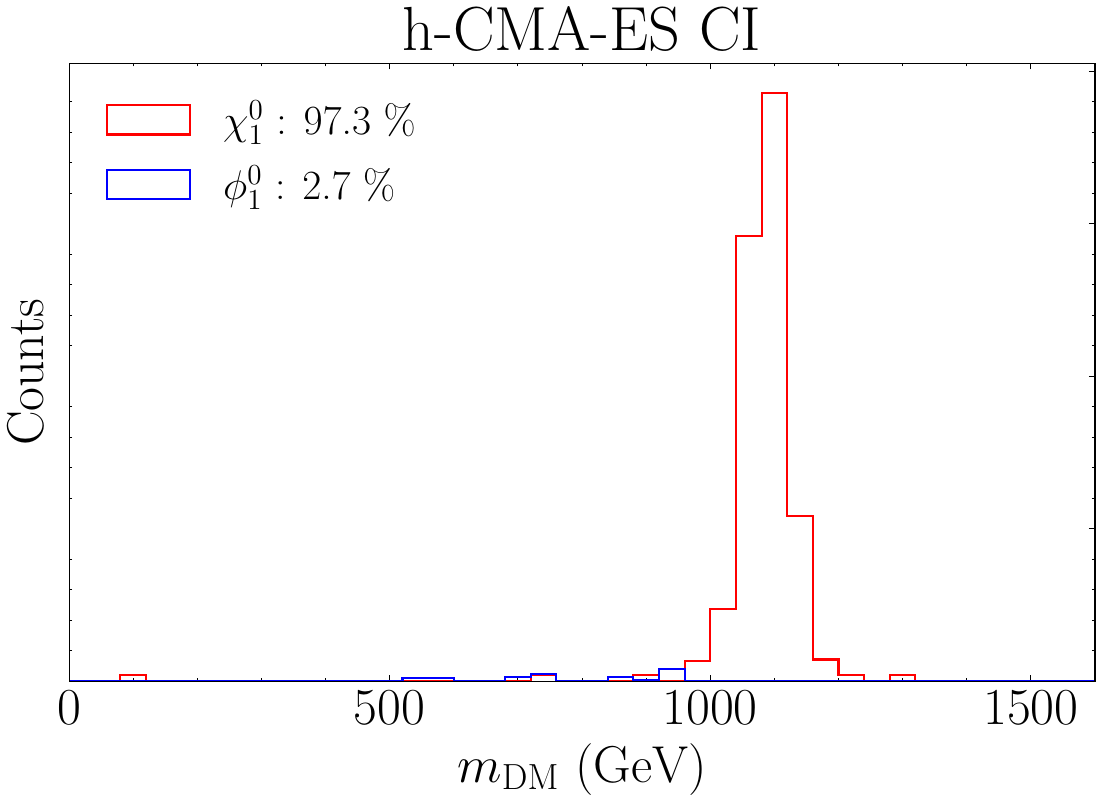}}\\
\caption{DM nature and mass for valid points found by h-CMA-ES scans. On the left not using CI parametrization and on the right using CI parametrization.}
\label{fig:dm_mass_appendix}
\end{figure}

\begin{figure}[H]
\centering

\subfloat{\includegraphics[width = 0.47\textwidth]{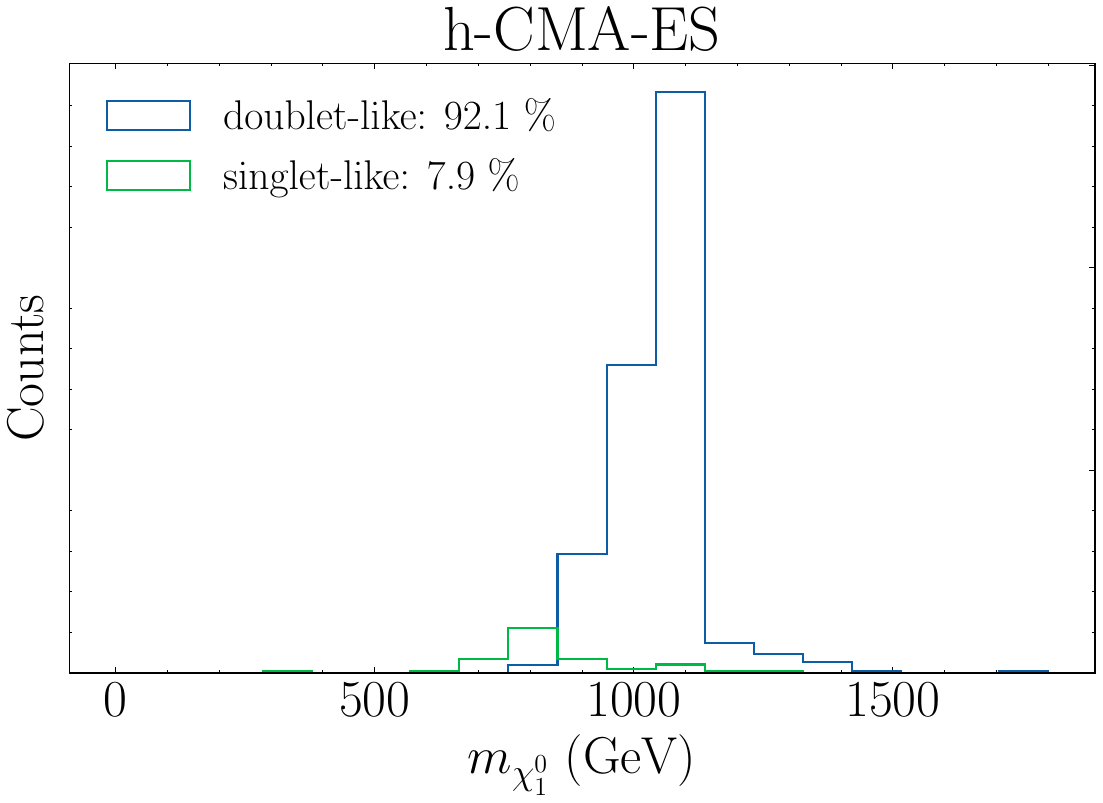}}
\subfloat{\includegraphics[width = 0.47\textwidth]{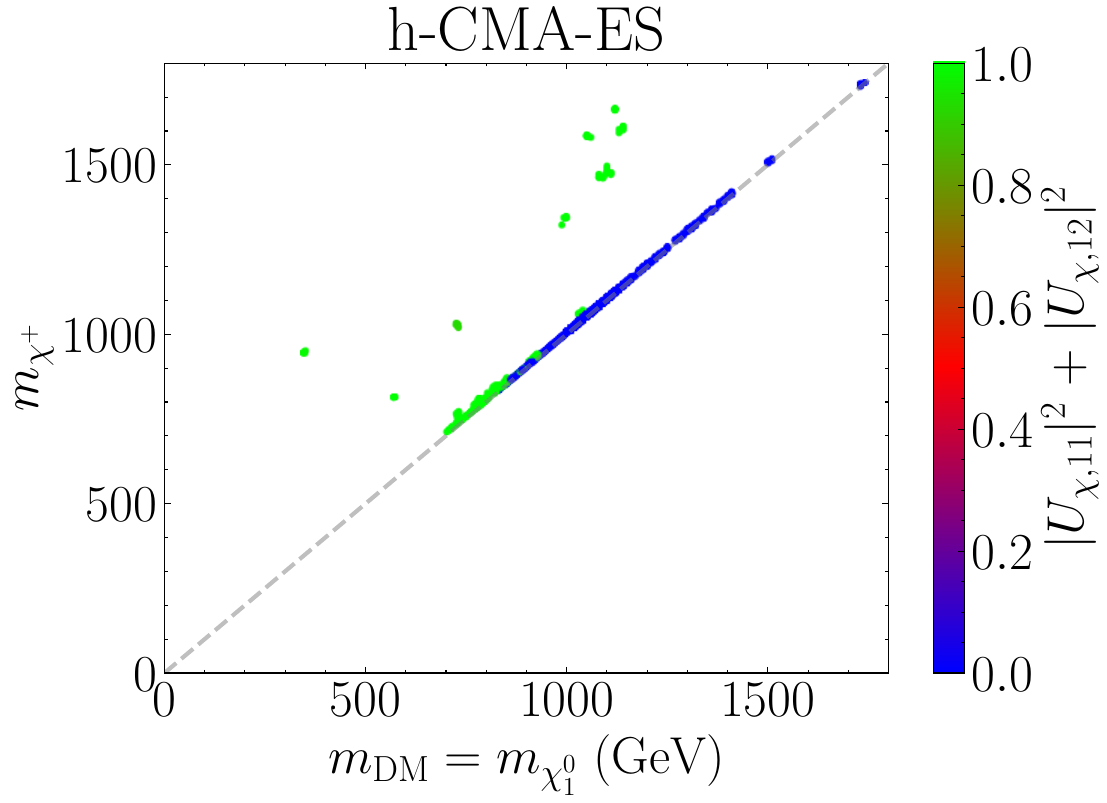}}\\
\subfloat{\includegraphics[width = 0.47\textwidth]{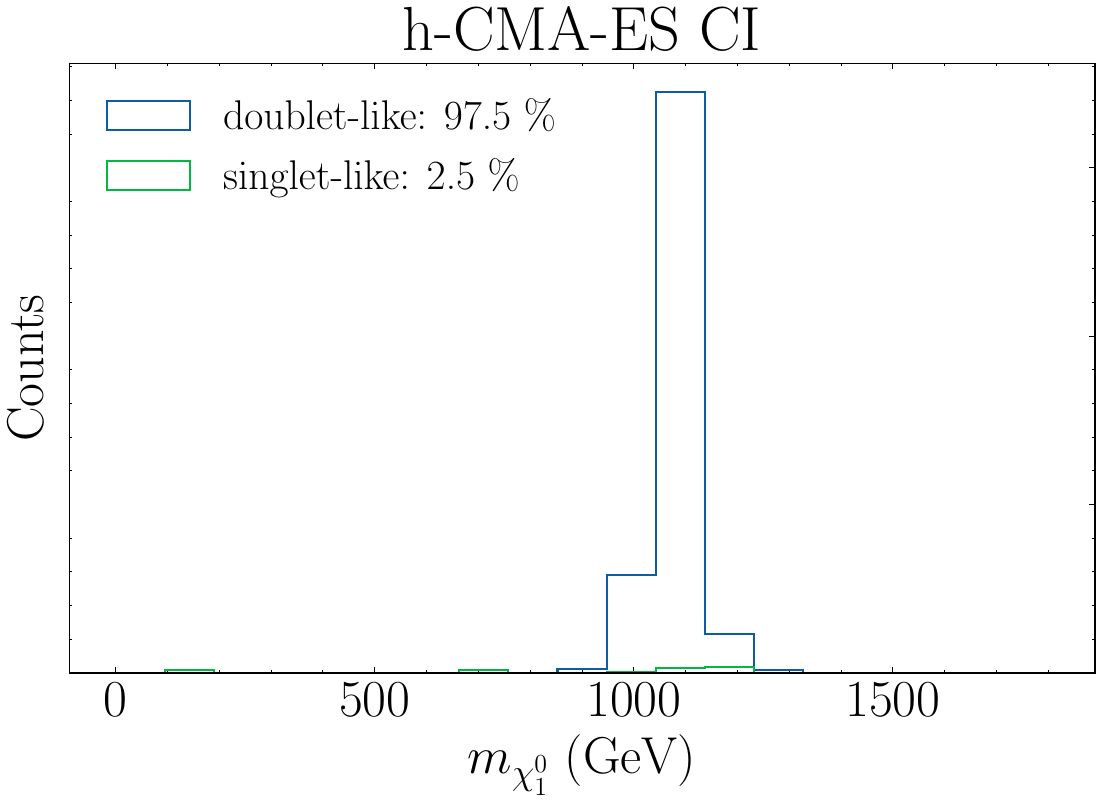}}
\subfloat{\includegraphics[width = 0.47\textwidth]{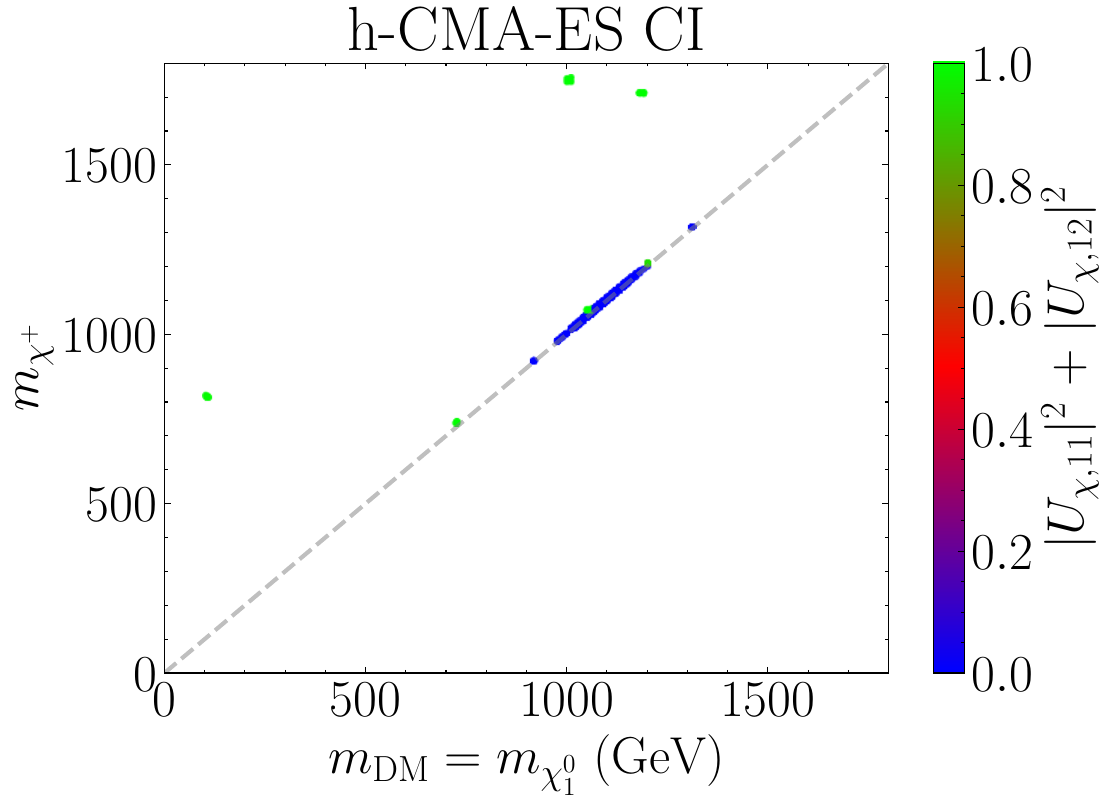}}\\

\caption{Mixing and mass split of fermionic DM for h-CMA-ES. Top: without CI parametrization, bottom: with CI parametrization. On the right: fermionic DM mass plotted against charged fermion state and colour coded by the singlet amount of the DM fermion.}
\label{fig:fermionic_dm_mixing_appendix}
\end{figure}

\section{Scalar Dark Matter Results}
\label{sec:appendix_scalar_dm}

For scalar DM, results from h-NSGA-III for the mixing with singlet and doublet states are presented in~\cref{fig:scalar_mixing} plotted against the DM and the mass of heavy charged scalar. Here, we see mostly doublet-like ($\sim 70\%$) with $\sim 30\%$ singlet-like solutions found. The algorithms did not find DM scalar with high mixing between singlet and doublet states. Solutions for scalar DM masses are mostly in the interval $500 \text{ GeV} \lesssim m_{\text{DM}} = m_{\phi^{0}_1} \lesssim 1000 \text{ GeV}$.

\begin{figure}[H]
\centering
\subfloat{\includegraphics[width = 0.47\textwidth]{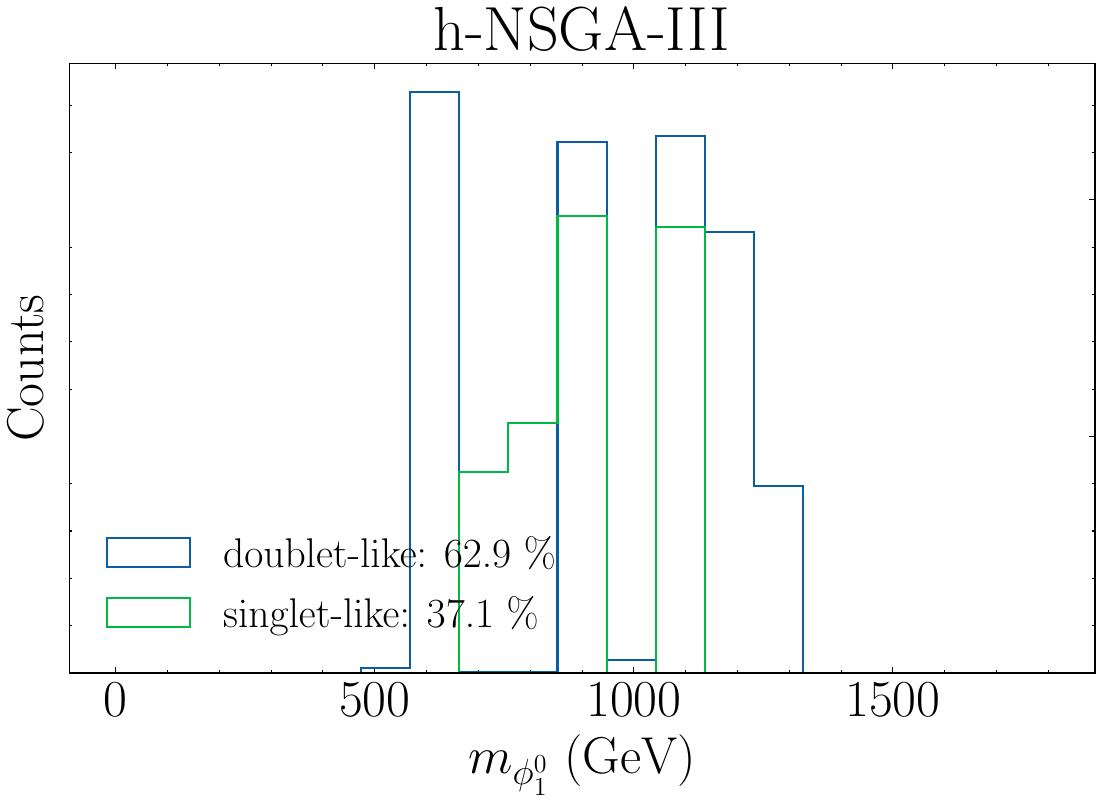}} \quad
\subfloat{\includegraphics[width = 0.47\textwidth]{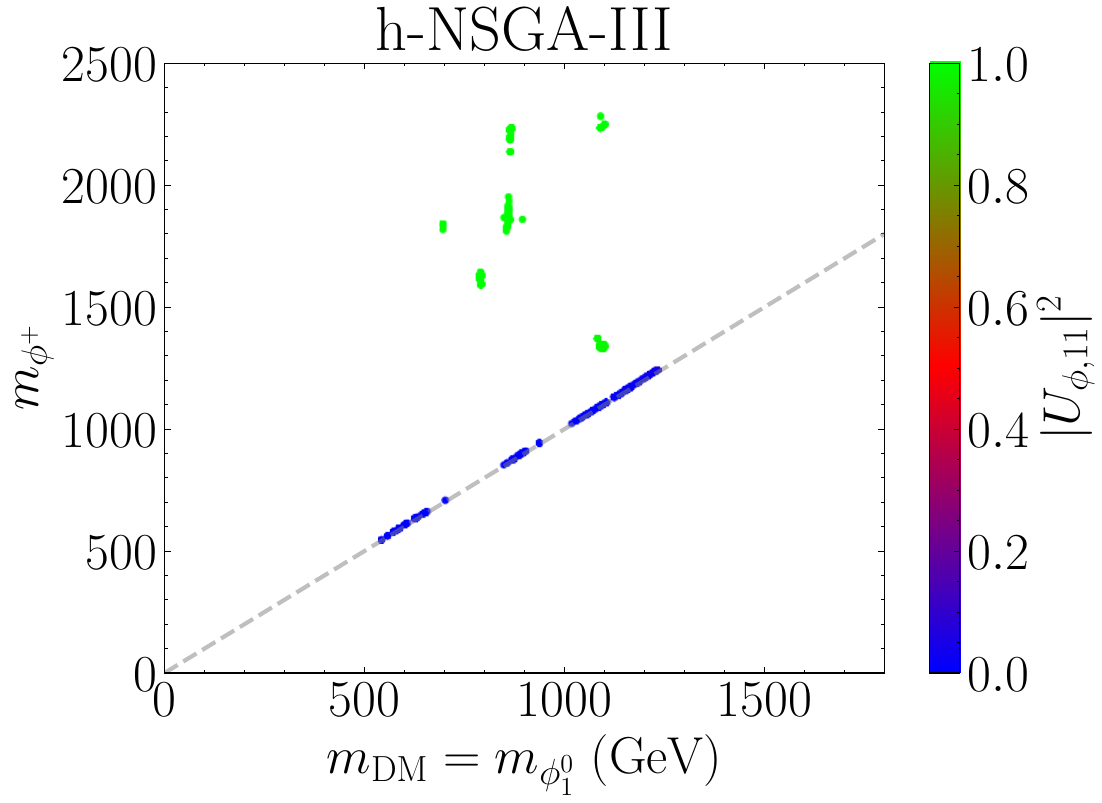}}\\
\subfloat{\includegraphics[width = 0.47\textwidth]{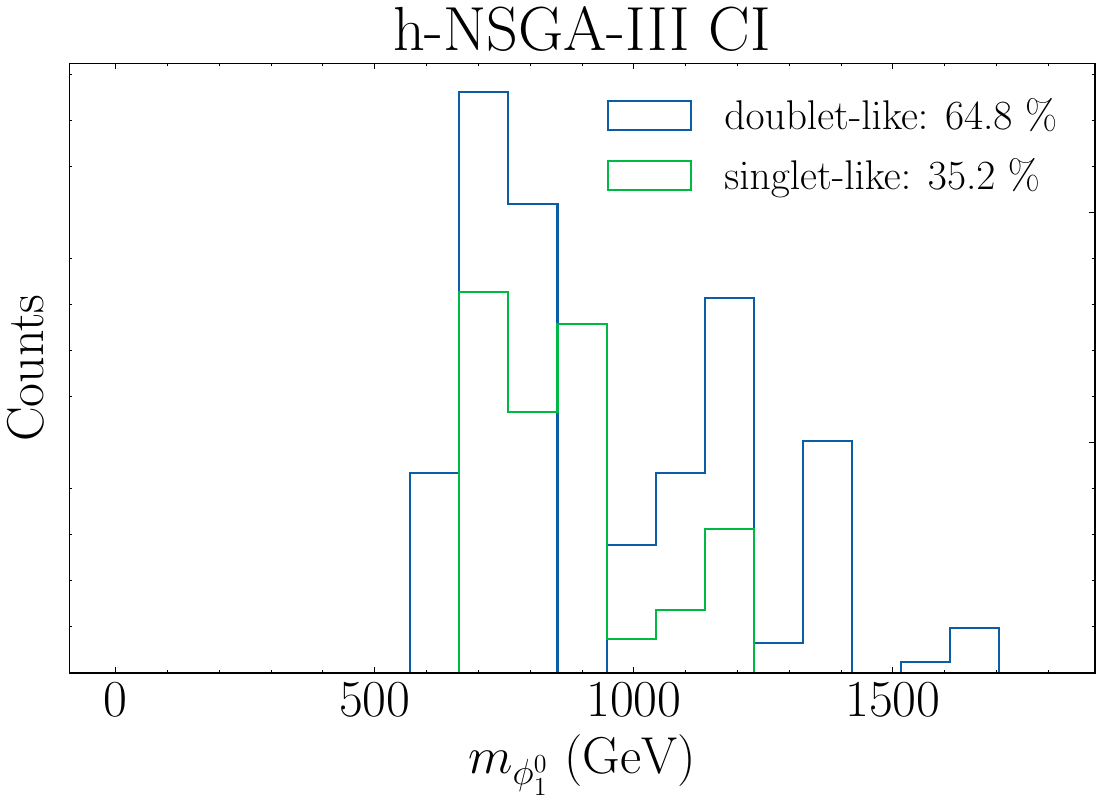}} \quad
\subfloat{\includegraphics[width = 0.47\textwidth]{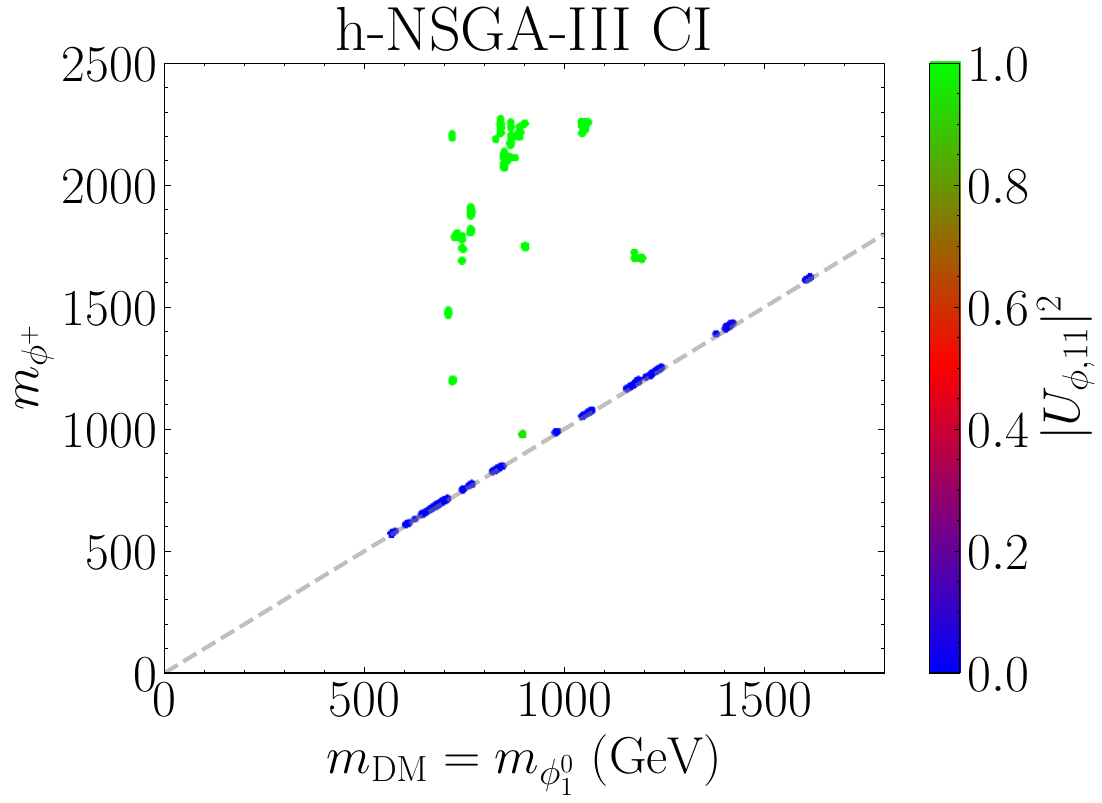}}
\caption{Mixing and split of scalar DM for h-NSGA-III. Top: without CI parametrization, bottom: with CI parametrization. On the right: scalar DM mass plotted against charged scalar state and colour coded by the amount of mixing between DM and the singlet neutral heavy scalar.}
\label{fig:scalar_mixing}
\end{figure}

\bibliographystyle{apsrev4-1}
\bibliography{references} 

\end{document}